\documentclass[12pt, final]{article}

    \usepackage[USenglish]{babel}
    \usepackage[utf8]{inputenc}
    \usepackage[T1]{fontenc}

    \usepackage{amsfonts}
    \usepackage{amssymb}
    \usepackage{amsmath}
    \usepackage{amsthm}
    \usepackage{dsfont}

    \usepackage{graphicx}
    \usepackage{booktabs}
    \usepackage{multirow}
    \usepackage{array}
    \usepackage{caption}
    \usepackage{subcaption}
    \usepackage[flushleft,online,para]{threeparttable}
    \usepackage{multirow}

    \usepackage{parskip}                       

    \usepackage{floatrow}
    \usepackage{tabularx}
        \newcolumntype{Z}{>{\centering\arraybackslash}X}
        \newcolumntype{L}{>{\raggedright\arraybackslash}X}

    \usepackage{dcolumn}
        \newcolumntype{d}[1]{D{.}{.}{#1}}

    \usepackage{rotating}                      
    \usepackage{lscape}
    \usepackage{pdflscape}

    \usepackage[round,longnamesfirst]{natbib}

    \usepackage{xcolor}
        \definecolor{darkblue}{rgb}{0,0,0.5}
    \usepackage{hyperref}
        \hypersetup{
            colorlinks = true,
            linkcolor = darkblue,
            citecolor = darkblue,
            pdfborder = 0 0 0,
            pdfdisplaydoctitle = true,
            pdfhighlight = /N,
            pdfpagelayout = OneColumn,
            pdfpagemode = UseNone,
            pdfstartview = {FitH},
            pdfauthor = {{FF \& DP}},
            pdftitle = {{ReinforcementLearning}},
            pdfsubject = {{}}
        }

    \usepackage[textsize=footnotesize, colorinlistoftodos, textwidth=4cm, obeyDraft]{todonotes}

    \usepackage{geometry}
    \usepackage{setspace}
    \usepackage[bottom, multiple]{footmisc}    

    \usepackage{verbatim}
    \usepackage[normalem]{ulem}     
    \usepackage{mathptmx}

    \usepackage[toc,page]{appendix}

    \newtheorem{definition}{Definition}
    \newtheorem{remark}{Remark}
    \newtheorem{proposition}{Proposition}
    
    \newtheorem{example}{Example}
    \newtheorem{lemma}{Lemma}
    \newtheorem{corollary}{Corollary}
    \newtheorem{assumption}{Assumption}

\usepackage{sectsty}

\usepackage{amsmath,amsthm,amssymb,amsfonts}
\usepackage{hyperref,lineno,graphicx,subcaption}
\usepackage{natbib}
\setcitestyle{authoryear}

\usepackage{xcolor}
\usepackage{geometry}
\usepackage{newtxmath}
\usepackage{pgfplots}
\usepackage{mathptmx}
\usepackage{titletoc}
\usepackage[flushleft,online,para]{threeparttable}

\numberwithin{equation}{section}
\numberwithin{proposition}{section}
\numberwithin{corollary}{section}
\numberwithin{lemma}{section}
\numberwithin{remark}{section}

\DeclareSymbolFont{newfont}{OML}{cmm}{m}{it}
\DeclareMathSymbol{\epsilon}{3}{newfont}{15}

\hypersetup{
	colorlinks,
	linkcolor={red!50!black},
	citecolor={blue!50!black},
	urlcolor={blue!80!black}
}

\title{Reinforcing RCTs with Multiple Priors \\ while Learning about External Validity \thanks{We thank Susan Athey, Marina Dias, Pat Kline, Chiara Motta, Chao Qin and Vira Semenova for useful comments. We also thank Mengsi Gao for excellent research assistance. Usual disclaimer applies.}}

\author{Frederico Finan \thanks{Department of Economics, 508-1 Evans Hall, Berkeley, California 94720-3880. Email: ffinan@berkeley.edu; and BREAD, IZA, NBER}\\UC Berkeley \and Demian Pouzo\thanks{Department of Economics, 508-1 Evans Hall, Berkeley, California 94720-3880. Email: dpouzo@econ.berkeley.edu }\\UC Berkeley }

\date{September 2024}

\begin{document}

\maketitle
\thispagestyle{empty}

\begin{abstract}
This paper introduces a framework for incorporating prior information into the design of sequential experiments. These sources may include past experiments, expert opinions, or the experimenter’s intuition. We model the problem using a multi-prior Bayesian approach, mapping each source to a Bayesian model and aggregating them based on posterior probabilities. Policies are evaluated on three criteria: learning the parameters of payoff distributions, the probability of choosing the wrong treatment, and average rewards. Our framework demonstrates several desirable properties, including robustness to sources lacking external validity, while maintaining strong finite sample performance.\\

\vspace{.2cm}

\end{abstract}

\newpage

\setcounter{page}{1}



\onehalfspacing

\section{Introduction}

Adaptive experiments offer several potential advantages over standard randomized controlled trials (RCTs), including shorter experiment duration and more optimized treatments. This can lead to more effective outcomes and lower experimental costs. Such experimentation methods are increasingly used across various fields, including clinical research and marketing. Although their application in economics is still limited, adaptive experiments hold significant promise for improving RCTs aimed at identifying optimal policy variants.

When deciding to conduct adaptive experiments, policymakers or researchers often have some prior knowledge about the effectiveness of the treatments. This information might come from past experiments, pilot studies, or expert opinions. Depending on its external validity, this prior knowledge could allow for less experimentation. This raises two relatively unexplored questions: how does one incorporate this previous knowledge into an adaptive experiment? And in doing so, will this allow the experiment to stop sooner without increasing the risk of choosing the wrong treatment? This paper provides a tractable, but novel framework for how to incorporate prior sources of information into the design of a sequential experiment. Our framework is sufficiently general so that priors can include previous experiments, expert opinions, or the experimenter’s own introspection.

\paragraph{Setup} We consider a policymaker who has to decide how to assign a set of treatments sequentially to an eligible population and when to stop the experiment. Subjects arrive in stages and at the beginning of each stage, the policymaker must first decide whether to stop the experiment. If she stops the experiment, she then assigns what she thinks is the best treatment to all subsequent subjects. But if the policymaker decides to continue the experiment, she assigns treatment just to the new arrivals and then moves onto a new stage. At each stage, the policymaker knows the history of previous treatment assignments and the corresponding realized outcomes, but does not know the probability distributions of potential outcomes, which she tries to learn about using the observed data. The policymaker does, however, have prior information about these distributions, which can arise from many sources, including her own introspection and knowledge, previous experiments, or expert opinions.


As the policymaker gathers more data from own experiment, she uses Bayes' rule to update each of her prior sources and then takes a weighted average of each source's posterior where the weights depend on how well the sources fit the observed data. On the basis of these beliefs, the policymaker then decides whether to stop the experiment and which treatment to assign. By incorporating potentially useful information, our policymaker may be able to stop the experiment sooner, thereby generating efficiency gains without increasing the risk of adopting the incorrect treatment.

In settings, such as this one, in which the policymaker must learn the truth, it is common not to use the optimal assignment rule. This rule (i.e. the one that maximizes her \emph{subjective} payoff) can have undesirable properties, such as failing to learn the correct treatment effects or being hard to compute and implement.\footnote{To illustrate this point, consider a simple model with two treatments, A and B. For simplicity, suppose the policymaker knows that the average effect of treatment A is zero. The policymaker, however, does not know the true average effect of treatment B and incorrectly believes that it is negative. In this simple example, an optimal policy is to never assign treatment B; and without feedback, the policymaker will never update her (incorrect) prior that treatment B is bad. While this assignment rule is optimal from the perspective of the policymaker, it is undesirable from an objective point of view. This example also illustrates the need for experimentation because such a situation would not occur if the policy rule involved some degree of experimentation.} As a result, the literature on multi-armed bandits have studied different heuristic rules such as $\epsilon$-greedy \citep{Watkins:1989} and Thompson Sampling \citep{Thompson:1933} and its refinements (e.g. Upper Confidence Bounds \citep{LaiRobbins:1985}, or exploration sampling \citep{KasySautmann:2021,Russo2016}). We take a different approach and study a large class of policy rules that encompass, among others, the aforementioned examples. Importantly, we find that the only feature of the policy rule that matters for performance is the exploration structure -- a sequence quantifying the amount of experimentation that occurs under a given policy rule at each stage of the experiment.


\paragraph{Performance Criteria} Given that optimality from the perspective of the policymaker may not be desirable,  we evaluate our class of assignment rules on the basis of three regularly-used outcomes that are considered to be important from the point of view of an outside observer. Specifically, we explore whether the policymaker learns the true average treatment effects and at what rate. We also consider the likelihood that the policymaker does not choose the most beneficial treatment arm when deciding to stop the experiment. The third outcome measures the average payoff of the policymaker. Unlike the other two criteria, which are statistical in nature (i.e. they describe statistical properties of the experiment and its assignment rule), this outcome captures how much subjects benefit in net from the experiment both during and afterwards. When evaluated along these criteria, we can show, both theoretically and via Monte Carlo simulations, that our setup exhibits several nice finite sample properties, including robustness to incorrect priors.

\paragraph{Main Findings} We show that our policymaker will learn the average treatment effects, in the sense that her posterior mean of the potential outcome distribution concentrates around the true mean, and it does so at a rate of $1/(\sqrt{t} h^2_t)$, where $t$ is the number of stages and $h_{t}$ is the amount of experimentation. That this concentration result holds was not, ex ante, obvious: in contrast to a standard randomized control trial setting, the policy functions in our setup are quite general and can depend on the entire history of play, thus creating time-dependence in the data. Nevertheless, by exploiting the concept of the exploration structure and Azuma-Hoeffding type concentration inequalities for Martingales, we not only obtain the rate of $1/(\sqrt{t} h^2_t)$, but we can also characterize and quantify how this rate depends on the initial parameters of the setup.

Importantly, we are able to show that our aggregation method exhibits an attractive robustness property: Our model discards sources that do not extrapolate well to the current experiment, thereby exhibiting robustness to sources of information that are not externally valid. To aggregate her multiple priors, our policymaker uses a Bayesian approach that weights each prior according to the posterior probability that a particular model best fits the observed data within the class of sources being considered. Thus, if relative to the other priors, one of the policymaker's priors (about the average effects of the treatments) puts ``low probability'' on the true mean, then our approach will place close to zero weight on this source when aggregating across sources. Consequently, this prior will have little to no effect on the policymaker's decisions or the learning rate. Similarly, sources whose priors put high probability on the truth receive higher weights that can approach one in finite samples. This feature gives rise to an oracle type property wherein our concentration rates are close to those associated to the best source (the one with priors more concentrated around the truth) provided the other sources are sufficiently separated from this one.


Besides assigning treatments, our policymaker also has to consider when to stop the experiment and subsequently, what treatment to adopt. Both the duration of the experiment and adopting the correct treatment can have important welfare consequences. In our setup, the policymaker works with a class of stopping rules that stops the experiment when the average effect of a treatment is sufficiently above the others.  This class of rules resembles the standard test of two means, but takes into account the fact that the data are not IID and accounts for the presence of prior information.  Of course, whenever we stop an experiment, we worry about the possibility of making a mistake (i.e. not choosing the most beneficial treatment). We characterize the bounds on the probability of making a mistake for our setup. We show that these bounds decay exponentially fast with the length of the experiment, and that they are non-increasing in the degree of experimentation and in the size of the treatment effects. Moreover, we propose stopping rules that for any given tolerance level will yield a lower probability of making a mistake.

Finally, we also compute bounds for the rate at which the average observed outcomes converges to the maximum expected outcome. We show that the rate of convergence for these bounds are governed by an ``exploitation versus exploration'' trade-off. If we increase the degree of experimentation (less exploitation, more exploration) our data become more independent and the underlying uncertainty decreases. However, by exploring more, we are also increasing the bias associated with not choosing the optimal treatment. Unfortunately, these bounds are sufficiently complicated that we cannot characterize analytically the ``optimal'' degree of experimentation. Nevertheless, the results do suggest that pure experimentation (as in the case of an RCT) is unlikely to be optimal, and we verify this numerically in a series of simulations.

\paragraph{Application to Debt Refinancing} To further illustrate our procedure, we ran an experiment with the largest private bank in Argentina to enhance the marketing of their debt-refinancing program. This initiative came at a critical time, as Argentina was experiencing a surge in household debt, exacerbated by high inflation and interest rates. The bank aimed to engage over 300,000 delinquent clients, primarily from lower-income households, through targeted email campaigns. In February 2024, the bank conducted an RCT with a sample of around 15,491 clients to determine if displaying monthly payments instead of the high-interest rates in emails would increase engagement. Contrary to expectations, clients were more likely to click on the link when the interest rate was displayed, with response rates varying significantly across different regions of Argentina.

In June 2024, as interest rates started to decline, the bank launched a second, adaptive experiment using our algorithm. The experiment was conducted in three regions and used the results of the previous experiment  as initial priors.  The results showed that our multi-prior approach proved effect as fewer emails were needed to achieve significant results compared to traditional randomized controlled trials. In the capital city of Buenos Aires, for instance, the experiment was stopped after 294 emails, far fewer than would be needed in traditional methods. 

\paragraph{Contributions to the Literature} Our paper relates to three strands of the literature. First, we speak to an extensive multi-disciplinary literature on adaptive experimental design. Much of the focus of this literature has been on the multi-arm bandit problem, which considers how best to assign experimental units sequentially across treatment arms. Depending on the objective function, numerous studies have proposed a variety of alternative algorithms that, on average, outperform the static assignment mechanisms of traditional RCTs.\footnote{See \cite{AtheyImbens:2019} for a survey of machine learning techniques as it applies to experimental design and problems in economics.}  In this paper, we focus less about constructing an alternative policy function than about on how to introduce information from different sources for a given class of policy functions. By doing so, the fundamental `earn vs learn' tradeoff that characterizes the multi-arm bandit problem is not only a function of sampling variability in target data, but also uncertainty over the data generating process of the source data. To our knowledge, this is the first paper to introduce multiple priors into the design of an adaptive experiment.

Much of the literature on multi-armed bandits has focused on deriving bounds on expected regret for specific solution heuristics.\footnote{For example, related to bounds on regret, see \cite{AG} and \cite{russo2016information} for regret bounds for Thompson sampling; or \cite{cesa2006prediction} for a broad discussion about multi-armed bandit problems and bounds on regret.} Instead, we focus on alternative performance criteria, such as average outcomes, the probability of making a mistake, and concentration rates for posterior means, which to the best of our knowledge have not been formalized in a multi-prior multi-arm Bayesian bandit framework.\footnote{Average outcomes is related to regret. However, we do not provide bounds for the expected value, but instead provide exponential inequalities for the tail probability. There are  classical results related to the probability of making a mistake stemming from the foundational work by \cite{chernoff} and \cite{wald}. }   Moreover, the results we derive are for a general class of solution heuristics, not a specific one. For these reasons, even though we do not view the technical results as the primary contribution of the paper, we do believe that they might be of independent interest even in standard multi-arm bandit problems. Furthermore, we view our paper as complementary to this existing literature, as techniques tailored for particular solution heuristics can be combined with our multi-prior Bayesian setting to obtain sharper theoretical guarantees.

By introducing issues of externality validity into the multi-arm bandit problem, our study also connects to the literature on measuring the generalizability of experiments.  In general, scholars have taken three approaches for assessing external validity. One common approach is to measure how well treatment effect heterogeneity extrapolates to `left out' study sites. Under the assumption that study site characteristics are independent of potential outcomes, a number of studies applying alternative estimators have interpreted the out-of-sample prediction errors as a measure or test of external validity.\footnote{ See for example \cite{Dehejia_etal:2021}, \cite{Stuart_etal:2011}, \cite{Buchanan_etal:2018}, \cite{ImaiRatkovi:2013}, \cite{Hotz_etal:2005} and the references cited therein.}  A related approach uses local average treatment effects across different complier populations to test for evidence of external validity (e.g. \cite{angristfv:2013, Kowalski:2016, Bisbee_etal:2017}). The general idea being that if differences in observable characteristics across subgroups explain differences in treatment effect heterogeneity then we can make some claim for external validity. A third common approach adopted in the meta-analysis literature is the use of hierarchical models to aggregate treatment effects across different study sites. A byproduct of this framework is a ``pooling factor'' across study sites that has a natural interpretation of generalizability. The factor compares the sampling variation of a particular study site to the underlying variation in treatment heterogeneity: the higher the measure, the larger the sampling error and the less informative the study site is about the overall treatment effect (e.g. \cite{Vivalt:2020}, \cite{GelmanCarlin:2014}, \cite{GelmanPardoe:2006}, \cite{Meager:2020}).\footnote{The first and third approaches --- and hence our paper as well --- relates to a burgeoning sub-branch of machine learning called transfer learning (see \cite{Pan:2010} for a survey) wherein a model developed for a task is re-used as the starting point for a model on a second task. Even though elements of our problem are conceptually similar, to the best of our knowledge both our setup and approach are different to those considered in transfer learning.}

Our paper contributes to these approaches in two ways. First, we provide a formal definition for a subjective Bayesian model to be externally invalid using a Kullblack-Leibler (KL) divergence criteria. Importantly, our definition offers a way to quantify or rank external invalidity among models. Second, we provide a link between this ranking of external invalidity and our aggregation method. We show that, as $t$ diverges, the weights are only positive for the least externally invalid models, allowing us to interpret these weights as measures of external validity.

While it is natural to interpret our measure of external validity in the context of other experiments, our setup is agnostic as to the source of the information and its level of uncertainty. Whether the policymaker’s priors come from previous experiments, observational studies, or expert opinions is immaterial for our setup. In this respect, our study also relates to a nascent, but growing literature measuring the extent to which experts can forecast experimental results (e.g. \cite{DellaVignaPope:2018,DellaVigna_etal:2020}). Our paper provides a method for incorporating these forecasts in the design of policy evaluations in a manner that is robust to misspecified priors or behavioral biases \citep{VivaltColville:2021}.

Finally, our paper is related to the Hierarchical Bayes methodology and, albeit tangentially, to the ambiguity aversion literature. We employ the multi-prior Bayesian updating formalism from the ambiguity aversion literature in dynamic settings (e.g., \cite{al2009ambiguity,epstein2007learning}) to describe the problem of a policymaker (henceforth, PM) with access to different sources of information. However, the literature has emphasized dynamic consistency within optimal, non-myopic agents. Our paper is unrelated to these issues, as we are not concerned with optimality, and therefore not with consistency either. As explained above, it is not obvious to us that optimality (from the point of view of the PM) is a desirable property in this context. Our paper simply aims to provide a method for incorporating prior information under a class of heuristic procedures, which encompasses commonly used methods and offers certain appealing theoretical guarantees, such as learning the true treatment effects and controlling the probability of making a mistake when stopping the experiment.

Our methodology is also related to the hierarchical Bayesian methodology. Our PM acknowledges (model) uncertainty but is not averse to it, as she \emph{averages} across different models. In Appendix \ref{Appsec:Hbayes}, we show and discuss a type of \emph{certainty equivalence result}, which states that the updating problem we propose is mathematically equivalent to a certain empirical hierarchical Bayesian model (EHB). This result relies on an insight akin to the classical certainty equivalence result wherein, in a framework with risk, a risk-neutral agent makes decisions as if there is no risk. However, an important feature of our updating problem that differentiates it from classical hierarchical Bayesian models is that the weights are updated iteratively as more information arises, shifting towards models that best fit the data.\footnote{This last observation presents a dichotomy between learning—using relevant information from the data—and uncertainty aversion—being robust to potential model misspecification. In Appendix \ref{Appsec:Hbayes}, we discuss this further and present a possible extension that merges these ideas.} We prefer the "model uncertainty" interpretation to the EHB because we believe it better describes the motivation behind the PM's problem.\footnote{Another, perhaps less important, reason to prefer the "model uncertainty" interpretation is that the EHB interpretation is only valid if the PM is uncertainty neutral; it breaks down if the PM is ambiguity averse. See Appendix \ref{Appsec:Hbayes} for more details.}

\paragraph{Organization of the Paper} The structure of the paper proceeds as follows. In Section \ref{sec:setup}, we set up the problem. We present two versions of the setup, one for the general model and the other for a  Gaussian model. In Section \ref{sec:analytical_results}, we provide analytical results for the  Gaussian model.  We then illustrate the main analytical results by simulation in Section \ref{sec:simulations}. In Section \ref{Sec:App}, we illustrate our procedure using data from a charitable giving experiment. Section \ref{sec:conclusions} concludes.

\section{Setup}\label{sec:setup}

In this section, we describe the problem our policymaker (PM) aims to solve. Our PM's problem consists of three parts: the experiment, the policy functions, and the learning framework.

\subsection{The Experiment} The PM has to decide how to assign a treatment to a given unit (e.g. individuals or firms) and when to stop the experiment. We define an experiment by a number of instances $T \in \mathbb{N}$; a discrete set of observed characteristics of the unit, $\mathbb{X}$; a set of treatments $\mathbb{D} : = \{ 0,...,M\} $; and the set of potential outcomes. For now, we do not include a payoff function.  

At this point, it is useful to introduce some notation. For each $(d,x) \in \mathbb{D} \times \mathbb{X}$, let $Y_{t}(d,x) \in \mathbb{R}$ denote the potential outcome associated with treatment $d$ and characteristic $x$ in instance $t$; let $Y_{t}(d): = (Y_{t}(d,x))_{x \in \mathbb{X}}$.  Let $D_{t}(x) \in \mathbb{D}$ be the treatment assigned to the unit with characteristic $x$ in instance $t$. The \emph{observed} outcome of the unit with characteristic $x$ in instance $t$ is $Y_{t}(D_{t}(x),x)$.

The experiment has the following timing. At each instance, $t \in \{1,...,T\}$, the PM is confronted with $|\mathbb{X}| < \infty$ units, one for each value of the observed characteristic. 
At the beginning of the period, the PM decides whether to stop the experiment.
\begin{itemize}
  \item If the PM decides to stop the experiment,
   \begin{itemize}
     \item she chooses a treatment assignment at instance $t$ that will be applied to all subsequent units.
   \end{itemize}

  \item If the PM does not stop the experiment,
  \begin{itemize}
    \item she chooses a treatment assignment for each unit $x$ at time $t$.
    \item Nature draws potential outcomes, $Y_{t}(d,x)$, for each unit.
    \item The PM only observes the outcome corresponding to the assigned treatment, i.e. $Y_{t}(D_{t}(x),x)$.
  \end{itemize}
\end{itemize}

We impose the following restriction on the data generating process for the potential outcomes.
\begin{assumption}\label{ass:IID}
	For each $t \in \{1,...,T\}$ and each $x \in \mathbb{X}$, $(Y_{t}(d,x))_{d \in \mathbb{D} }$ is drawn IID and $Y(d,x) \sim P(\cdot | d,  x) \in \Delta(\mathbb{R})$.
\end{assumption}


	Assumption \ref{ass:IID} implies that units do not self-select across instances, i.e., the types of unit treated in instance $t$ are the same as the types treated in instance $t'$. Implicit in this assumption and framework is also the absence of any selection into treatment or attrition, which is reasonable to assume for most experimental settings.
	
	Finally, the assumption that the PM is confronted with $|\mathbb{X}| < \infty$ units, one for each value of the observed characteristic, is made out of convenience: it is straightforward to extended our theory to situations where the PM receives a random number of units, including zero, for each characteristic, provided this random number is exogenous. However, to extend the assumption of discrete covariates --- $|\mathbb{X}| < \infty$ ---  to continuous ones is non-trivial. For learning in multi-arm bandits with continuous covariates, we refer the reader to \cite{dimakopoulou2017estimation} and references therein, as well as to \cite{qinrusso} where the authors adapt the Thompson Sampling algorithm to handle a potentially non-stationary sequence of covariates influencing the arms' performance.
	

\paragraph{The parameter of interest.} For each instance $t$, the learning setup gives raise to a subjective PDF over each of the $(d,x)$-outcomes given by $\int p_{\theta}(\cdot) \mu^{\alpha}_{t}(d,x)(d\theta)$ that the PM uses to form recommendations and decisions. While this setup is sufficiently general to allow for any parameter of interest, we focus on the average treatment effect setup wherein the PM wants to learn
$$\theta(d,x): = E_{P(\cdot|d,x)}[Y(d,x)],~\forall (d,x) \in \mathbb{D} \times \mathbb{X}.$$

\subsection{The Policy Rule}
\label{subsec:policyrule}

The policy rule associated with this experiment defines the behavior of the PM. We define it as a sequence of two policy functions that, at each instance $t$, determine the probability of stopping the experiment and the probability of treatment for each $x \in \mathbb{X}$.

The first policy function, $(y^{t-1} , d^{t-1}) \mapsto \sigma_{t}(  y^{t-1} , d^{t-1}    )(x) \in [0,1]$, specifies the probability of stopping the experiment for unit $x\in \mathbb{X}$ given the observed history $y^{t-1} , d^{t-1}$. The second policy function, $(y^{t-1} , d^{t-1}) \mapsto \delta_{t}(y^{t-1},d^{t-1})(\cdot|x)  \in \Delta(\mathbb{D})$, specifies the probability distribution over treatments for each $x \in \mathbb{X}$; i.e., $\delta_{t}(Y^{t-1},D^{t-1})(d|x)$ is the probability that $x \in \mathbb{X}$ receives treatment $d$ given the past history. When there is no risk of confusion, we will omit the dependence on the history.

The policy rule defines two consecutive stages: a first stage of exploitation \emph{and} exploration and a second stage of pure exploitation, in which the PM has stopped the experiment and has selected what she believes to be the best treatment. How the PM regulates the trade-off between exploitation and exploration in the first stage will be key for the results presented in Section \ref{sec:analytical_results}. With this in mind, we now define a \textbf{structure of exploration  for the policy rule $(\delta_{t})_{t}$} as two positive-valued sequences $(h_{t},\omega_{t})_{t}$ such that for any $(d,x) \in \mathbb{D} \times \mathbb{X}$ and any $t \geq 0$, $\omega_{t}(d,x) \in [0,1]$, $h_{t}(\cdot|x) \in \Delta(\mathbb{D})$, and
\begin{align}
	\mathbf{P} \left(  t^{-1} \sum_{s=1}^{t} \delta_{s}(d|x) \geq h_{t}(d|x)  \right) \geq 1 - \omega_{t}(d,x).
\end{align}
We call $(h_{t})_{t}$ the \textbf{degree of exploration} of the policy rule and $(1-\omega_{t})_{t}$ the \textbf{likelihood of exploration} of the policy rule. By providing a lower bound on the (average) propensity score, the structure of exploration quantifies the extent to which experimentation  occurs under the policy rule $(\delta_{t})_{t}$. This structure is the \emph{only} feature of the policy rule that matters for our performance criteria. We present these results formally in Section \ref{sec:analytical_results}.\footnote{The structure of exploration is not unique (e.g. $\omega_{t} = 0$ and $h_{t} = 0$ or $\omega_{t}=1$ and $h_{t}=0$ are both explorations structures), however, the results in Section \ref{sec:analytical_results} provide a criteria for ranking the different structures.}

Supplemental material \ref{app:PolicyRulesExamples} presents several commonly-used policy rules --- and their associated exploration structure --- in the context of the Gaussian learning framework, which we describe next.

\subsection{The Gaussian Learning Model}
\label{sec:specific}

The PM does not know the probability distribution of potential outcomes $P$, but does have prior beliefs about it. This prior knowledge can come from many sources: the PM's own prior knowledge, expert opinions, or past experiments. Importantly, we allow for multiple sources, in case the PM is unwilling or unable to discard one in favor of the others.  If her prior sources of information extrapolate to the current experiment, then she should use them because they contain relevant information. But if some sources are not externally valid, then incorporating them in her assignment of treatment may lead to incorrect decisions, at least in finite samples. Thus, our PM not only faces the question of whether to incorporate the different sources, but how to aggregate them as well. We formalize this ``external validity dilemma'' by using a multiple prior Bayesian model.

Formally, for each $(d,x) \in \mathbb{D} \times \mathbb{X}$, the PM has a family of PDFs indexed by a finite dimensional parameter $\theta \in \Theta$,  $\mathcal{P}_{d,x} : = \{ p_{\theta} \colon \theta \in \Theta  \}$, that describes what she believes are plausible descriptions of the true probability of the potential outcome $Y(d,x)$. The PM also has $L+1$ prior beliefs, $(\mu^{o}_{0}(d,x))_{o=0}^{L}$, regarding which elements of $\mathcal{P}_{d,x}$ are more likely; these prior beliefs summarize the prior knowledge obtained from the $L+1$ different sources.

For each $(d,x) \in \mathbb{D} \times \mathbb{X}$, the family $\mathcal{P}_{d,x}$ and the collection of prior beliefs give rise to $L+1$ subjective Bayesian models for $P(\cdot|d,x)$. Given the observed data of past treatments and outcomes, at instance $t \geq 1$, the PM observes the realized outcome $Y_{t}(D_{t}(x),x)$ and the treatment assignment $D_{t}(x)$. Using Bayesian updating, she then forms posterior beliefs for each model, which we denote by $\mu^{o}_{t}(d,x)$. 
We observe that the belief is updated using observed data, $(Y_{t}(D_{t}(x),x),D_{t}(x))$ and using $p_{\theta}$ as the PDF of $Y_{t}(D_{t},x)$ given $D_{t}(x) = d$, this feature is analogous to the missing data problem featured in experiments under the frequentist approach.\footnote{Because the PM already knows the probability of $D_{t}(x)$, she does not need to include it as part of the Bayesian updating problem.}


In what follows, we will assume that the PM takes subjective models within the Gaussian family (see Section \ref{app:General.Learning.Model} in the Supplemental Material for the general setup). This assumption, and the corresponding conjugate priors, imply that the posterior belief is fully characterized by a finite dimensional object, which is more tractable. 
Formally, for each $(d,x) \in \mathbb{D} \times \mathbb{X}$, $\mathcal{P}_{d,x}$ is a family of Gaussian PDFs given by $\{ \phi(\cdot; \theta , 1) \colon \theta \in \mathbb{R}   \}$ and the prior for every source is also assumed to be Gaussian with mean $\zeta^{o}_{0}(d,x)$ and variance $1/\nu^{o}_{0}(d,x)$.\footnote{ Throughout, $\phi (\cdot;\theta,\sigma^{2})$ is the Gaussian PDF with mean $\theta$ and variance $\sigma^{2}$.} The quantity $\nu^{o}_{0}(d,x)$ can be interpreted as the number of units with characteristics $x$ that were assigned treatment $d$ in a past experiment. The higher the $\nu^{o}_{0}(d,x)$, the more certain source $o$ is about $\phi (\cdot;\zeta^{o}_{0}(d,x),1)$ being the correct model. Throughout, we will assume $(\zeta^{o}_{0},\nu^{o}_{0})_{o=0}^{L}$ are non-random.

Given the observed data of past treatments and observed outcomes, at instance $t$ the posterior belief, $\mu_{t}$,  is also Gaussian with mean and inverse of the variance given by the following recursion:
\begin{align}\notag
	\zeta^{o}_{t}(d,x) =   &   \frac{ 1\{  D_{t}(x) = d  \}   }{  \nu^{o}_{t-1}(d,x) + 1\{  D_{t}(x) = d  \}   } Y_{t}(d,x) + \frac{ 	\nu^{o}_{t-1}(d,x)  }{ 	 \nu^{o}_{t-1}(d,x) + 1\{  D_{t}(x) = d  \}   } \zeta^{o}_{t-1}(d,x)   \\ 
	\label{eqn:zeta.o}
	= &  \frac{ 	J_{t}(d,x)   }{ 	f_{t}(d,x)   + \nu^{o}_{0}(d,x) /t  }  + \frac{ 	\nu^{o}_{0}(d,x) /t  }{ 	f_{t}(d,x)   + \nu^{o}_{0}(d,x) /t  } \zeta^{o}_{0}(d,x)\\
	\notag
	&~where \\ \label{eqn:nu.o}
	\nu^{o}_{t}(d,x) = & N_{t}(d,x)   + \nu^{o}_{0}(d,x),~f_{t}(d,x) : = N_{t}(d,x)/t\\
	&~and~J_{t}(d,x):= t^{-1} \sum_{s=1}^{t} Y_{s}(d,x) 1\{ D_{s}(x)  = d \}  .
\end{align}
From these expressions, we can see how Gaussianity simplifies the dynamics of the problem. We only need to analyze $(\zeta^{o}_{t}(d,x),\nu^{o}_{t}(d,x))_{t=0}^{T}$, a finite dimensional object, as opposed to $(\mu^{o}_{t}(d,x))_{t=0}^{T}$, an infinite dimensional object that is quite intractable. Note however that even with the Gaussianity assumption, our setup remains quite general in practice as we describe in the following remark.


\begin{remark}
	(1) Since the PM cares about learning the ATE, this model is sufficiently general to encompass the canonical RCT setup for estimation of average treatment effects, even when the potential outcomes are not necessarily Gaussian. To see this, note that even if the PM's subjective model for potential outcomes is misspecified (i.e. she incorrectly assumes that $Y(d,x)$ is Gaussian) the PM can still accurately learn the true average effect because, for each $(d,x)$, there always exists a $\theta$ such that $\theta = E_{P(.|d,x)}[Y(d,x)]$. We show this is the case in Section \ref{sec:zeta.concentration}.

	(2) Our results and methodology extend to any subjective model whose posterior beliefs can be fully described by low finite-dimensional objects. For instance, in cases where  $Y(d,x) \in \{0,1\}$, they extend to the Bernoulli-Beta model wherein the $t$ instance posterior is given by a Beta density with parameters given by $(\sum_{s=1}^{t} 1\{ D_{s}(x) =d  \} Y_{s}(d,x) +  \nu^{o}_{0}(d,x) \zeta^{o}_{0}(d,x)   , \sum_{s=1}^{t} 1\{ D_{s}(x) =d  \} (1-Y_{s}(d,x)) + \nu^{o}_{0}(d,x) (1-\zeta^{o}_{0}(d,x))  ) $. More generally, our methodology can be extended to the entire exponential family --- which includes the models considered here and more (see \cite{schlaifer1961applied} for examples), however, the interpretation of $\zeta_{t}(d,x)$ may change.  $\triangle$
\end{remark}

\paragraph{Model Aggregation \& External Validity. }
For each $(d,x) \in \mathbb{D}\times \mathbb{X}$ and faced with $L+1$ distinct subjective Bayesian models, $\{ \langle \mathcal{P}_{d,x},   \mu^{o}_{0}(d,x) \rangle \}_{o=0}^{L} $, our PM has to aggregate this information. There are different ways to do this; we choose one that at each instance $t$, averages the posterior beliefs of each model using as weights the posterior probability that model $o$ best fits the observed data within the class of models being considered, 
i.e.,
\begin{align}
	\mu^{\alpha}_{t}(d,x) : =  \sum_{o=0}^{L} \alpha^{o}_{t}(d,x) \mu^{o}_{t}(d,x) \label{eqn:mu.alpha}
\end{align}
where
\begin{align*}
	\alpha^{o}_{t}(d,x) : =  \frac{ \int \prod_{s=1}^{t}  \phi(Y_{s}(d,x) ; \theta , 1 ) ^{1\{ D_{s}(x) =d  \}}  \phi (\theta; \zeta^{o}_{0}(d,x) , 1/\nu^{o}_{0}(d,x))(\theta) d\theta   }{ \sum_{o=0}^{L} \int \prod_{s=1}^{t}  \phi (Y_{s}(d,x) ; \theta , 1)  ^{1\{ D_{s}(x) =d  \}}  \phi (\theta; \zeta^{o}_{0}(d,x) , 1/\nu^{o}_{0}(d,x))(\theta) d\theta    }.
\end{align*}

We can interpret $\alpha^{o}_{t}(d,x)$ as the PM's subjective probability that source $o$ for $(d,x)$ is more externally valid for her current experiment. To expound on this last point, we introduce the concept of degree of external validity of a given source withing the Gaussian learning model, and then we relate this concept to the behavior of $(\alpha^{o}_{t}(d,x))_{o=0}^{L}$. But first, it useful to introduce some nomenclature. We call $|\zeta^{o}_{0}(d,x) - \theta(d,x) |$ the \textbf{bias of source $o$} and $\nu^{o}_{0}(d,x)$ the \textbf{degree of conviction of source $o$} --- since a higher $\nu^{o}_{0}(d,x)$ implies a lower prior variance. As such, we can interpret $|\zeta^{o}_{0}(d,x) - \theta(d,x) | \sqrt{\nu^{o}_{0}(d,x)}$ as the \textbf{ degree of stubbornness of source $o$}. 
 


\begin{definition}[Degree of external validity]\label{def:DEV}
	For each $(d,x) \in \mathbb{D}\times \mathbb{X}$ the degree of external validity (DEV) of source $o$ is given by
	\begin{align*}
		\mathbb{EV}_{(d,x)}(o) : = - \nu^{o}_{0}(d,x) (\theta(d,x) - \zeta^{o}_{0}(d,x))^{2} + \log \nu^{o}_{0}(d,x).
	\end{align*}
\end{definition}

The value of this quantity is that it allows us to define a partial ordering over sources given by the next definition 

\begin{definition}[Externally valid sources]\label{def:EV}
	For $(d,x)$, a source $o$ is more externally valid than source a $o'$ if has a higher degree of external validity, i.e.,
	\begin{align*}
			\mathbb{EV}_{(d,x)}(o) > 	\mathbb{EV}_{(d,x)}(o').
		\end{align*}
We denote this as $o \succ_{(d,x)} o'$. A source $o$ is externally valid, if $\nu^{o}_{0}(d,x) (\theta(d,x) - \zeta^{o}_{0}(d,x))^{2}  = 0$ and $\nu^{o}_{0}(d,x) = + \infty$.
\end{definition}

According to this definition the degree of external validity of a given source depends on its degree of stubbornness in a decreasing manner, i.e., the more stubborn the source is, the lower its DEV is. The DEV also depends on the degree of conviction, but its dependence is more nuanced. To see this, observe that $\frac{d\mathbb{EV}_{(d,x)}(o)}{ d \nu^{o}_{0}(d,x) } = - (\theta(d,x) - \zeta^{o}_{0}(d,x))^{2} + 1/\nu^{o}_{0}(d,x)$. Thus, the effect of the degree of conviction on the DEV depends on the level of the bias, and it does so in an intuitive way. To see this, first note that for unbiased sources, a higher degree of conviction can only increase the DEV. However, for biased sources, the effect is not uniform: For low initial values of conviction an increase on this quantity will increase the DEV, but for high enough initial levels of conviction an increase of it will lower the DEV; i.e., the source is becoming more stubborn, re-affirming the bias. Finally, we note that the ``cutoff" conviction level is inversely proportional to the level of the bias: A higher bias imply a larger range of initial conviction levels for which $\frac{d\mathbb{EV}_{(d,x)}(o)}{ d \nu^{o}_{0}(d,x) } < 0$. This behavior of the DEV with respect to conviction can be achieved with functional forms other than the current one. In particular, one can replace the $\log(.)$ in the definition by other increasing and concave function and still be able to obtain the aforementioned behavior. The particular choice of $\log(.)$ in this case stems from the fact that the prior and subjective model are both Gaussians.

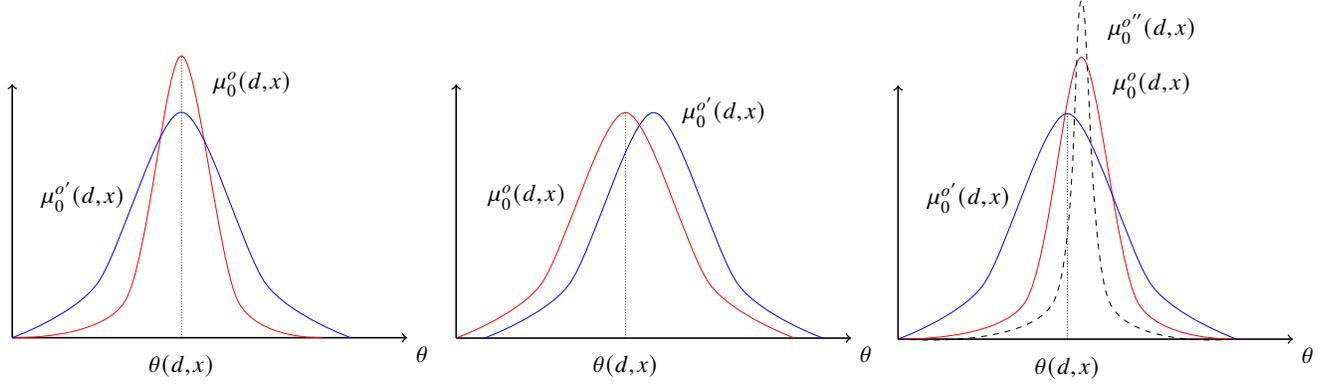
\begin{figure}
	
	\begin{subfigure}[t]{0.2\textwidth}
		\hspace{-7cm}
		\scalebox{.75}{\begin{tikzpicture}
				
				\draw[thick,->] (0,0) -- (7,0) node[anchor=north west] {$\theta$};
				\draw[thick,->] (0,0) -- (0,4.5) node[anchor=south east] {~};
				
				
				\draw [red] plot [smooth] coordinates {(0,0) (2,0.65) (3,5) (4,0.65) (5.5,0)};

				\draw [blue] plot [smooth] coordinates {(0,0) (1.5,0.95) (3,4) (4.5,0.95) (6,0)};
				
				
				\draw [densely dotted] plot  coordinates {(3,0) (3,5)};
				
				\coordinate (C) at (3,-0.5);
				\coordinate (D) at (1.25,2.5);
				\coordinate (E) at (4.25,4.5);

				\node[yshift=0cm] at (C) {$\theta(d,x)$};
				\node[yshift=0cm] at (D) {$\mu^{o'}_{0}(d,x)$};
				\node[yshift=0cm] at (E) {$\mu^{o}_{0}(d,x)$};
		\end{tikzpicture}	}
	\end{subfigure}
	
	\vspace{-4.25cm}
	\hspace{-2.25cm} 
	\begin{subfigure}[b]{0.2\textwidth}
		
		\scalebox{.75}{	\begin{tikzpicture}
				\draw[thick,->] (0,0) -- (7,0) node[anchor=north west] {$\theta$};
				\draw[thick,->] (0,0) -- (0,4.5) node[anchor=south east] {~};
				
				
				\draw [blue] plot [smooth] coordinates {(0.5,0) (2,0.95) (3.5,4) (5,0.95) (6.5,0)};
				
				\draw [red] plot [smooth] coordinates {(0,0) (1.5,0.95) (3,4) (4.5,0.95) (6,0)};
				
				
				\draw [densely dotted] plot  coordinates {(3,0) (3,4)};
				
				\coordinate (C) at (3,-0.5);
				\coordinate (D) at (1.25,2.5);
				\coordinate (E) at (4.75,4.0);

				\node[yshift=0cm] at (C) {$\theta(d,x)$};
				\node[yshift=0cm] at (D) {$\mu^{o}_{0}(d,x)$};
				\node[yshift=0cm] at (E) {$\mu^{o'}_{0}(d,x)$};
		\end{tikzpicture}}
		
	\end{subfigure}
	
	\vspace{-5.15cm}
	\hspace{9.5cm} 
	\begin{subfigure}[b]{0.2\textwidth}
		
		\scalebox{.75}{	\begin{tikzpicture}
				\draw[thick,->] (0,0) -- (7,0) node[anchor=north west] {$\theta$};
				\draw[thick,->] (0,0) -- (0,4.5) node[anchor=south east] {~};
				
				
				\draw [red] plot [smooth] coordinates {(0,0) (2.25,0.65) (3.25,5) (4.25,0.65) (6,0)};
				
				\draw [black,dashed] plot [smooth] coordinates {(0,0) (2.75,0.65) (3.25,6) (3.75,0.65) (6,0)};
				
				\draw [blue] plot [smooth] coordinates {(0,0) (1.5,0.95) (3,4) (4.5,0.95) (6,0)};
				
				
				\draw [densely dotted] plot  coordinates {(3,0) (3,4)};
				
				\coordinate (C) at (3,-0.5);
				\coordinate (D) at (1.25,2.5);
				\coordinate (E) at (4.5,4.5);
				\coordinate (F) at (4.5,5.5);
				
				\node[yshift=0cm] at (C) {$\theta(d,x)$};
				\node[yshift=0cm] at (D) {$\mu^{o'}_{0}(d,x)$};
				\node[yshift=0cm] at (E) {$\mu^{o}_{0}(d,x)$};
				\node[yshift=0cm] at (F) {$\mu^{o''}_{0}(d,x)$};
		\end{tikzpicture}}
		
	\end{subfigure}
	\caption{Different Degrees of External Validity}
	\label{fig:ev}
\end{figure}

Figure \ref{fig:ev} illustrates this discussion. The plot on the left presents the case where $o \succ_{(d,x)} o'$ because even though both sources are unbiased, source $o$ has a higher level of conviction. In the middle plot, $o \succ_{(d,x)} o'$, but now the the level of conviction is the same whereas the bias is lower for source $o$. Finally, the right plot illustrates the nuanced role of conviction: Source $o''$ (dashed black line) has a small bias but this gets amplified by a very high degree of conviction (i.e., it is highly stubborn), rendering it less externally valid than source $o'$ (blue solid line) which is unbiased and with low degree of conviction; however source $o$ (solid red line) has the same small bias as source $o''$ but a degree of conviction that is lower than $o''$ but higher than $o'$, rendering \emph{more} externally valid that source $o'$ and $o''$.

In sum, definitions \ref{def:DEV} and \ref{def:EV} extend the concept of external validity to a Bayesian framework. In the classical --- non-Bayesian --- setup, bias defines whether a sources is  externally validity (unbiased) or not (bias), as stated in Definition \ref{def:EV}. However, this view of external validity is perhaps too narrow within a (Gaussian) Bayesian framework with a conviction level that may not be infinite. In this framework, external validity ceases to be a \emph{qualitative} notion and becomes a \emph{quantitative} notion which depends both on the bias and the level of conviction. This is precisely what our DEV measure captures.

%

The next proposition provides a link between this ranking of external invalidity and our weights $(\alpha^{o}_{t}(d,x))_{o=0}^{L}$.  It shows that under some technical regularity assumptions, the weights are only positive for the least externally invalid models as $t$ diverges, provided $(d,x)$ is played sufficiently often.

\begin{proposition}\label{pro:alpha.asymptotics.general}
For any $(d,x) \in \mathbb{D}\times \mathbb{X}$,
\begin{enumerate}
		\item $\lim_{\mathbb{EV}_{(d,x)}(o)   \rightarrow - \infty}  \alpha^{o}_{t}(d,x) = 0$.
		\item If $\inf_{t} t^{-1} \sum_{s=1}^{t} \delta_{s}(d|x) > 0 $, then
\begin{align*}
	\alpha^{o}_{t}(d,x) = \frac{    e^{ 0.5 \mathbb{EV}_{(d,x)}(o) }    }  {  \sum_{o'=0}^{L}   e^{ 0.5 \mathbb{EV}_{(d,x)}(o') }    }  + o_{\mathbf{P}}(1) .
\end{align*} 
		In particular, if $o \succ_{(d,x)} o'$, then $\alpha^{o'}_{t}(d,x) < \alpha^{o}_{t}(d,x) + o_{\mathbf{P}}(1) $, and if $o$ is the externally valid source, then $\alpha^{o}_{t}(d,x) = 1 - o_{\mathbf{P}}(1) $.\footnote{The definition of externally valid source forces us to work with $+\infty$. We follow the standard convention that $a/(+\infty) = 0$ for any real number $a$.}
\end{enumerate}

\end{proposition}

\begin{proof}
	See Appendix \ref{app:General.Bound.alpha}.
\end{proof}

Part (1) of the proposition shows that $\alpha^{o}_{t}$ offers certain robustness properties against externally-invalid models. If a source has a high degree of external invalidity --- i.e., a very negative $\mathbb{EV}_{(d,x)}(o)$ ---, then the associated weight of that source is approximately 0. Part (2) offers a sharper characterization of this robustness property, but asymptotically. It links the degree of external validity and the weight each source will have in the limit. In particular, sources that are less externally valid will receive less weight, and externally valid sources will get weight approaching one.\footnote{In Section \ref{Appsec:Hbayes} we discuss alternative interpretations of and potential extensions to our model.}

\section{Analytical Results}\label{sec:analytical_results} \label{sec:finite}

As mentioned above, the object of interest is the average effect of each treatment, and, at each instance $t$ and for each $ (d,x) \in \mathbb{D} \times \mathbb{X}$, the PM will estimate it using a subjective average treatment effect given by
\begin{align*}
	\zeta^{\alpha}_{t}(d,x) : =	\int y \int_{\Theta}  p_{\theta}(y)  \mu^{\alpha}_{t}(d,x)(d\theta) dy 
	= : \sum_{o=0}^{L}   \alpha^{o}_{t}(d,x) \zeta^{o}_{t}(d,x).
\end{align*}
The middle expression is simply the mean of outcome computed with respect to the PM's subjective PDF of $y$ constructed using the aggregate beliefs at time $t$, $\int_{\Theta}  p_{\theta}(\cdot)  \mu^{\alpha}_{t}(d,x)(d\theta)$. The right-most expression shows that such mean is the weighted average of the posterior mean for each source.

In this section, we establish some finite sample properties of this quantity, such as the rate at which it concentrates around the true average effect. Before we do so, a bit of housekeeping is required. Moving forward, we will omit $x$ from the notation and derive our results for $|\mathbb{X}| = 1$. Given our assumptions, we can learn the fundamentals for each $x \in \mathbb{X}$ by treating them as separate and independent problems. Thus, we can extend all our results to the case of $|\mathbb{X}| >1$ by treating the relevant quantities (e.g. $\theta(d)$, $Y(d)$, etc.) as vectors of dimension $|\mathbb{X}|$. Furthermore, to derive the results below we will need some assumptions on the (true) distribution of the potential outcomes,
\begin{assumption}\label{ass:sub-gauss}
	There exists a $\upsilon < \infty$ such that for any $\lambda >0$ and any $d \in \mathbb{D}$, $E[e^{ \lambda (Y(d) - \theta(d))} ] \leq e^{ \upsilon \sigma(d)^{2} \lambda^{2}}$ where $\sigma(d)^{2} : = Var(Y(d))$.
\end{assumption}
This assumption imposes that $Y(d)$ is sub-gaussian, which, in effect, ensures that the probability $Y(d)$ takes large values decays at the same rate as the Normal does. Sub-gaussianity plays two roles in our results. First, it ensures that some higher moments, like the variance, exist. Second, and more importantly, it is used to derive how fast the average outcome concentrates around certain population quantities (see Lemma \ref{lem:Azzuma} in the Appendix \ref{app:ConcentrationInequalities}). We could relax this assumption, but at the cost of getting slower concentration rates; see Remark \ref{rem:sub-gaussian} in that appendix for more details.

Before presenting these results formally, it is useful to present our general approach for how we derived them. As we discussed above, a key of object of interest is $\zeta^{\alpha}_{t} = \sum_{o=0}^{L} \alpha^{o}_{t} \zeta^{o} _{t}$, the subjective average effect of treatment at instance $t$. Most of our results hinge on understanding how this object concentrates around the true expected value $\theta$.

For each treatment $d$, the randomness of $\zeta^{\alpha}_{t}(d)$ comes from two quantities: the frequency of play, $f_{t}(d) = t^{-1} \sum_{s=1}^{t}  1\{D_{s} =d \}  $ and the treatment-outcome average, defined as $J_{t}(d) := t^{-1} \sum_{s=1}^{t} 1\{D_{s} =d \} Y_{s}(d).$ Hence, to derive the concentration rate of $\zeta^{\alpha}_{t}(d)$, we first need to understand how $f_{t}(d)$ and $J_{t}(d)$ concentrate. For $J_{t}(d)$, we can employ exponential inequalities (see  Lemma \ref{lem:Azzuma} in the Appendix \ref{app:ConcentrationInequalities}) to determine how fast the treatment-outcome average concentrates around $f_{t}(d) \theta(d)$.  Given the time dependent aspect of the data, the concentration inequalities are obtained for martingale differences as opposed to the standard IID data. The case of $f_{t}(d)$ is a bit more nuanced because we care not only about how fast it concentrates around the average propensity score, $t^{-1} \sum_{s=1}^{t} \delta_{s}(d)$, but also about how far the average propensity score is from zero (i.e. the degree of exploration). Our structure of exploration enables us to separate the problem into two parts: we use exponential inequalities for martingale differences to determine the concentration rate and the structure of exploration to assess how far the average propensity score is from zero.

The next important step is to understand how the concentration rates of $f_{t}(d)$ and $J_{t}(d)$ translate into the concentration rate of $\zeta^{\alpha}_{t}(d)$ and how the parameters of the model and the exploration structure affect this rate. To do this, take any $\gamma , \eta >0$ where $\gamma$ and $\eta$ quantify the concentration rate of $J_{t}(d)$  around $f_{t}(d)\theta(d)$ and $f_{t}(d)$ around $t^{-1}\sum_{s=1}^{t} \delta_{s}(d)$ respectively. Given this, Lemma \ref{lem:bound.zeta.alpha} in the Appendix \ref{app:zeta.alpha.bound} shows that, given an exploration structure $(h_{t},\omega_{t})_{t}$,
\begin{align}
	| \zeta^{\alpha}_{t}(d)  - \theta(d) |  \leq \Gamma( \gamma, h_{t}(d) - \eta, |\zeta_{0}(d) - \theta(d) | , \nu_{0}(d)  ) \label{eqn:zeta.alpha.Gamma.bound}
\end{align}
where $h_{t}(d)$ is the degree of exploration and  $\Gamma : \mathbb{R}_{+} \times [0,1] \times \mathbb{R}^{L+1} \times \mathbb{N}^{L+1} \rightarrow \mathbb{R}$ is a function defined in Appendix \ref{app:zeta.alpha.bound}. Thus, $\Gamma$ maps the concentration rate of $J_{t}(d)$ and $f_{t}(d)$ (given by $\gamma$ and $\eta$ respectively) to the concentration rate of the posterior mean around the true parameter. In fact, we show in Lemma \ref{lem:properties.Gamma} in the Appendix \ref{app:zeta.alpha.bound} that $\Gamma$ is increasing in the first argument and decreasing in the second one, thereby implying that a faster concentration rate of $J_{t}(d)$ and $f_{t}(d)$ translate into a faster concentration rate of the posterior mean. Moreover, $\Gamma$ also quantifies how a higher degree of exploration translates into a faster concentration rate, as well as how the source's parameters, $(\zeta_{0}(d),\nu_{0}(d))$, affect this rate.

\subsection{ Concentration bounds on the Posterior Mean}
\label{sec:zeta.concentration}

The next proposition establishes the rate at which the posterior mean concentrates around the true expected outcome.

\begin{proposition}\label{pro:concentration.alpha.zeta}
	For any $d \in \{0,...,M\}$, any $t \in \mathbb{N}$ and any $\varepsilon \geq 0$ such that $t h_{t}(d)^{2} \geq  \varepsilon   $,
	\begin{align*}
		\mathbf{P} \left(  |  \zeta^{\alpha} _{t}(d) - \theta(d) | >   \Gamma  \left( \sqrt{\frac{2\upsilon \varepsilon }{ h_{t}(d)^{2} t}} \sigma(d) , 0.5 h_{t}(d) , |\zeta_{0}(d) - \theta(d)| , \nu_{0}(d)   \right)    \right)  \leq  4 (e^{   -  \varepsilon    } + \omega_{t}(d) ).
	\end{align*}
\end{proposition}

\begin{proof}
	See Appendix \ref{app:zeta.concentration}.
\end{proof}

The intuition behind the proof relies on the arguments discussed above that explain how the concentration rate of the posterior mean depends on two factors: the concentration rates of the random quantities $J_{t}(d)$ and $f_{t}(d)$ and how these get distorted by the function $\Gamma$. More precisely, we show that by employing concentration inequalities for Martingale difference sequences (see Lemma \ref{lem:Azzuma} in Appendix \ref{app:ConcentrationInequalities}), $J_{t}(d)$ and $f_{t}(d)$  are (up to constants) within $\gamma = \sqrt{\delta/t} $ and $\eta = h_{t}(d)  \sqrt{\delta/t}   $ of their respective population values with probability higher than $1-4e^{-\delta/h^{2}_{t}(d)}$ for any $\delta>0$. To obtain the result, we simply plug these quantities into expression \ref{eqn:zeta.alpha.Gamma.bound}, while noting that $h_{t}(d) - \eta \geq 0.5 h_{t}(d)$ for large enough $t$ and that $\varepsilon = \delta/h^{2}_{t}(d)$.

Through the term $\Gamma$ and the probability bound, the proposition illustrates the effect of the structure of exploration, $(h_{t},\omega_{t})_{t}$, on the concentration rates.  In particular, $\Gamma$ is of order $O \left( \frac{(t h^{2}_{t}(d))^{-1/2}  }{h_{t}(d) + t^{-1} }   \right)$ (see Lemma \ref{lem:properties.Gamma}(3) in the Appendix \ref{app:zeta.alpha.bound}). Hence, for policy functions with $h_{t}(d) \geq \epsilon > 0$ (e.g. $\epsilon$-greedy) the concentration rate is of order $t^{-1/2}$, but for those with $h_{t}(d) = o(1)$ then the concentration rate is slower and consistency of the posterior mean to the truth is only ensured if $\sqrt{t} h^{2}_{t}(d) \rightarrow \infty$.

Our method for aggregating multiple priors offers an attractive feature regarding our concentration rates. Sufficiently stubborn models, i.e. $|\zeta^{o}_{0}(d) - \theta(d) | \sqrt{\nu^{o}_{0}(d)}$ is sufficiently large,  will have close to zero effect on the concentration rate of $ \zeta^{\alpha}_{t}(d)$, as they are essentially dropped from the weighted average. This implies an \emph{oracle} property in the sense that the concentration rate becomes arbitrary close to the least stubborn model, provided there is enough separation between the stubbornness of this model and the others.  We formalize this property in the next corollary.

\begin{corollary}\label{cor:OracleRobust}
	Take any $(t,d,\varepsilon)$ as in Proposition \ref{pro:concentration.alpha.zeta}. Furthermore, let model $o=0$ denote the least stubborn model and suppose that for any given $\delta>0$, 
	there exists a $C$ such that $\min_{o \ne 0} |\zeta_{0}^{o}(d) - \theta(d)|\sqrt{\nu^{o}_{0}(d)} \geq C$. Then,
	\begin{align*}
		\mathbf{P} \left( | \zeta^{\alpha} _{t}(d) - \theta(d) | >   \Omega  \left( \sqrt{  \frac{ 2 \upsilon \varepsilon  } { h_{t}(d)^{2}  t}   } \sigma(d), 0.5 h_{t}(d) , |\zeta^{o}_{0}(d) -\theta(d)| , \nu^{o} _{0}(d)   \right)   + \delta  \right)  \leq 4 (e^{ - \varepsilon  } + \omega_{t}(d))
	\end{align*}
	
\end{corollary}

\begin{proof}
	See Appendix \ref{app:zeta.concentration}.
\end{proof}

The function $\Omega$, which is formally defined in Appendix \ref{app:zeta.o.bound},  acts as $\Gamma$ but for one model; i.e., for any $o \in \{0,...L\}$ and any $\gamma \geq 0$, assuming  $J_{t}(d)$ and $f_{t}(d)$ are within $\gamma$ of their population analogues,
\begin{align*}
	| \zeta^{o} _{t}(d) - \theta(d) |  \leq \Omega(\gamma, h_{t}(d) - \gamma,   |\zeta^{o}_{0}(d) - \theta(d)| , \nu^{o} _{0}(d)   ).
\end{align*}
Thus, the expression inside the probability in the corollary is in fact the concentration rate of the least stubborn model.

We summarize the implications of the previous proposition in the following remark and illustrate them numerically in Section \ref{sec:simulations}.

\begin{remark}[Properties of the Concentration Rate]\label{remark:cb}

The following properties are based on Lemma \ref{lem:properties.Omega} in Appendix \ref{app:zeta.o.bound}.

	\begin{enumerate}
		\item  All else equal, the concentration rate decreases as the bias increases; it also decreases with the degree of stubbornness, i.e. $|\zeta^{o}_{0}(d) - \theta(d) | \sqrt{ \nu^{o}_{0}(d) }$. The concentration rate is fastest when the bias is zero.
		
		\item For confident models, the concentration rate increases with the degree of conviction, i.e. $\nu^{o}_{0}(d)$ increases. The intuition behind this result is as follows: If $\nu^{o}_{0}(d)$ increases but $|\zeta^{o}_{0}(d) - \theta(d) | \sqrt{ \nu^{o}_{0}(d) }$ remains constant --- equal to 0, in particular ---, then necessarily, the model is becoming more convinced about a prior that is unbiased, thereby implying a faster convergence rate.
		
		\item The effects of the degree of stubbornness and conviction on the concentration rate decrease as $t$ increases.
		
		
		\item An increase in the degree of the exploration, $h_{t}(d)$, improves the concentration rate. This comes from the fact that $h_{t}(d) \mapsto  \Omega \left( \sqrt{  \frac{2 \upsilon \varepsilon  } {h^{2}_{t}(d) t}   } \sigma(d) , 0.5 h_{t}(d) , |\zeta^{o}_{0}(d) - \theta(d)| , \nu^{o}_{0}(d)  \right) $  is decreasing (see Lemma \ref{lem:properties.Omega} in the Appendix \ref{app:zeta.o.bound}). Intuitively, increasing  $h_{t}(d)$ implies having more observations to estimate $\theta(d)$ ---  ``more information'' about treatment $d$ implies a faster concentration rate. $\triangle$
	\end{enumerate}
\end{remark}

\subsection{Probability of making a mistake}
\label{sec:PoM}

In this section, we provide bounds on the probability of making a mistake. To do so, we need to first define the policy rule for stopping the experiment, $\sigma$, since this rule does govern the probability of making mistakes when stopping the experiment. A desirable property for this rule is that, for a given tolerance level $\beta \in (0,1)$ chosen by the PM, the probability of making a mistake when stopping the experiment is no larger than $\beta$. We propose the following rule:

\begin{example}[Threshold Stopping Rule]\label{exa:StoppingRule}
	For any positive-valued non-increasing sequence $(\gamma_{t})_{t}$ and $B \in \mathbb{N}$, the stopping rule parameterized by $((\gamma_{t})_{t},B)$ is such that, for any $t \geq B$,
\begin{align*}
	\sigma_{t}(Y^{t-1},D^{t-1}) = 1,~iff~\max_{d} \{     \min_{m \ne d} \zeta^{\alpha}_{t-1}(d)  -  \zeta^{\alpha}_{t-1}(m)  - c_{t-1}(\gamma_{t-1},d,m) \}  > 0,
\end{align*}
and if $t < B$, $\sigma_{t}(Y^{t-1},D^{t-1}) = 0$, where, for any $d,m \in \{0,...,M\}$ and any $o \in \{0,...,L\}$,
\begin{align*}
	c_{t}(\gamma_{t},d,m) :  = c_{t} (\gamma_{t}, d)  + c_{t} (\gamma_{t}, m)  : =  \left( \sum_{o=0}^{L}    \frac{ \alpha^{o}_{t}(d) \sqrt{t} \gamma_{t}(d) }{ N_{t}(d) + \nu^{o}_{0}(d)   } + \sum_{o=0}^{L}  \frac{ \alpha^{o}_{t}(m) \sqrt{t} \gamma_{t}(m) }{ N_{t}(m) + \nu^{o}_{0}(m)   } \right) 
\end{align*}
where $N_{t}(d) : = \sum_{s=1}^{t} 1\{ D_{s} = d \}$.

Loosely speaking, the rule proposes to stop the experiment after $B$ instances and as soon as the distance between the highest average posterior and the rest --- measured by $\max_{d}     \min_{m \ne d} (  \zeta^{\alpha}_{t}(d)   -    \zeta^{\alpha}_{t}(m) )$ --- is far enough from zero, where ``far enough'' is essentially measured by the cutoff  $c_{t}(\gamma_{t},d,m)$. While the expression for this cutoff is a bit involved, we highlight one key aspect of it. The appropriate scaling is given by $\sqrt{t}/(N_{t}(d) + \nu^{o}_{0}(d))$ as opposed to the standard scaling of $1/\sqrt{t}$ one obtains in hypothesis testing. The difference lies on the fact that our methodology uses prior information and thus leverages the ``prior observations" represented by $\nu^{o}_{0}(d)$. Models with high levels of confidence (a high $\nu^{o}_{0}(.)$) will have lower cutoffs, prompting the rule to stop sooner. This effect naturally decreases as the sample size $t$ increases, but for finite sample sizes it could still be sizable. $\triangle$
\end{example}

Suppose treatment $M$ has the largest expected effect, i.e.,  $\Delta: = \theta(M) -  \max_{d \ne M}  \theta(d) > 0 $. We define a mistake as recommending a treatment different than $M$ at the instance $t$ in which the experiment was stopped. Because recommendations are based on the PM's posteriors, a mistake is given by
\begin{equation*}
	\max_{d \ne M}    \zeta^{\alpha}_{\tau}(d)  -    \zeta^{\alpha}_{\tau}(M)     > 0,
\end{equation*}
where $\tau$ indicates when the experiment is stopped, i.e., is the first instance after $B$ such that $  \max_{d}    \min_{m \ne d}  \{  \zeta^{\alpha}_{t}(d) -  \zeta^{\alpha}_{t}(m)  - c_{t}(\gamma_{t},d,m) \} > 0$ where the cutoffs $c_{t}$ are defined in Example \ref{exa:StoppingRule}.\footnote{Once again, we will omit $x$ from the notation in what follows.}

The following proposition provides an upper bound for the probability of making a mistake associated with this stopping rule, when all sources are unbiased or are biased ``in the right direction".\footnote{By ``in the right direction" we mean the priors rank treatment $M$ as the highest one. For the general case where sources can be biased (in any direction), see Lemma \ref{lem:stopping.alpha} in the Appendix \ref{app:PoMM}.}


\begin{proposition}\label{pro:stopping.alpha}
	Suppose for each $d  \in \{0,...,M-1\}$, $\zeta_{0}(d) \leq \theta(d)$ and $\zeta_{0}(M) \geq \theta(M)$. Consider the stopping rule defined in Example \ref{exa:StoppingRule} with parameters $((\gamma_{t})_{t},B)$ then for any $t \geq B$,
	\begin{align}\label{eqn:PoMM.UpperBound}
		\mathbf{P} \left(  \max_{d\ne M} \{ \zeta^{\alpha}_{\tau}(d) -  \zeta^{\alpha}_{\tau}(M)   \} > 0  \cap \{\tau = t \}   \right)	\leq 2  \sum_{d=0}^{M}  e^{-0.5 \frac{  \gamma_{t}(d)^{2} }{  \upsilon \sigma(d)^{2} }  }.
	\end{align}
\end{proposition}

\begin{proof}
	See Appendix \ref{app:PoMM}.
\end{proof}

The intuition behind this proposition is as follows. Mistakes occur when, at some instance $t$ greater than $B$, the posterior mean of some treatment $d$ --- different from $M$ --- is ``much larger" than the others. This implies that the posterior has to be ``much larger'' than its population mean, $\theta(d)$, where ``much larger'' depends on the pre-specified cutoff. Hence, given our assumption on the priors, 
the probability of a mistake is essentially given by the probability that the outcome-treatment average exceeds its population value by an amount given by $\gamma_{t}/\sqrt{t}$. Lemma \ref{lem:Azzuma} in Appendix \ref{app:ConcentrationInequalities} provides the bound of $e^{-0.5 \frac{  \gamma_{t}^{2} }{  \upsilon \sigma(d)^{2} }  }$. 

In Proposition \ref{pro:stopping.alpha}, we assumed unbiased sources or that the priors ranked treatment $M$ as the highest. The next collorary proves that when some sources are biased, there still exists an oracle property akin to the one demonstrated for the concentration rates. In particular, we show that upper bound is arbitrary close to the unbiased source, provided the other sources are sufficiently biased.\footnote{A more general statement that relaxes the unbiased assumption of source $o=0$ is proven in Lemma \ref{lem:robust.PoMM} in Appendix \ref{app:PoMM}.}

%

\begin{corollary}\label{cor:robust.PoMM}
	Let $o=0$ denoted the unbiased source. There exists a $C$ such that, if $\min_{o \ne 0} |\zeta^{o}_{0}(.) - \theta(.)| \geq C$ and $\zeta^{0}_{0}(.) = \theta(.)$, then for any $t \geq B$,	
	\begin{align*}
		\mathbf{P} \left(  \max_{d\ne M} \{ \zeta^{\alpha}_{\tau}(d) -  \zeta^{\alpha}_{\tau}(M)   \} > 0  \cap \{ \tau = t \}   \right)	\leq 2 \sum_{d=0}^{M}   e^{  -0.5 \frac{ (\gamma_{t} )^{2} } { \upsilon \sigma(d)^{2}} } 
	\end{align*}
\end{corollary}

\begin{proof}
	See Appendix \ref{app:PoMM}.
\end{proof}

Proposition \ref{pro:stopping.alpha} also reveals how by properly choosing $((\gamma_{t})_{t} ,B)$, the probability of a mistake associated with the stopping rule is bounded by $\beta$, where $\beta \in (0,1)$ is any tolerance level. The next corollary presents such result.\footnote{For the general result allowing for biased sources, please see Lemma \ref{lem:PoMM.beta} in Appendix \ref{app:PoMM}.}


\begin{corollary}\label{cor:PoMM.beta}
	Suppose all the conditions of Proposition \ref{pro:stopping.alpha} hold, and, for any $t$, $\gamma_{t}(d) \geq 2 \sqrt{\upsilon} \sigma(d) A $ for all $d \in \mathbb{D}$  with $A$ such that
	\begin{align}\label{eqn:stopping-0}
	A \geq - \log \frac{\beta}{M+1}.
\end{align}
	Then
	\begin{align*}
	\max_{t \in \{B,...,T\}} 	\mathbf{P} \left(  \max_{d\ne M} \{ \zeta^{\alpha}_{\tau}(d) -  \zeta^{\alpha}_{\tau}(M)   \} > 0   \cap \{\tau = t\}  \right)	\leq \beta.
	\end{align*}
\end{corollary}

\begin{proof}
	See Appendix \ref{app:PoMM}.
\end{proof}

%
%

The choice of $(\gamma_{t})_{t}$ is so that the terms in the upper bound in Proposition \ref{pro:stopping.alpha} are less than $\beta$. The sequence $(\gamma_{t})_{t}$ has to bounded below by $2 \sqrt{\upsilon} \sigma(.) A$, the term $2 \sqrt{\upsilon} \sigma(.)$ arises from Assumption \ref{ass:sub-gauss}. In cases where $ \upsilon \sigma(.)$ is unknown, one can replace this value by a sample analog  --- the same way one estimates the standard deviations in the difference in means test --- or simply by an upper bound such as $\log \log t$, both will be valid for large $t$. The term $A$ is simply to ensure that $2 \sum_{d \in \mathbb{D}} e^{-A}$ is less than the tolerance level. Finally, we note that the sequence $(\gamma_{t})_{t}$ can be larger than $2 \sqrt{\upsilon} \sigma(.) A$. However, we do not recommend this, large values of $\gamma_{t}$ are undesirable because the larger the $\gamma_{t}$, the less likely it is to stop the experiment at any instance thereby implying longer --- and more costly --- experiments.

\subsection{Average Observed Outcomes}
\label{sec:avg.Y}

In this section, we characterize the behavior of the average outcome $t^{-1} \sum_{s=1}^{t} Y_{s}$. By  Lemma \ref{lem:Azzuma} in Appendix \ref{app:ConcentrationInequalities}, $ t^{-1} \sum_{s=1}^{t}  Y_{s}$ will concentrate around  a weighted average of $\theta(\cdot)$, with the time average of the propensity score as weights, i.e.,
\begin{equation*}
  t^{-1} \sum_{s=1}^{t}   \sum_{d=0}^{M} \theta(d) \delta_{s}(d).
\end{equation*}

Without further knowledge of $(\delta_{t})_{t}$, it is nearly impossible to characterize this average any further. However, for generalized \emph{$\epsilon$-greedy} policy functions, indexed by a non-random sequence $\boldsymbol{\Xi} : = (\Xi_{t})_{t}$:
\begin{align*}
	\delta_{t}(d) = \Xi_{t} (M+1)^{-1} + (1-\Xi_{t}) 1\{ d = \arg\max_{a} \zeta^{\alpha}_{t-1}(a)   \},~ \forall t \in \{1,...,T\},
\end{align*}
we can establish the following proposition for unbiased sources (the general result for biased sources can be found in Lemma \ref{lem:AvgOutcome.lower.1} in the Appendix \ref{app:avg.Y}).

 \begin{proposition}\label{pro:avg.Y}
	Suppose all sources are unbiased. For any $\gamma>0$ and any $t \in \{1,...,T\}$
	\begin{align*}
		\mathbf{P} \left(   \max_{d} \theta(d) - t^{-1} \sum_{s=1}^{t} Y_{s}    > - \mathcal{S}(t)   - 	\mathcal{E}(t,\gamma, \bar{\Xi}_{t}) -  \mathcal{B}(\bar{\Xi}_{t})   \right) \leq 5 e^{-\gamma}.
	\end{align*}
	where
	\begin{align*}
		\mathcal{S}(t,\gamma)	: = &  \sqrt{\frac{\gamma}{t}} \left(    \sqrt{2\upsilon} \sigma(d)   +  \frac{ ||\theta ||_{1} }{2}   \right)~and~\mathcal{E}(t,\gamma,\bar{\Xi}_{t}) : = 2 ||\theta||_{1}  \sqrt{1-\bar{\Xi}_{t}} \sqrt{ e^{\gamma} \sum_{d=0}^{M}  \frac{ \upsilon \sigma(d)^{2}  }{2t   \gamma^{2} } }\\
		with~\mathcal{B}(\bar{\Xi}_{t}) : = & ||\theta||_{1} \frac{\bar{\Xi}_{t} }{M+1},~with~\bar{\Xi}_{t} : =  t^{-1} \sum_{s=1}^{t} \Xi_{s}.
	\end{align*}
\end{proposition}

\begin{proof}
	See Appendix \ref{app:avg.Y}.
\end{proof}

Despite the length of the proposition, its parts are quite intuitive. The term $\mathcal{S}( t,  \gamma  )$  controls the stochastic error that arises from the difference between $t^{-1}\sum_{s=1}^{t} Y_{s} = \sum_{d=0}^{M} t^{-1}\sum_{s=1}^{t} Y_{s}(d) 1\{ D_{s} =d  \}  $ and its conditional expectation $ \sum_{d=0}^{M} t^{-1}\sum_{s=1}^{t} \theta(d)   \delta_{s}(d)    $. This term is essentially of order $O(  \sqrt{ \gamma/ t}    )$. The term $\mathcal{E}(t,\gamma, \bar{\Xi}_{t} )$ arises from choosing the wrong treatment in the ``exploitation" part because the policy function depends on $\zeta^{\alpha}$ and not $\theta$. It is decreasing on the quantity $\bar{\Xi}_{t}$, which regulates the trade-off between exploitation and exploration and can be viewed as the degree of exploration. A higher degree of exploration will result on more information about the treatment and in turn a smaller likelihood of choosing the wrong treatment. It is also decreasing on $t$, reflecting the fact that as instances pass, the likelihood of choosing the wrong treatment also decreases. Finally, the term $\mathcal{B}(\bar{\Xi}_{t})$ is a non-random bias that stems from the ``exploration'' part of the policy function: With probability $\bar{\Xi}_{t} (M+1)$ the treatment is chosen at random, producing $\sum_{d=0}^{M} \theta(d)/(M+1)$.


If $\bar{\Xi}_{t} = o(1)$, i.e., if the exploration part of the policy function vanishes, then $\gamma$ can be chosen to diverge with $t$, however slowly, and so, with probability approaching one,  $t^{-1} \sum_{s=1}^{t} Y_{s}  $ converges to $\max_{d}\theta(d)$. 

The term $\mathcal{E}(t,\gamma,\bar{\Xi}_{t})+\mathcal{B}(\bar{\Xi}_{t})$ illustrates the so-called ``exploration vs. exploitation'' tradeoff and how it is regulated by $\bar{\Xi}_{t}$. This tradeoff suggests a choice for $\bar{\Xi}_{t}$ that balances $\mathcal{B}(\bar{\Xi}_{t})$ and $\mathcal{E}(t,\gamma,\bar{\Xi}_{t})$. Unfortunately, such a choice is infeasible as both terms depend on unknown quantities. Nevertheless, we can conclude that $\boldsymbol{\Xi} =1$ --- the choice used in RCTs --- will typically not be optimal. In fact, as $t$ increases, the ``optimal'' $\Xi_{t}$ will decrease to $0$, favoring ``exploitation'' to ``exploration''. We explore the choice of $\boldsymbol{\Xi}$ further, when we simulate our model in the next section.

\section{Model Simulations}\label{sec:simulations}

In this section, we present Monte Carlo simulations of our model using the generalized $\epsilon$-Greedy policy rule presented in Section \ref{subsec:policyrule}. The purpose of these simulations is to highlight different aspects of our analytical results and to provide a sense of the tightness of our analytic bounds. We consider the case with only two treatment arms, $D\in\{0,1\}$, and assume that $Y(0) \sim N(1,1)$ and $Y(1) \sim N(1.3,1)$. We assess the performance of our model according to the three outcomes outlined in Section \ref{sec:analytical_results}: concentrations bounds, probability of making a mistake, and average earnings. We simulate each experiment 1000 times, with each experiment lasting at most 1000 instances.

\subsection*{Multiple Priors, External Validity, Robustness}

We begin by illustrating how our setup weights the different models over the course of the experiment. Recall that to aggregate across several distinct subjective Bayesian models, our setup will average the posterior beliefs of each model using as weights, $\alpha_t^o(d)$ -- the posterior probability that model $o$ best fits the observed data within the class of models being considered. We demonstrated in Proposition \ref{pro:alpha.asymptotics.general} for the general case, and Lemma \ref{lem:alpha.properties} for Gaussianity, that if there exists an externally valid model among externally \emph{invalid} models, then $\alpha_t^o(d)$ will approach one for the externally valid model. Conversely, $\alpha_t^o(d)$ will approach zero if models are far from the true $\theta(d)$.

To illustrate this property, we simulate our model under different sets of priors. For each simulation, we assume that our policymaker has two sets of priors about the potential outcomes distributions. One is her initial set of priors, which we will assume are correct (i.e. $\zeta^{o}_o = \theta$) but diffuse (i.e. $\nu$=1).  For the other set of priors, we consider four alternative scenarios varying in their degree of stubbornness.

\begin{figure}[h]
    \includegraphics[width=\textwidth]{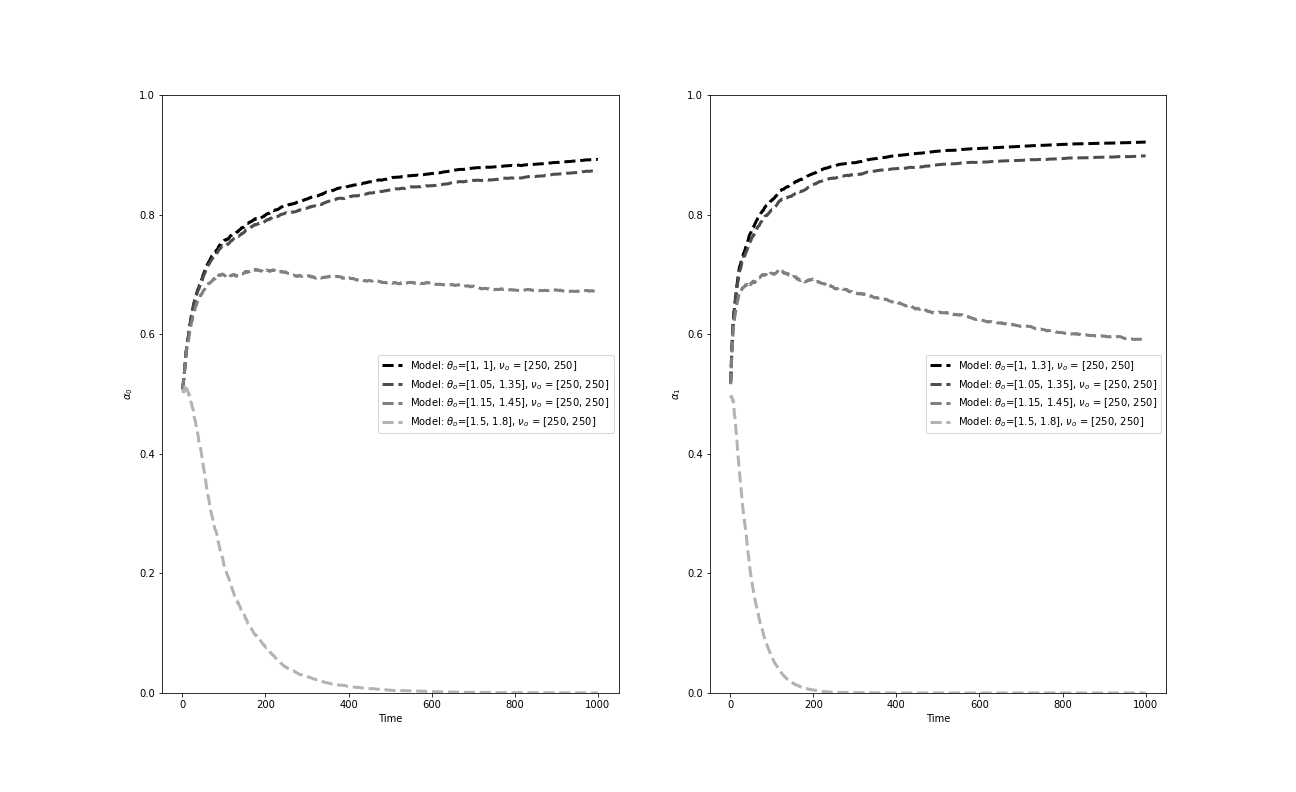}
    \caption{External Validity - $\alpha^o_t$}
    \label{fig:MultiPriors}
    \floatfoot{Notes: This figure plots $\alpha^0(d=0,x)$ (left plot) and $\alpha^0(d=1,x)$ (right plot) under two alternative sets of priors. For the confident model, the initial priors are: $\zeta^0_0=\zeta^1_0=\theta; \nu^0_0 = [1,1];\nu^1_0=[250,250]$. For the stubborn model, the initial priors are: $\zeta^0_0=\theta; \zeta^1_0=\theta+0.3; \nu^0_0 = [1,1];\nu^1_0=[250,250]$. These figures are based on 1,000 simulations using the following parameters: $\theta=[1,1.3]$, $\epsilon=0.5$.  }
\end{figure}

In Figure \ref{fig:MultiPriors}, we plot $\alpha_t^o(d)$ corresponding to the second set of priors over the course of the experiment. The graph on the left is for the $d=0$ arm and the one on the right is for the $d=1$ arm. Each line corresponds to a different set of priors, and the lighter the line, the more stubborn the prior. Starting with the top and darkest line, we see that $\alpha_t^o(d)$ increases over time putting more and more weight on an externally valid model. By the end of the experiment, $\alpha_t^o(d)$ is close to 95\% for both arms. As we consider more stubborn models, we can see that the corresponding $\alpha_t^o(d)$ becomes smaller. So much so that for extremely stubborn models (i.e. the lowest line) $\alpha_t^o(d)$ becomes essentially zero by the $600^{th}$ instance. This is why we interpret the parameter $\alpha_t^o(d)$ as a measure of external validity: the more externally valid the model, the higher the corresponding $\alpha_t^o(d)$.

\begin{figure}[h]
    \includegraphics[width=\textwidth]{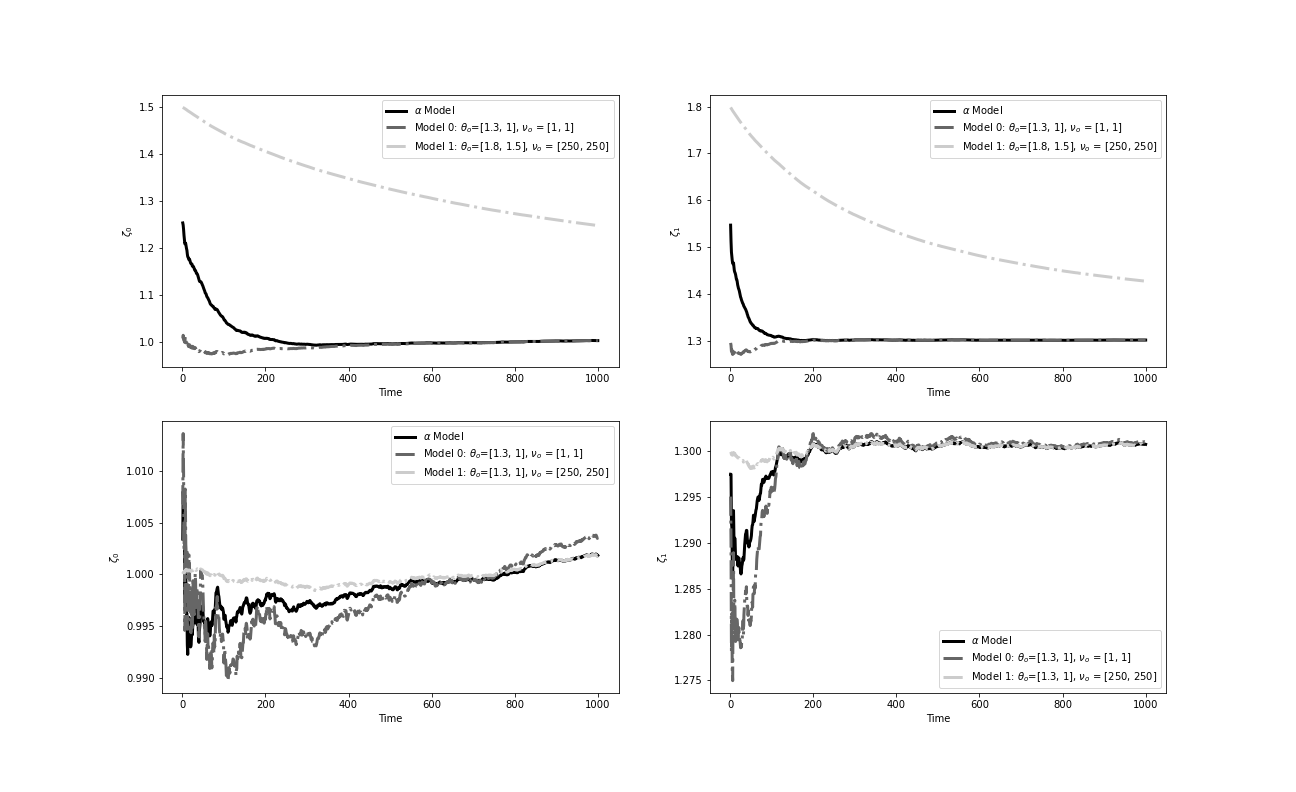}
    \caption{Posterior Beliefs Over Time, Holding Behavior Constant}
    \label{fig:zeta_beliefs}
    \floatfoot{Notes: This figure plots the policymakers posterior beliefs (i.e. $[\zeta^o_t(0,x),\zeta^o_t(1,x)]$) over time, distinguishing between two alternative sets of initial priors. In the top panel, one of the initial priors is stubborn; and in the bottom panel, one of the initial priors is confident. For the stubborn model, the initial priors are: $\zeta^0_0=\theta; \zeta^1_0=\theta+0.3; \nu^0_0 = [1,1];\nu^1_0=[250,250]$.  For the confident model, the initial priors are: $\zeta^0_0=\zeta^1_0=\theta; \nu^0_0 = [1,1];\nu^1_0=[250,250]$. These figures are based on 1,000 simulations using the following parameters: $\theta=[1,1.3]$, $\epsilon=0.5$. }
\end{figure}

An important feature of how we aggregate across models is that it generates a robustness property. Because $\alpha_t^o(d)$ will place less weight on models that are not externally valid, over time they will have limited influence on the PM's beliefs and consequent decisions. We illustrate this Figure \ref{fig:zeta_beliefs}. In the top graphs, we plot the policymaker's posterior beliefs about the mean of the potential outcome distributions over time. The plot distinguishes between three posterior means. The bottom (dashed) line corresponds to one set of priors, which we assume to be unbiased (i.e. $\zeta^{o}_o = \theta$), but diffuse.  The top (dash-dotted) line refers to an alternative set of priors, which contains some degree of stubbornness (i.e. $\zeta^{o}_o = \theta+.5, \nu=250$). The middle (solid) line comes from the combined model, which is a weighted average of the two sets of priors using $\alpha_t^o(d)$ as weights. We see that even though our policymaker starts with a stubborn prior, the combined model converges relatively quickly to the non-stubborn model. This is the result of both the oracle property – concentrating on the least stubborn model – and robustness property – putting less and less weight on sufficiently stubborn models.

In the bottom graphs, we consider the case in which the alternative model is confident. Thus, both sets of priors are unbiased; the alternative prior simply comes with a higher degree of conviction. Because both priors are correct, the combined model does not immediate converge to one of the models as we saw in case with stubborn priors. As we started in Lemma \ref{lem:alpha.properties}, our parameter $\alpha$ is more responsive to bias than conviction.

\subsection*{Concentration Bounds}

\begin{figure}[h]
    \includegraphics[width=1.1\textwidth]{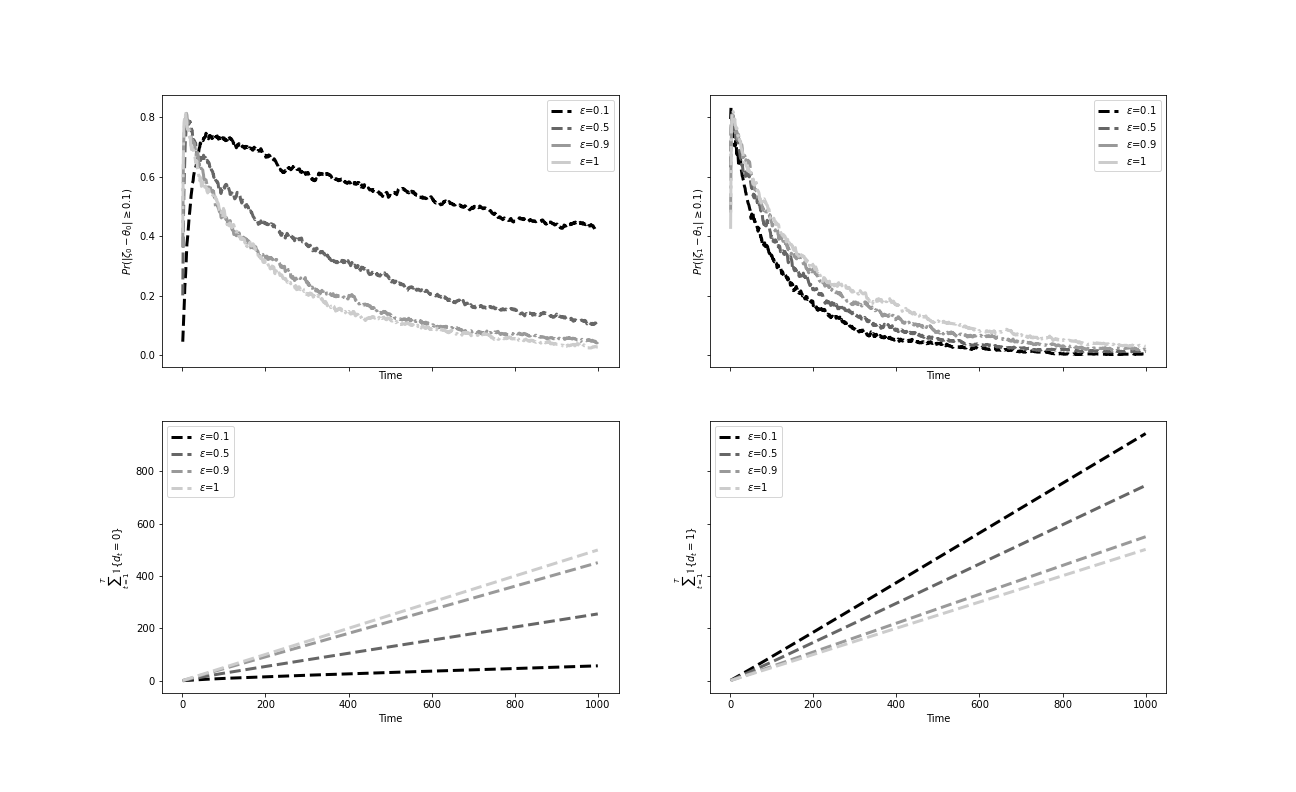}
    \caption{Concentration Bounds and Frequency of Play}
    \label{fig:CB_epsilon}
    \floatfoot{Notes: The top panel plots concentration bounds over time for different values of $\epsilon$. The bottom panel plots the number of times the experimental arm was played at time $t$ for different values of $\epsilon$. The graphs on the left correspond treatment arm $d=0$; the graphs on the right correspond to treatment arm $d=1$. These figures are based on 1,000 simulations using the following parameters: $\theta=[1,1.3]$; $\zeta_0^o=\theta$; $\zeta_1^o=\theta$; $\nu_0^o=[1,1]$; $\nu_1^o=[1,1]$. }
\end{figure}

\paragraph{Effects of $\epsilon$.} We now simulate our model's concentration bounds and some its key properties. Recall from Remark \ref{remark:cb} in Section \ref{sec:zeta.concentration}, the concentration rate increases with the parameter $\epsilon$. We demonstrate this property in the top panel of Figure \ref{fig:CB_epsilon}, in which we plot  concentration bounds for three different values of $\epsilon \in \{0.1,0.5,0.9\}$. That is, for a given $\epsilon$, we compute the difference over time between the policymaker’s posterior belief of the true mean, $\zeta^o_t(d)$,  and the true mean, $\theta(d)$. We then plot the probability that these differences are greater than 0.1.  For these simulations, we assume that our policymaker has correct, but diffuse priors (i.e. $\zeta^o_0 =\theta$ and $\nu^o_0=[1,1]$).

In the top panel, we see that except for early on, our concentration bounds decrease over time and in the case of $\zeta^o_t(0)$ decrease faster, the higher the $\epsilon$. For instance, after 1000 instances, $Pr(\zeta^o_t(0)-\theta(0)>0.1)$ is almost zero for the case of $\epsilon=0.9$, but is still close to 0.5 for $\epsilon=0.1$. For the other treatment arm, the patterns are reversed. All three lines decrease relatively quickly, with the lower $\epsilon$ lines decreasing faster.

The intuition for these patterns is straightforward and speaks to the point about frequency of play in Remark \ref{remark:cb}. When the PM selects a treatment arm, she will only learn about the distribution of potential outcomes for that arm. As she become more confident in which arm is better, she will play the other arm only when forced to by the $\epsilon$-greedy algorithm. In this case, the higher the $\epsilon$ the more the PM will be forced to play treatment $d=0$ and the more she learns about $\theta(0)$. We can see this clearly in the bottom panel, which depicts the cumulative number of times the treatment has been played over time by different values of $\epsilon$'s. As we compare the two panels, the more we play a particular arm, the more we learn about it, and the sooner our beliefs converge to the truth.

\begin{figure}[h]
\center
    \includegraphics[width=\textwidth]{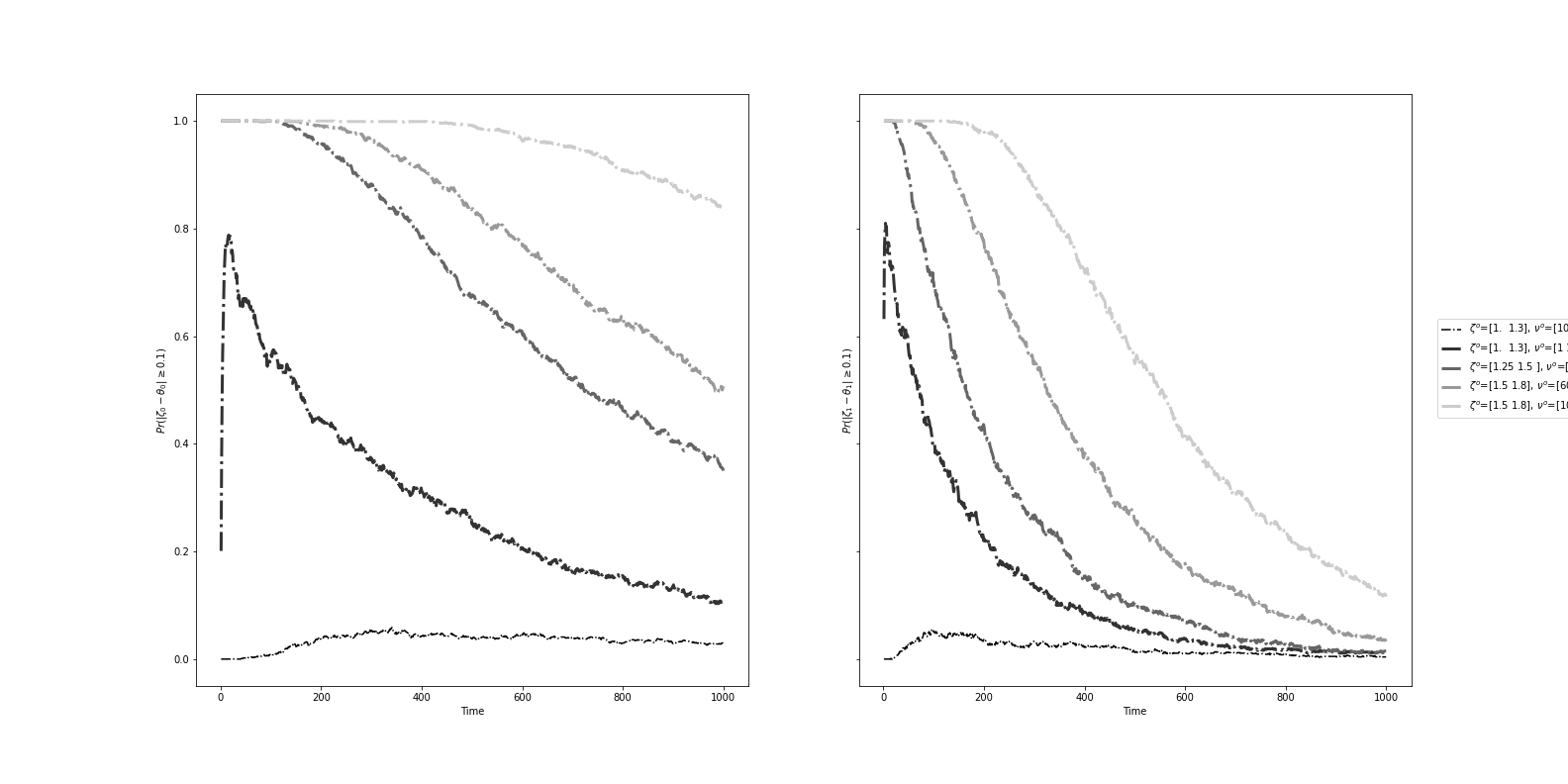}
    \caption{Concentration Bounds by Model Stubbornness}
    \label{fig:CB_Priors}
    \floatfoot{Notes: The figure plots concentration bounds over time for different degrees of model stubbornness. The lines in these plots appear in descending order of stubbornness, with the top line being most stubborn and the bottom line being the most confident. The graphs on the left correspond treatment arm $d=0$; the graphs on the right correspond to treatment arm $d=1$. These figures are based on 1,000 simulations using the following parameters: $\theta=[1,1.3]$, $\epsilon=0.5$. The initial priors are specified in the legend.}
\end{figure}

\paragraph{Effects of Priors.} In Figure \ref{fig:CB_Priors}, we investigate the effects of different priors on the concentration bounds. We plot different concentration bounds for priors with different degrees of stubbornness and confidence. For example, in the bottom two lines, we consider two unbiased priors, but with different levels of confidence. According to Remark \ref{remark:cb}, concentration rates increase as the degree of conviction increases and this is precisely what we see. It is also the case, that the concentration rate decreases faster with less stubborn models. We can see this pattern clearly by comparing the top two lines. By comparing the two middle lines, we can also see that conditional on the degree of stubbornness, the higher the bias, the slower the concentration rate. Lastly, as before, the concentration rates for $\theta(1)$ tend to be faster than those for $\theta(0)$ because of the frequency of play.

\subsection*{Probability of Making a Mistake}

In Section \ref{sec:PoM}, we defined a mistake as recommending a treatment arm different from the one that yields the largest expected effect at the instance in which the experiment was stopped. In Figure \ref{fig:stopping}, we plot the average stopping period (left axis) and the probability of making a mistake at that stopping period (right axis) by $\epsilon$. It is clear from the graph that the more we experiment across treatment arms (i.e., higher $\epsilon$), the faster we stop the experiment. This makes sense. As we experiment more, the data become more IID and we are able to better learn the true means of the potential outcome distributions. According to these simulations, the degree of experimentation does not have to be particularly high. Even though at low levels of $\epsilon$ the experiment lasts for almost its entire duration, the drop off is fairly quick. Once $\epsilon$ is greater than 0.5, the difference gained in stopping periods from additional experimentation is minimal.

Shorter stopping periods do not come at the cost of making more mistakes. This result is to some extent an artifact of our stopping rule, whose parameters control the probability of type I errors. As the graph depicts, the probability of making a mistake varies little with $\epsilon$ and is always below 1\%.

\begin{figure}[h]
    \includegraphics[width=.7\textwidth]{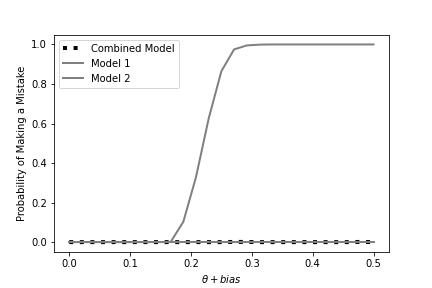}
    \caption{Probability of Making a Mistake by Model Bias}
    \label{fig:bias}
    \floatfoot{Notes: The figure plots the probability of making a mistake at the stopping period by the degree of bias in model 1's initial priors. These figures are based on 1,000 simulations using the following parameters: $\theta=[1,1.3]$; $\nu^0_0=\nu^1_0=[250,250]$; $\zeta^0_0=[\theta(0)+bias,\theta(1)-bias]$ ; $\zeta^1_0=\theta$, $\epsilon=0.5$. }
\end{figure}

In Figure \ref{fig:bias}, we explore how the initial priors affect the probability of making a mistake. We again consider two sets of priors, both with $\nu_0=[250,250]$. One, however, is confident with $\zeta^o_0 = \theta$, whereas the other is stubborn, with $\zeta^o_0 = [\theta(0)+\delta,\theta(1)-\delta]$, where $\delta$ is indicated by a point on the x-axis. For $\delta \in (0,0.15)$, the priors are biased, but have a proper ranking of the treatment arms. For $\delta > 0.15$, the priors are not only biased, but reverse the ranking of the arms. On the y-axis, we plot the probability of making a mistake associated with each set of priors and for the combined model.

We can see that for $\delta \in (0,0.15)$, the probability of making mistake is small, less than 1\%, for all three models. But once $\delta > 0.15$, and the ranking of treatment arms are reversed, the probability of making a mistake for the stubborn model increases significantly and approaches 1 by $\delta \geq 0.3$. Importantly, the probability of making a mistake for the combined model mirrors the one for the confident model, which again illustrates the robustness property of $\alpha^o_t(d)$.

\subsection*{Expected Earnings}

\begin{figure}[h]
    \includegraphics[width=.7\textwidth]{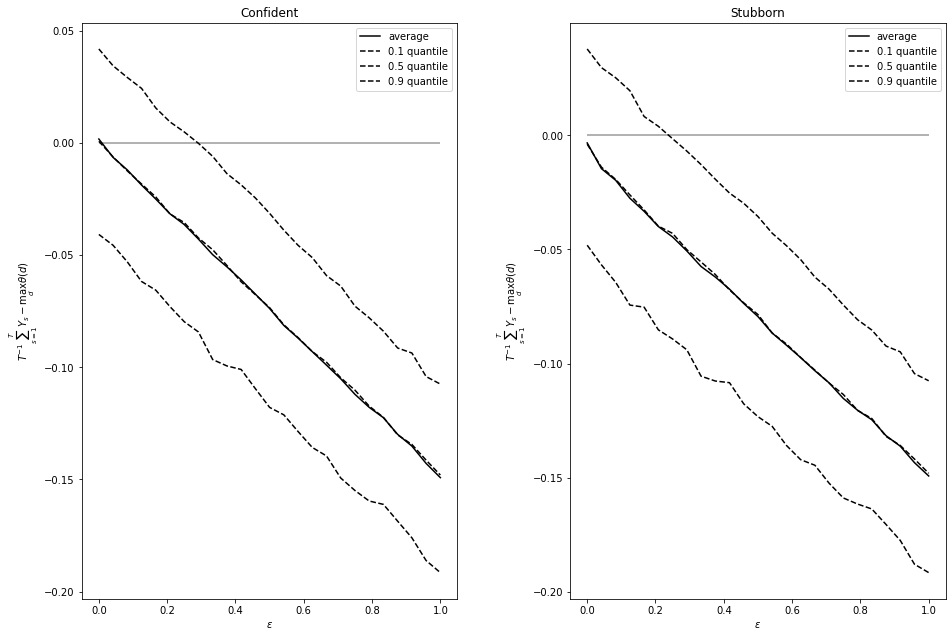}
    \caption{Relative Average Earnings During the Experiment}
    \label{fig:earnings}
    \floatfoot{Notes: This figure plots by $\epsilon$, the average earnings net of maximal earnings. These figures are based on 1,000 simulations using the following parameters: $\theta=[1,1.3]$; $\zeta_0^o=\theta$; $\zeta_1^o=\theta$; $\nu_0^o=[1,1]$; $\nu_1^o=[1,1]$.}
\end{figure}

The final outcome we evaluate is expected earnings. According to Proposition \ref{pro:avg.Y}, the distance between the average outcomes and maximum expected outcome is decreasing in $\epsilon$. In Figure \ref{fig:earnings}, we plot by $\epsilon$, the difference between the policymaker's average impact and the maximum expected outcome, $\max_d\theta(d)$, for an experiment that lasts 1000 instances. The figure also distinguishes between our two familiar sets of priors, a confident one and a stubborn one.

Two important observations emerge from this figure. First, there is a steep negative monotonic relationship between expected earning and $\epsilon$. In fact, the 10\% quantile of the average earnings distribution for $\epsilon=0.10$ lies above the 90\% quantile of the average earnings distribution for $\epsilon=0.90$. Second, if we compare across the two plots, we can see that starting off with a stubborn prior affects average earnings, but only minimally. Again, this result is a product of the robustness property that our model aggregation approach provides.

\begin{figure}[h]
    \includegraphics[width=.7\textwidth]{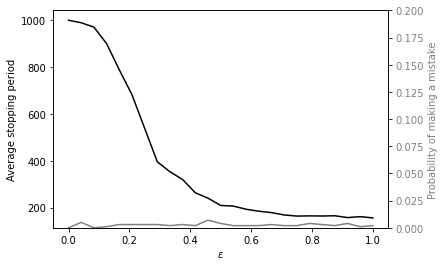}
    \caption{Stopping Period and Probability of Making a Mistake}
    \label{fig:stopping}
    \floatfoot{Notes: This figure plots the average stopping period (left axis) and the probability of making a mistake at the stopping period (right axis) by $\epsilon$. These figures are based on 1,000 simulations using the following parameters: $\theta=[1,1.3]$; $\zeta_0^o=\theta$; $\zeta_1^o=\theta$; $\nu_0^o=[1,1]$; $\nu_1^o=[1,1]$; $B=100$.  }
\end{figure}

The fact that average earnings declines with experimentation does not imply that our policymaker should set $\epsilon$ close to zero. Because as we saw in Figure \ref{fig:stopping}, lower $\epsilon$'s result in longer experiments, which can come with costs. Moreover, as we show in Proposition \ref{pro:stopping.alpha}, the upper bound the probability of making a mistake is weakly smaller for higher levels of $\epsilon$. Thus, to properly capture the experimentation versus exploitation tradeoff inherent in multi-armed bandit problems, we need to specify a payoff function.

We consider the following payoff function:
\begin{align}\label{eq:payoffs}
	\Pi^{I}_{\beta,c}  =  & \sum_{d=0}^{M}  \sum_{t=0}^{T^{\ast}}  \beta^{t}  1\{ D_{t} = d \}   ( Y_{t}(d)  - c_1) +   \sum_{t=T^{\ast}+1}^{\infty} \beta^{t} 1\{ D_{T^{\ast}} = d \}  (\theta(d)-c_2)\\
  =  & \sum_{d=0}^{M}  \sum_{t=0}^{T^{\ast}}  \beta^{t}  1\{ D_{t} = d \}   ( Y_{t}(d)  - c_1) +   \frac{ \beta^{T^{\ast}+1}} {1-\beta}  1\{ D_{T^{\ast}} = d \}  (\theta(d)-c_2) 	
\end{align}
where $c_1$ indicates the costs of running the experiment, $c_2$ cost of administering the treatment, $\beta^t$ represents a discount factor, and $T^{\ast}$ denotes the stopping period. This payoff function comprises of two parts. The first part is the earnings during the experiment net of cost. The second part captures the expected future benefits under the chosen treatment, net of cost.

\begin{figure}[h]
    \includegraphics[width=.7\textwidth]{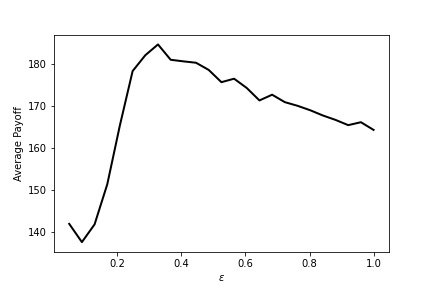}
    \caption{Experimentation versus Exploitation -- Expected Payoffs }
    \label{fig:payoffs}
    \floatfoot{Notes: This figure plots by $\epsilon$, the expected payoffs as defined by Equation \ref{eq:payoffs}. These figures are based on 1,000 simulations using the following parameters: $\theta=[1,1.3]$; $\zeta_0^o=\theta$; $\zeta_1^o=\theta$; $\nu_0^o=[1,1]$; $\nu_1^o=[1,1]$; $B=100$; $\beta^t =0.994$; $c=1.15$; $\lambda = 1,100$. }
\end{figure}

In Figure \ref{fig:payoffs}, we compute the payoff function for our model simulations by different values of $\epsilon$. In contrast with the previous figure, we see that the average payoffs are increasing with $\epsilon$ until approximately $\epsilon=0.38$, at which point the payoffs start to decline. While this ``optimal'' value of $\epsilon$ is clearly a function of an arbitrary set of parameter choices, our conjecture is that the inverted u-shape relationship is likely to hold more generally, suggesting that some combination of experimentation and exploitation is optimal.

\section{Debt Refinancing Experiment}\label{Sec:App}

In this section, we present a real-world experiment to show that by incorporating multiple priors, our policymaker can stop the experiment sooner without significantly increasing the probability of making a mistake. This results in large performance gains relative to a standard RCT.

In January 2024, we partnered with a major private bank in Argentina to implement our algorithm for marketing their debt-refinancing program. At that time, Argentina was experiencing a surge in household debt due to high inflation and interest rates. Consequently, the bank, one of the country's largest financial service providers, had over 300,000 clients who were delinquent on their loans and targeted for debt refinancing. These clients are located across the country and are predominantly from lower-income households.

The bank markets these refinancing loans via email, informing clients about the basic terms of the contract and providing a hyperlink to connect them with a bank representative. Getting clients to click on this link is a crucial first step in the refinancing process. As a result, the bank continuously experiments with the content and presentation of these emails to maximize engagement.

In February 2024, the bank ran a randomized control trial on a random sample of around 15,491 clients. The bank was interested in learning whether clients were more likely to click on the refinancing link and refinance their loans if the email displayed the monthly payments associated with their loan instead of the interest rate. With interest rates hovering above 100\%, the bank thought that an alternative framing might help avoid with any potential sticker shock (see Appendix Figure \ref{fig:debt_rct} for the treatment and control messaging).    

We present the results of this experiment in Table \ref{tab:debt_tab1}. Each column is a separate regression, corresponding to a particular region of Argentina. In columns 1-6, the dependent variable is whether the client clicked on the link; whereas, in columns 7-12, the dependent variable is whether the individual refinanced their loan. For ease of interpretation, the dependent variables have been scaled by 100. The independent variable ($1{Interest Rate}$) is an indicator for whether the email provided the interest rate. Otherwise, the email provided just the implied monthly payments (and no interest rate).

From the table, we see that in contrast to the bank’s priors, clients are much more likely to click on the link when the interest rate is displayed as opposed to the monthly payments. For example, in the west region of Buenos Aires (BA Zona Oeste) clients are 4.6 percentage points more likely to click on the refinancing link under the interest rate treatment. This effect is more than a doubling of the percentage of clients who clicked on the hyperlink in the control group (i.e., 3.049 percent). The effects are also quite heterogeneous across the different regions, ranging from the 4.6 percentage points BA Zone Oeste to a statistically insignificant 0.31 percentage points BA Zona Norte. It is also the case that clicks are necessary, but not sufficient, for getting the client to refinance. Only when the treatment effects on clicks are sufficiently large, as is the case in BA Zona Oeste or NE, do we see a corresponding treatment effect on refinance rates.   

\textit{Adaptive Experiment.} In June 2024, as interest rates began to fall below 50\%, the bank was interested in rerunning their experiment, but adaptively and using our algorithm. This second experiment provided an ideal setting in which to pilot our algorithm. Even though the previous experiment had provided support for an interest rate messaging, it was no longer clear whether the previous treatment effects would extrapolate to the new period given how much the macroeconomy had changed. Moreover, with all the treatment effect heterogeneity across regions, our algorithm offered a robust approach to run the experiment by region, without sacrificing information from the other regions. 

For this adaptive experiment, our experimental design consisted of the same two treatment arms as before (interest rate vs monthly payments) but sent out sequentially in batches of approximately forty emails. In the context of our setup, we assumed the policymaker (i.e., the bank) wanted to learn about the average click rate of each treatment arm. For this experiment, we decided to focus on click rate instead of refinancing rates, because the feedback loop for click rates is significantly shorter (1 to 2 days versus 1 to 2 weeks) and the relationship between click rates and refinance rates is both quite clear and stable. For our policy functions, we decided to use the epsilon greedy algorithm with $\epsilon=0.20$. We chose this value of epsilon based on simulations using data from the previous experiment. The stopping rule is the one specified in our setup, above. We set parameters of the stopping rule such that the probability of making a mistake was no larger than 1 percent.  We required that each experiment last for at least 4 batches, but no longer than 14 batches (per the bank's requirement), unless our algorithm stopped the experiment sooner.

The experiment was conducted three regions of Argentina, the capital city of Buenos Aires (CABA), the southern region of Buenos Aires (BA Zona Sur), and the northwest part of the country. Together these regions account for 20 percent of the population of Argentina. For priors, we used the average click rates estimated by treatment using data from the first experiment, but as a stress test of our algorithm excluded the own region (e.g. for the experiment in CABA we did not include the priors from CABA). We did include a ``diffuse prior'', which sets the average click rate of each treatment arm equal to zero and the variance close to zero. As we described in our setup, each of these other regions is a model that the policymaker will form posterior beliefs over as she accumulates more information from clients living in BA Zona Sur or CABA. She then aggregates these beliefs to determine how to allocate her batch of emails.

In Table \ref{tab:debt_tab2}, we report the results of the experiment. The table reports for each prior source by treatment arm: the posterior belief, the posterior probability the source fits the observed data, and the inverse of the variance. At the bottom of the table, we also report by treatment arm: the policymaker's subjective, aggregated, posterior beliefs at the end of the experiment, the average click rates among clients, and the number of emails sent.

In CABA, the experiment was stopped after sending out 294 emails. Of these, 84.4 percent (248 emails) included information about the interest rate, and 5.2 percent of recipients clicked on the message. In contrast, among the 46 emails that provided information on monthly payments, only 2.1 percent of clients clicked. This difference in click rates (3.07 percent) is slightly larger than the difference in beliefs (2.73 percent) because the initial priors for the control arms were higher.

The results highlight two key aspects of our approach. First, with a 1 percent probability of making a mistake, the policymaker could stop the experiment even though the difference in average click rates between the treatment arms was just 3.07 percent. In a standard randomized controlled trial (RCT), at least 600 emails would be needed to detect this difference with 95 percent confidence. If we had used the same epsilon-greedy algorithm without incorporating prior information, more than 2,000 emails would have been required to stop the experiment.\footnote{This calculation assumes that the sample means remain constant with additional batches.} By incorporating prior information, we effectively increased our sample size, allowing us to stop the experiment earlier.

Second, the table shows how our method combines different data sources. Given the similarities in initial priors and the speed with which we stopped the experiment, the weights assigned to the different regions are similar. The highest weight for the interest rate treatment appears in the Northwest Region, which, according to Definition 1, is the region most externally valid to CABA.

The results in the southern region of Buenos Aires (BA Zona Sur) were qualitatively similar to those in CABA. The experiment stopped after 366 emails. In this region, a standard RCT would have required at least 750 emails, and an epsilon-greedy algorithm without priors would have needed over 2,800 emails. However, the experiment in the Northwest region of Argentina was only stopped because the bank decided to end it. Unlike the other regions, clients in this region responded more to the monthly payment messaging. The average click rate for the monthly payment emails was 5.1 percent, compared to just 3.9 percent for the interest rate emails. Since our priors had suggested the opposite, the experiment did not stop within 14 batches, which would have also been the case in a simple RCT. In hindsight, it would have been interesting to see the experiment's results if we had included priors that had the opposite predictions.

\section{Conclusions}\label{sec:conclusions}

This paper presents a conceptual framework for how to incorporate prior sources of information into the design of a sequential experiment. An obvious issue is how to handle the potential lack of external validity of each of these sources. We address this issue by first presenting a formal definition of external validity that can be used to differentiate sources with different degrees of external invalidity and second, by showing that our framework is robust to including externally-invalid sources. This last property relaxes the burden on the policymaker of having to correctly choose relevant sources of information based on limited ex-ante information. As ``stubborn'' sources are harder to discard, we believe it is useful to incorporate many priors, including versions that are diffuse.

For the common problem of learning about average treatment effects, we show that our framework offers several nice properties. As we illustrated for the case of a debt refinancing program, these properties translate into substantial gains in performance --- such as reducing the duration of experiment and increasing the average payoffs while keeping an acceptable probability of making a mistake ---  over both standard RCTs and adaptive experiments.

\bibliography{references}

\begin{thebibliography}{}

\bibitem[Agrawal and Goyal, 2017]{AG}
Agrawal, S. and Goyal, N. (2017).
\newblock Near-optimal regret bounds for thompson sampling.
\newblock {\em J. ACM}, 64(5).

\bibitem[Angrist and Fernández-Val, 2013]{angristfv:2013}
Angrist, J.~D. and Fernández-Val, I. (2013).
\newblock {\em ExtrapoLATE-ing: External Validity and Overidentification in the
  LATE Framework}, volume~3 of {\em Econometric Society Monographs}, pages
  401--434.
\newblock Cambridge University Press.

\bibitem[Athey and Imbens, 2019]{AtheyImbens:2019}
Athey, S. and Imbens, G. (2019).
\newblock Machine learning methods economists should know about.

\bibitem[Banerjee et~al., 2015a]{Banerjee_etalb:2015}
Banerjee, A., Duflo, E., Goldberg, N., Karlan, D., Osei, R., Parient{\'e}, W.,
  Shapiro, J., Thuysbaert, B., and Udry, C. (2015a).
\newblock A multifaceted program causes lasting progress for the very poor:
  Evidence from six countries.
\newblock {\em Science}, 348(6236).
\newblock Publisher Copyright: {\textcopyright} 2015 by the American
  Association for the Advancement of Science; all rights reserved.

\bibitem[Banerjee et~al., 2015b]{Banerjee_etal:2015}
Banerjee, A., Karlan, D., and Zinman, J. (2015b).
\newblock Six randomized evaluations of microcredit: Introduction and further
  steps.
\newblock {\em American Economic Journal: Applied Economics}, 7(1):1--21.

\bibitem[Bisbee et~al., 2017]{Bisbee_etal:2017}
Bisbee, J., Dehejia, R., Pop-Eleches, C., and Samii, C. (2017).
\newblock Local instruments, global extrapolation: External validity of the
  labor supply–fertility local average treatment effect.
\newblock {\em Journal of Labor Economics}, 35(S1):S99--S147.

\bibitem[Buchanan et~al., 2018]{Buchanan_etal:2018}
Buchanan, A.~L., Hudgens, M.~G., Cole, S.~R., Mollan, K.~R., Sax, P.~E., Daar,
  E.~S., Adimora, A.~A., Eron, J.~J., and Mugavero, M.~J. (2018).
\newblock Generalizing evidence from randomized trials using inverse
  probability of sampling weights.
\newblock {\em Journal of the Royal Statistical Society: Series A (Statistics
  in Society)}, 181(4):1193--1209.

\bibitem[Cesa-Bianchi and Lugosi, 2006]{cesa2006prediction}
Cesa-Bianchi, N. and Lugosi, G. (2006).
\newblock {\em Prediction, learning, and games}.
\newblock Cambridge university press.

\bibitem[Chabrier et~al., 2016]{Chabrier_etal:2016}
Chabrier, J., Cohodes, S., and Oreopoulos, P. (2016).
\newblock What can we learn from charter school lotteries?
\newblock {\em Journal of Economic Perspectives}, 30(3):57--84.

\bibitem[Chernoff, 1959]{chernoff}
Chernoff, H. (1959).
\newblock Sequential design of experiments.
\newblock {\em The Annals of Mathematical Statistics}, 30(3):755--770.

\bibitem[Dehejia et~al., 2021]{Dehejia_etal:2021}
Dehejia, R., Pop-Eleches, C., and Samii, C. (2021).
\newblock From local to global: External validity in a fertility natural
  experiment.
\newblock {\em Journal of Business \& Economic Statistics}, 39(1):217--243.

\bibitem[DellaVigna et~al., 2020]{DellaVigna_etal:2020}
DellaVigna, S., Otis, N., and Vivalt, E. (2020).
\newblock Forecasting the results of experiments: Piloting an elicitation
  strategy.
\newblock {\em AEA Papers and Proceedings}, 110:75--79.

\bibitem[DellaVigna and Pope, 2018]{DellaVignaPope:2018}
DellaVigna, S. and Pope, D. (2018).
\newblock Predicting experimental results: Who knows what?
\newblock {\em Journal of Political Economy}, 126(6):2410--2456.

\bibitem[Dimakopoulou et~al., 2017]{dimakopoulou2017estimation}
Dimakopoulou, M., Zhou, Z., Athey, S., and Imbens, G. (2017).
\newblock Estimation considerations in contextual bandits.
\newblock {\em arXiv preprint arXiv:1711.07077}.

\bibitem[Epstein and Schneider, 2003]{EPSTEIN20031}
Epstein, L.~G. and Schneider, M. (2003).
\newblock Recursive multiple-priors.
\newblock {\em Journal of Economic Theory}, 113(1):1--31.

\bibitem[Garcia and Saavedra, 2022]{Garcia_Saavedra:2022}
Garcia, S. and Saavedra, J. (2022).
\newblock Conditional cash transfers for education.
\newblock Working Paper 29758, National Bureau of Economic Research.

\bibitem[Gelman and Carlin, 2014]{GelmanCarlin:2014}
Gelman, A. and Carlin, J. (2014).
\newblock Beyond power calculations: Assessing type s (sign) and type m
  (magnitude) errors.
\newblock {\em Perspectives on Psychological Science}, 9(6):641--651.
\newblock PMID: 26186114.

\bibitem[Gelman and Pardoe, 2006]{GelmanPardoe:2006}
Gelman, A. and Pardoe, I. (2006).
\newblock Bayesian measures of explained variance and pooling in multilevel
  (hierarchical) models.
\newblock {\em Technometrics}, 48(2):241--251.

\bibitem[Imai and Ratkovic, 2013]{ImaiRatkovi:2013}
Imai, K. and Ratkovic, M. (2013).
\newblock {Estimating treatment effect heterogeneity in randomized program
  evaluation}.
\newblock {\em The Annals of Applied Statistics}, 7(1):443 -- 470.

\bibitem[Joseph~Hotz et~al., 2005]{Hotz_etal:2005}
Joseph~Hotz, V., Imbens, G.~W., and Mortimer, J.~H. (2005).
\newblock {Predicting the efficacy of future training programs using past
  experiences at other locations}.
\newblock {\em Journal of Econometrics}, 125(1-2):241--270.

\bibitem[Karlan and List, 2007]{Karlan_List:2007}
Karlan, D. and List, J.~A. (2007).
\newblock Does price matter in charitable giving? evidence from a large-scale
  natural field experiment.
\newblock {\em American Economic Review}, 97(5):1774--1793.

\bibitem[Karlan and List, 2020]{Karlan_List:2020}
Karlan, D. and List, J.~A. (2020).
\newblock How can bill and melinda gates increase other people's donations to
  fund public goods?
\newblock {\em Journal of Public Economics}, 191:104296.

\bibitem[Kasy and Sautmann, 2021]{KasySautmann:2021}
Kasy, M. and Sautmann, A. (2021).
\newblock Adaptive treatment assignment in experiments for policy choice.
\newblock {\em Econometrica}, 89(1):113--132.

\bibitem[Klibanoff et~al., 2005]{KMM2005}
Klibanoff, P., Marinacci, M., and Mukerji, S. (2005).
\newblock A smooth model of decision making under ambiguity.
\newblock {\em Econometrica}, 73(6):1849--1892.

\bibitem[Kowalski, 2016]{Kowalski:2016}
Kowalski, A.~E. (2016).
\newblock {Doing More When You're Running LATE: Applying Marginal Treatment
  Effect Methods to Examine Treatment Effect Heterogeneity in Experiments}.
\newblock NBER Working Papers 22363, National Bureau of Economic Research, Inc.

\bibitem[Lai and Robbins, 1985]{LaiRobbins:1985}
Lai, T. and Robbins, H. (1985).
\newblock Asymptotically efficient adaptive allocation rules.
\newblock {\em Advances in Applied Mathematics}, 6(1):4--22.

\bibitem[Meager, 2020]{Meager:2020}
Meager, R. (2020).
\newblock Aggregating distributional treatment effects: Abayesian hierarchical
  analysis of the microcredit literature.

\bibitem[Pan and Yang, 2010]{Pan:2010}
Pan, S.~J. and Yang, Q. (2010).
\newblock A survey on transfer learning.
\newblock {\em IEEE Transactions on Knowledge and Data Engineering},
  22(10):1345--1359.

\bibitem[Qin and Russo, 2022]{qinrusso}
Qin, C. and Russo, D. (2022).
\newblock Adaptivity and confounding in multi-armed bandit experiments.
\newblock {\em arXiv}.

\bibitem[Robbins, 1992]{robbins1992empirical}
Robbins, H.~E. (1992).
\newblock {\em An empirical Bayes approach to statistics}.
\newblock Springer.

\bibitem[Russo, 2016]{Russo2016}
Russo, D. (2016).
\newblock Simple bayesian algorithms for best arm identification.

\bibitem[Russo and Van~Roy, 2016]{russo2016information}
Russo, D. and Van~Roy, B. (2016).
\newblock An information-theoretic analysis of thompson sampling.
\newblock {\em The Journal of Machine Learning Research}, 17(1):2442--2471.

\bibitem[Schlaifer and Raiffa, 1961]{schlaifer1961applied}
Schlaifer, R. and Raiffa, H. (1961).
\newblock {\em Applied statistical decision theory}.

\bibitem[Stuart et~al., 2011]{Stuart_etal:2011}
Stuart, E.~A., Cole, S.~R., Bradshaw, C.~P., and Leaf, P.~J. (2011).
\newblock The use of propensity scores to assess the generalizability of
  results from randomized trials.
\newblock {\em Journal of the Royal Statistical Society: Series A (Statistics
  in Society)}, 174(2):369--386.

\bibitem[Thomke, 2020]{thomke2020}
Thomke, S. (2020).
\newblock {\em Experimentation Works: The Surprising Power of Business
  Experiments}.
\newblock Harvard Business Review Press.

\bibitem[Thompson, 1933]{Thompson:1933}
Thompson, W.~R. (1933).
\newblock On the likelihood that one unknown probability exceeds another in
  view of the evidence of two samples.
\newblock {\em Biometrika}, 25(3-4):285--294.

\bibitem[Vivalt, 2020]{Vivalt:2020}
Vivalt, E. (2020).
\newblock {How Much Can We Generalize From Impact Evaluations?}
\newblock {\em Journal of the European Economic Association}, 18(6):3045--3089.

\bibitem[Vivalt and Coville, 2021]{VivaltColville:2021}
Vivalt, E. and Coville, A. (2021).
\newblock How do policy-makers update their beliefs?

\bibitem[Wald, 1945]{wald}
Wald, A. (1945).
\newblock Sequential tests of statistical hypotheses.
\newblock {\em The Annals of Mathematical Statistics}, 16(2):117--186.

\bibitem[Watkins, 1989]{Watkins:1989}
Watkins, C. J. C.~H. (1989).
\newblock {\em Learning from Delayed Rewards}.
\newblock PhD thesis, King's College, Cambridge, UK.

\end{thebibliography}


\begin{thebibliography}{}

\bibitem[Agrawal and Goyal, 2017]{AG}
Agrawal, S. and Goyal, N. (2017).
\newblock Near-optimal regret bounds for thompson sampling.
\newblock {\em J. ACM}, 64(5).

\bibitem[Al-Najjar and Weinstein, 2009]{al2009ambiguity}
Al-Najjar, N.~I. and Weinstein, J. (2009).
\newblock The ambiguity aversion literature: a critical assessment.
\newblock {\em Economics \& Philosophy}, 25(3):249--284.

\bibitem[Angrist and Fernández-Val, 2013]{angristfv:2013}
Angrist, J.~D. and Fernández-Val, I. (2013).
\newblock {\em ExtrapoLATE-ing: External Validity and Overidentification in the
  LATE Framework}, volume~3 of {\em Econometric Society Monographs}, pages
  401--434.
\newblock Cambridge University Press.

\bibitem[Athey and Imbens, 2019]{AtheyImbens:2019}
Athey, S. and Imbens, G. (2019).
\newblock Machine learning methods economists should know about.

\bibitem[Bisbee et~al., 2017]{Bisbee_etal:2017}
Bisbee, J., Dehejia, R., Pop-Eleches, C., and Samii, C. (2017).
\newblock Local instruments, global extrapolation: External validity of the
  labor supply–fertility local average treatment effect.
\newblock {\em Journal of Labor Economics}, 35(S1):S99--S147.

\bibitem[Buchanan et~al., 2018]{Buchanan_etal:2018}
Buchanan, A.~L., Hudgens, M.~G., Cole, S.~R., Mollan, K.~R., Sax, P.~E., Daar,
  E.~S., Adimora, A.~A., Eron, J.~J., and Mugavero, M.~J. (2018).
\newblock Generalizing evidence from randomized trials using inverse
  probability of sampling weights.
\newblock {\em Journal of the Royal Statistical Society: Series A (Statistics
  in Society)}, 181(4):1193--1209.

\bibitem[Cesa-Bianchi and Lugosi, 2006]{cesa2006prediction}
Cesa-Bianchi, N. and Lugosi, G. (2006).
\newblock {\em Prediction, learning, and games}.
\newblock Cambridge university press.

\bibitem[Chernoff, 1959]{chernoff}
Chernoff, H. (1959).
\newblock Sequential design of experiments.
\newblock {\em The Annals of Mathematical Statistics}, 30(3):755--770.

\bibitem[Dehejia et~al., 2021]{Dehejia_etal:2021}
Dehejia, R., Pop-Eleches, C., and Samii, C. (2021).
\newblock From local to global: External validity in a fertility natural
  experiment.
\newblock {\em Journal of Business \& Economic Statistics}, 39(1):217--243.

\bibitem[DellaVigna et~al., 2020]{DellaVigna_etal:2020}
DellaVigna, S., Otis, N., and Vivalt, E. (2020).
\newblock Forecasting the results of experiments: Piloting an elicitation
  strategy.
\newblock {\em AEA Papers and Proceedings}, 110:75--79.

\bibitem[DellaVigna and Pope, 2018]{DellaVignaPope:2018}
DellaVigna, S. and Pope, D. (2018).
\newblock Predicting experimental results: Who knows what?
\newblock {\em Journal of Political Economy}, 126(6):2410--2456.

\bibitem[Dimakopoulou et~al., 2017]{dimakopoulou2017estimation}
Dimakopoulou, M., Zhou, Z., Athey, S., and Imbens, G. (2017).
\newblock Estimation considerations in contextual bandits.
\newblock {\em arXiv preprint arXiv:1711.07077}.

\bibitem[Epstein and Schneider, 2003]{EPSTEIN20031}
Epstein, L.~G. and Schneider, M. (2003).
\newblock Recursive multiple-priors.
\newblock {\em Journal of Economic Theory}, 113(1):1--31.

\bibitem[Epstein and Schneider, 2007]{epstein2007learning}
Epstein, L.~G. and Schneider, M. (2007).
\newblock Learning under ambiguity.
\newblock {\em The Review of Economic Studies}, 74(4):1275--1303.

\bibitem[Gelman and Carlin, 2014]{GelmanCarlin:2014}
Gelman, A. and Carlin, J. (2014).
\newblock Beyond power calculations: Assessing type s (sign) and type m
  (magnitude) errors.
\newblock {\em Perspectives on Psychological Science}, 9(6):641--651.
\newblock PMID: 26186114.

\bibitem[Gelman and Pardoe, 2006]{GelmanPardoe:2006}
Gelman, A. and Pardoe, I. (2006).
\newblock Bayesian measures of explained variance and pooling in multilevel
  (hierarchical) models.
\newblock {\em Technometrics}, 48(2):241--251.

\bibitem[Imai and Ratkovic, 2013]{ImaiRatkovi:2013}
Imai, K. and Ratkovic, M. (2013).
\newblock {Estimating treatment effect heterogeneity in randomized program
  evaluation}.
\newblock {\em The Annals of Applied Statistics}, 7(1):443 -- 470.

\bibitem[Joseph~Hotz et~al., 2005]{Hotz_etal:2005}
Joseph~Hotz, V., Imbens, G.~W., and Mortimer, J.~H. (2005).
\newblock {Predicting the efficacy of future training programs using past
  experiences at other locations}.
\newblock {\em Journal of Econometrics}, 125(1-2):241--270.

\bibitem[Kasy and Sautmann, 2021]{KasySautmann:2021}
Kasy, M. and Sautmann, A. (2021).
\newblock Adaptive treatment assignment in experiments for policy choice.
\newblock {\em Econometrica}, 89(1):113--132.

\bibitem[Klibanoff et~al., 2009]{KLIBANOFF2009930}
Klibanoff, P., Marinacci, M., and Mukerji, S. (2009).
\newblock Recursive smooth ambiguity preferences.
\newblock {\em Journal of Economic Theory}, 144(3):930--976.

\bibitem[Kowalski, 2016]{Kowalski:2016}
Kowalski, A.~E. (2016).
\newblock {Doing More When You're Running LATE: Applying Marginal Treatment
  Effect Methods to Examine Treatment Effect Heterogeneity in Experiments}.
\newblock NBER Working Papers 22363, National Bureau of Economic Research, Inc.

\bibitem[Lai and Robbins, 1985]{LaiRobbins:1985}
Lai, T. and Robbins, H. (1985).
\newblock Asymptotically efficient adaptive allocation rules.
\newblock {\em Advances in Applied Mathematics}, 6(1):4--22.

\bibitem[Meager, 2020]{Meager:2020}
Meager, R. (2020).
\newblock Aggregating distributional treatment effects: Abayesian hierarchical
  analysis of the microcredit literature.

\bibitem[Pan and Yang, 2010]{Pan:2010}
Pan, S.~J. and Yang, Q. (2010).
\newblock A survey on transfer learning.
\newblock {\em IEEE Transactions on Knowledge and Data Engineering},
  22(10):1345--1359.

\bibitem[Qin and Russo, 2022]{qinrusso}
Qin, C. and Russo, D. (2022).
\newblock Adaptivity and confounding in multi-armed bandit experiments.
\newblock {\em arXiv}.

\bibitem[Robbins, 1992]{Robbins1992}
Robbins, H.~E. (1992).
\newblock {\em An Empirical Bayes Approach to Statistics}, pages 388--394.
\newblock Springer New York.

\bibitem[Russo, 2016]{Russo2016}
Russo, D. (2016).
\newblock Simple bayesian algorithms for best arm identification.

\bibitem[Russo and Van~Roy, 2016]{russo2016information}
Russo, D. and Van~Roy, B. (2016).
\newblock An information-theoretic analysis of thompson sampling.
\newblock {\em The Journal of Machine Learning Research}, 17(1):2442--2471.

\bibitem[Schlaifer and Raiffa, 1961]{schlaifer1961applied}
Schlaifer, R. and Raiffa, H. (1961).
\newblock {\em Applied statistical decision theory}.

\bibitem[Stuart et~al., 2011]{Stuart_etal:2011}
Stuart, E.~A., Cole, S.~R., Bradshaw, C.~P., and Leaf, P.~J. (2011).
\newblock The use of propensity scores to assess the generalizability of
  results from randomized trials.
\newblock {\em Journal of the Royal Statistical Society: Series A (Statistics
  in Society)}, 174(2):369--386.

\bibitem[Thompson, 1933]{Thompson:1933}
Thompson, W.~R. (1933).
\newblock On the likelihood that one unknown probability exceeds another in
  view of the evidence of two samples.
\newblock {\em Biometrika}, 25(3-4):285--294.

\bibitem[Vivalt, 2020]{Vivalt:2020}
Vivalt, E. (2020).
\newblock {How Much Can We Generalize From Impact Evaluations?}
\newblock {\em Journal of the European Economic Association}, 18(6):3045--3089.

\bibitem[Vivalt and Coville, 2021]{VivaltColville:2021}
Vivalt, E. and Coville, A. (2021).
\newblock How do policy-makers update their beliefs?

\bibitem[Wald, 1945]{wald}
Wald, A. (1945).
\newblock Sequential tests of statistical hypotheses.
\newblock {\em The Annals of Mathematical Statistics}, 16(2):117--186.

\bibitem[Watkins, 1989]{Watkins:1989}
Watkins, C. J. C.~H. (1989).
\newblock {\em Learning from Delayed Rewards}.
\newblock PhD thesis, King's College, Cambridge, UK.

\end{thebibliography}

\section*{Appendix: Figures \& Tables}


\begin{figure}[ht]
    \centering
    \begin{subfigure}{0.45\textwidth}
        \centering
        \includegraphics[width=\textwidth]{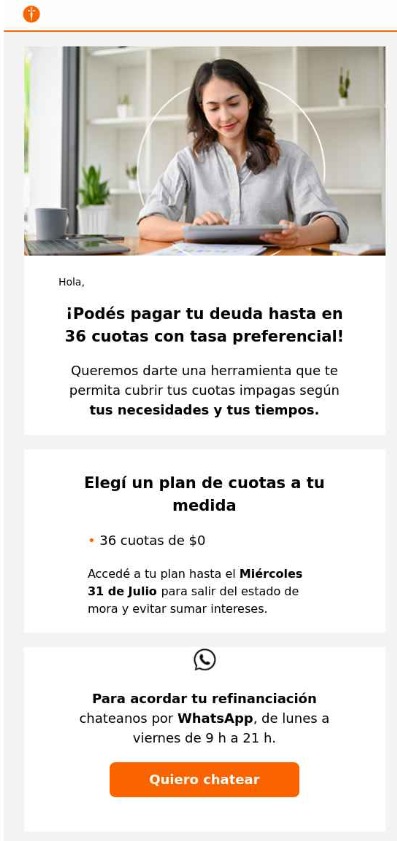}
        \caption{Monthly Payments}
        \label{fig:first}
    \end{subfigure}
    \hfill
    \begin{subfigure}{0.45\textwidth}
        \centering
        \includegraphics[width=\textwidth]{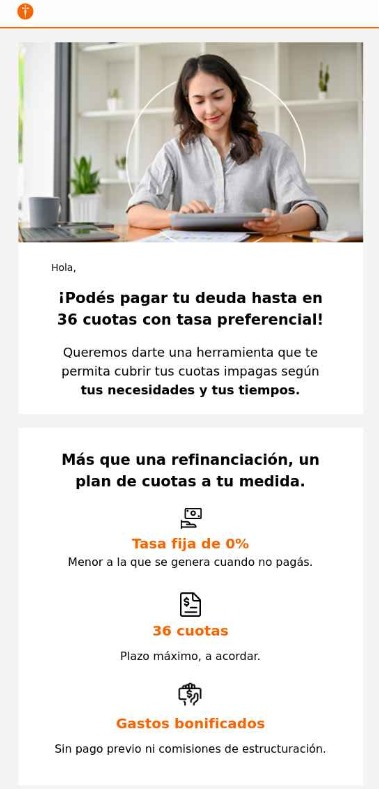}
        \caption{Interest Rate}
        \label{fig:second}
    \end{subfigure}
    \caption{Monthly Payment Messaging vs Interest Rate Messaging}
    \label{fig:debt_rct}
\end{figure}

\pagebreak
\clearpage


\begin{table}[ht]
  \centering
  \small
    \begin{threeparttable}
    \caption{Debt Refinancing Experiment - RCT}
\begin{tabular}{lcccccc}
\toprule
                 &                  BA Zona Norte & BA Zona Oeste & BA Zona Sur &       CABA &         NE &         NW \\
                 &                            (1) &           (2) &         (3) &        (4) &        (5) &        (6) \\
& \multicolumn{6}{c}{Panel A: Clicked on email hyper link} \\
\cmidrule(lr){2-7}
$1\{Interest Rate\}$ &                          0.311 &         4.619 &       3.409 &      1.311 &      3.886 &      2.934 \\
                 &                        (1.307) &      (1.056)* &    (0.940)* &    (0.872) &   (0.826)* &   (0.751)* \\
   Constant term &                          4.251 &         3.049 &       3.295 &       4.17 &      2.483 &      3.442 \\
                 &                       (0.955)* &      (0.601)* &    (0.556)* &   (0.596)* &   (0.434)* &   (0.444)* \\
     Sample Size &                            995 &          1759 &        2121 &       2404 &       2498 &       3316 \\
                 &                                &               &             &            &            &            \\
& \multicolumn{6}{c}{Panel B: Refinanced Loan} \\
\cmidrule(lr){2-7}
$1\{Interest Rate\} $&                         -0.919 &         2.113 &       0.378 &      0.047 &      1.302 &      0.478 \\
                 &                        (0.800) &      (0.661)* &     (0.567) &    (0.516) &   (0.502)* &    (0.453) \\
   Constant term &                          2.013 &         0.976 &        1.55 &      1.597 &      0.931 &      1.484 \\
                 &                       (0.665)* &      (0.343)* &    (0.385)* &   (0.374)* &   (0.268)* &   (0.295)* \\
     Sample Size &                            995 &          1759 &        2121 &       2404 &       2498 &       3316 \\
\bottomrule
\end{tabular}
  \label{tab:debt_tab1}%

  \begin{tablenotes}
    \scriptsize{
      Notes:  This table reports the results of the debt refinancing randomized control trial.  In Panel A, we report the results on click rates and in Panel, we report the results on refinancing rates. Each column reports the results of separate regressions, corresponding to a particular region of Argentina. For ease of interpretation, the dependent variables have been scaled by 100. The independent variable ($1{Interest Rate}$) is an indicator for whether the email provided the interest rate.  Robust standard errors are reported in parentheses. * indicates that the estimated coefficient is statistically different from zero at 95 percent confidence. }
    \end{tablenotes}
    \end{threeparttable}
\end{table}%

\begin{table}[ht]
  \centering
  \small
    \begin{threeparttable}
    \caption{Debt Refinancing Experiment - Multi-prior Adaptive Experiment}
\begin{tabular}{lcccccc}

\toprule
 \multicolumn{7}{c}{REGION: CABA} \\
 \hline 
 &     \multicolumn{3}{c}{Monthly Payments} &  \multicolumn{3}{c}{Interest Rates}   \\
        \cmidrule(lr){2-4}\cmidrule(lr){5-7}
       Sources &    $\zeta^{l}_{t}$ &  $\alpha^{l}_{t}$     &  $\nu^{l}_{t}$      &    $\zeta^{l}_{t}$     &     $\alpha^{l}_{t}$   &    $\nu^{l}_{t}$        \\
\midrule
   Buenos Aires- Otras regiones &          0.037   &      0.167 &     1284.0 &      0.055 &       0.172    &     1408.0 \\
       Buenos Aires- Zona Norte &          0.041   &      0.162 &      493.0 &      0.048 &       0.157    &      796.0 \\
       Buenos Aires- Zona Oeste &          0.030   &      0.167 &      866.0 &      0.072 &       0.159    &     1187.0 \\
         Buenos Aires- Zona Sur &          0.032   &      0.167 &     1078.0 &      0.064 &       0.167    &     1337.0 \\
                             NE &          0.025   &      0.168 &     1335.0 &      0.062 &       0.170    &     1457.0 \\
                             NW &          0.034   &      0.169 &     1731.0 &      0.062 &       0.174    &     1879.0 \\
                        Diffuse &          0.022   &      0.000 &       46.0 &      0.052 &       0.000    &      248.0 \\
        \cmidrule(lr){2-4}\cmidrule(lr){5-7}
$\zeta^{\alpha}_t$              &      \multicolumn{3}{c}{0.033}        &   \multicolumn{3}{c}{ 0.061}          \\
Outcome Averages              &      \multicolumn{3}{c}{ 0.022 }        &   \multicolumn{3}{c}{0.052}          \\
Sample Size              &      \multicolumn{3}{c}{46}        &   \multicolumn{3}{c}{248}          \\
\hline
                                &                  &            &            &            &                &            \\
 \multicolumn{7}{c}{REGION: BUENOS AIRES - ZONA SUR} \\
 \hline 
 &     \multicolumn{3}{c}{Monthly Payments} &  \multicolumn{3}{c}{Interest Rates}   \\
        \cmidrule(lr){2-4}\cmidrule(lr){5-7}
       Sources &         $\zeta^l_t$     &  $\alpha^l_t$     &  $\nu^l_t$      &       $\zeta^l_t$     &     $\alpha^l_t$   &    $\nu^l_t$        \\
\midrule
   Buenos Aires- Otras regiones &          0.037   &      0.168 &     1291.0 &      0.053 &       0.174    &     1473.0 \\
       Buenos Aires- Zona Norte &          0.040   &      0.161 &      500.0 &      0.044 &       0.160    &      861.0 \\
       Buenos Aires- Zona Oeste &          0.030   &      0.167 &      873.0 &      0.068 &       0.150    &     1252.0 \\
                           CABA &          0.041   &      0.167 &     1180.0 &      0.052 &       0.176    &     1590.0 \\
                             NE &          0.025   &      0.169 &     1342.0 &      0.059 &       0.168    &     1522.0 \\
                             NW &          0.034   &      0.169 &     1738.0 &      0.060 &       0.172    &     1944.0 \\
                        Diffuse &          0.019   &      0.000 &       53.0 &      0.042 &       0.000    &      313.0 \\                        
        \cmidrule(lr){2-4}\cmidrule(lr){5-7}
$\zeta^{\alpha}_t$              &      \multicolumn{3}{c}{0.034}        &   \multicolumn{3}{c}{ 0.056}          \\
Outcome Averages              &      \multicolumn{3}{c}{ 0.019 }        &   \multicolumn{3}{c}{0.042}          \\
Sample Size              &      \multicolumn{3}{c}{53}        &   \multicolumn{3}{c}{313}          \\
\hline
                                &                  &            &            &            &                &            \\
 \multicolumn{7}{c}{REGION: Northwest} \\
 \hline 
 &     \multicolumn{3}{c}{Monthly Payments} &  \multicolumn{3}{c}{Interest Rates}   \\
        \cmidrule(lr){2-4}\cmidrule(lr){5-7}
       Sources &         $\zeta^l_t$     &  $\alpha^l_t$     &  $\nu^l_t$      &       $\zeta^l_t$     &     $\alpha^l_t$   &    $\nu^l_t$        \\
\midrule
   Buenos Aires- Otras regiones &          0.039   &      0.170 &     1336.0 &      0.049 &       0.183    &     1971.0 \\
       Buenos Aires- Zona Norte &          0.044   &      0.161 &      545.0 &      0.042 &       0.160    &     1359.0 \\
       Buenos Aires- Zona Oeste &          0.033   &      0.165 &      918.0 &      0.059 &       0.138    &     1750.0 \\
         Buenos Aires- Zona Sur &          0.035   &      0.168 &     1130.0 &      0.055 &       0.161    &     1900.0 \\
                           CABA &          0.042   &      0.170 &     1225.0 &      0.049 &       0.187    &     2088.0 \\
                             NE &          0.027   &      0.166 &     1387.0 &      0.054 &       0.171    &     2020.0 \\
                        Diffuse &          0.051   &      0.000 &       98.0 &      0.039 &       0.000    &      811.0 \\
        \cmidrule(lr){2-4}\cmidrule(lr){5-7}
$\zeta^{\alpha}_t$              &      \multicolumn{3}{c}{0.037}        &   \multicolumn{3}{c}{ 0.051}          \\
Outcome Averages              &      \multicolumn{3}{c}{ 0.051 }        &   \multicolumn{3}{c}{0.039}          \\
Sample Size              &      \multicolumn{3}{c}{98}        &   \multicolumn{3}{c}{811}          \\
\bottomrule
\end{tabular}
  \label{tab:debt_tab2}%

  \begin{tablenotes}
    \scriptsize{
      Notes:  This table reports the results of the debt refinancing experiment using our multi- prior adaptive algorithm.  In column 2-4, we report outcomes for the monthly payment experimental arm, and in columns 5-7 we report the outcomes for the interest rate experimental arm. For the stopping rule, we set $B=200$ and a probability of making a mistake below 1\%. We used an $\epsilon$-greedy algorithm that set $\epsilon=0.20$.}
    \end{tablenotes}
    \end{threeparttable}
\end{table}%

\pagebreak
\newpage
\clearpage

\renewcommand\thesection{A.\arabic{section}}
\renewcommand\thesubsection{\thesection.\arabic{subsection}} \setcounter{section}{0}

\appendix
\setcounter{page}{1}

\begin{center}
	\Huge{Online Appendix}
\end{center}

\small
\section{Notation and some definitions}
\label{app:notation}
For any set $S$, let $\Delta(S)$ be the set of Borel probability measures over $S$. 

For each $t \in \mathbb{N}$ and each $(d,x) \in \mathbb{D}\times \mathbb{X}$, let $D^{t}(x) : = (D_{1}(x),...,D_{t}(x))$, $Y^{t}(x) : =  (Y_{1}(D_{1}(x),x),...,Y_{t}(D_{t}(x),x))$ and $Y^{t}(d,x) : = (Y_{1}(d,x),...,Y_{t}(d,x))$. Let $(Y^{t-1}(x),D^{t-1}(x)) \mapsto \delta_{t}(Y^{t-1},D^{t-1})(.|x) \in \Delta(\mathbb{D})$ be the treatment assignment policy rule and $(Y^{t-1}(x),D^{t-1}(x)) \mapsto \sigma_{t}(Y^{t-1},D^{t-1})(x) \in [0,1]$ be the stopping policy rule.  When there is no risk of confusion we will simply use $\delta_{t}(.|x)$ and $\sigma_{t}(x)$ to denote these rules. 

We define the probability measure $\mathbf{P}$ that is used in the probability statements in our proofs. Formally, let $\mathbf{P}$ be a probability measure over histories $((Y^{T}(d,x) )_{(d,x) \in \mathbb{D}\times \mathbb{X}} , ( D^{T}(x) )_{x \in \mathbb{X}}  )$ (and easily extended to infinite histories)  constructed as follows: By assumption,  for all $(d,x) \in \mathbb{D}\times \mathbb{X}$, $Y_{1}(d,x)$ is IID drawn from $P(.|d,x)$ and $D_{1}(x) \sim \delta_{1}(.|x)$. For any $t > 1$, given the past history $((Y^{t-1}(d,x) )_{(d,x) \in \mathbb{D}\times \mathbb{X}} , ( D^{t-1}(x) )_{x \in \mathbb{X}}  )$, with probability $\sigma_{t}(x)$ the experiment is stopped and $D_{t}(x)$ is the same for all subsequent instances; with probability $1-\sigma_{t}(x)$ the experiment is not stopped and $D_{t}(x) \sim \delta_{t}(.|x)$; $Y_{t}(d,x)$ is is IID drawn from $P(.|d,x)$.

\bigskip

For each $(d,x) \in \mathbb{D}\times \mathbb{X}$ and each $t \in \mathbb{N}$, let 
\begin{align}
&	N_{t}(d,x) : = \sum_{s=1}^{t} 1\{ D_{s}(x) = d \}~and~f_{t}(d,x) : = N_{t}(d,x)/t\\
&	\iota_{t}(d,x) : = t^{-1} \sum_{s=1}^{t} \delta_{s}(d|x) \\
&	 J_{t}(d,x)  : = t^{-1} \sum_{s=1}^{t} 1\{ D_{s}(x) = d \} Y_{s}(d,x).
 \end{align}

\section{Proof of Proposition \ref{pro:alpha.asymptotics.general}}
\label{app:General.Bound.alpha}

The proof of the proposition uses the following lemma, whose proof is relegated to the end of the section.

\begin{lemma}\label{lem:alpha.properties}
		For any $(d,x) \in \mathbb{D}\times \mathbb{X}$ and any $t \geq 1$,
	\begin{align*}
		\alpha^{o}_{t}(d,x) = \frac{       \phi( m_{t}(d,x) ; \zeta^{o}_{0}(d,x) , ( N_{t}(d,x)  + \nu^{o}_{0}(d,x)  )/( N_{t}(d,x)\nu^{o}_{0}(d,x) ) )       }  {  \sum_{o=0}^{L}   \phi( m_{t}(d,x) ; \zeta^{o}_{0}(d,x) , ( N_{t}(d,x)  + \nu^{o}_{0}(d,x)  )/( N_{t}(d,x)\nu^{o}_{0}(d,x) ) )    } ,
	\end{align*} 
	where $m_{t}(d,x) : = \sum_{s=1}^{t} 1\{ D_{s}(x) =d  \}  Y_{s}(d,x) /\sum_{s=1}^{t} 1\{ D_{s}(x) =d  \}$. 
\end{lemma}

\begin{proof}[Proof of Proposition \ref{pro:alpha.asymptotics.general}]

	(1) To show this part we note that $\nu^{o}_{0}(d,x) > 0$ and thus $\mathbb{EV}_{d,x}(o) \rightarrow - \infty$ iff  the bias diverges to plus infinity. From the characterization In Lemma \ref{lem:alpha.properties} and the fact that $   \phi( m_{t}(d,x) ; \zeta^{o}_{0}(d,x) , ( N_{t}(d,x)  + \nu^{o}_{0}(d,x)  )/( N_{t}(d,x)\nu^{o}_{0}(d,x) ) )   =    \phi( m_{t}(d,x) - \theta(d,x) ; \zeta^{o}_{0}(d,x) - \theta(d,x), ( N_{t}(d,x)  + \nu^{o}_{0}(d,x)  )/( N_{t}(d,x)\nu^{o}_{0}(d,x) ) )   $ will vanish if the bias diverges to plus infinity, the desired result is obtained. 
	
	(2)  Observe that
	\begin{align*}
		& \log   \phi( m_{t}(d,x) ; \zeta^{o}_{0}(d,x) , ( N_{t}(d,x)  + \nu^{o}_{0}(d,x)  )/( N_{t}(d,x)\nu^{o}_{0}(d,x) ) )  \\
		&	=  C - 0.5 \log \left( ( N_{t}(d,x)  + \nu^{o}_{0}(d,x)  )/( N_{t}(d,x)\nu^{o}_{0}(d,x) )  \right) - 0.5 \frac{\left( m_{t}(d,x) -  \zeta^{o}_{0}(d,x)  \right)^{2} }{  ( N_{t}(d,x)  + \nu^{o}_{0}(d,x)  )/( N_{t}(d,x)\nu^{o}_{0}(d,x) )  }
	\end{align*}
	where $C$ is some universal constant.  
	
	If $\inf_{t} t^{-1} \sum_{s=1}^{t} \delta_{s}(d|x) > 0 $, by LLN $N_{t}(d,x)$ will diverge with probability approaching 1 and $m_{t}(d,x) = \theta(d,x) + o_{\mathbf{P}}(1)$. Therefore, 
	\begin{align*}
		& \log   \phi( m_{t}(d,x) ; \zeta^{o}_{0}(d,x) , ( N_{t}(d,x)  + \nu^{o}_{0}(d,x)  )/( N_{t}(d,x)\nu^{o}_{0}(d,x) ) )  \\
		= & C - 0.5 \log \left( 1/\nu^{o}_{0}(d,x) )  \right)  - 0.5  \left( \theta(d,x) -  \zeta^{o}_{0}(d,x)  \right)^{2}  \nu^{o}_{0}(d,x) ) + o_{\mathbf{P}}(1) \\
		= & C + 0.5 \mathbb{EV}_{(d,x)}(o)  + o_{\mathbf{P}}(1) 
	\end{align*}
	where the last line follows from the expression in Definition \ref{def:DEV}. Hence, for any $o \in \{1,...,L\}$, 
	\begin{align*}
		\alpha^{o}_{t}(d,x) = \frac{    e^{ 0.5 \mathbb{EV}_{(d,x)}(o) }    }  {  \sum_{o'=0}^{L}   e^{ 0.5 \mathbb{EV}_{(d,x)}(o') }    }  + o_{\mathbf{P}}(1) .
	\end{align*} 
	
	Moreover, this expression implies 
	\begin{align*}
		\frac{\alpha^{o'}_{t}(d,x)}{\alpha^{o}_{t}(d,x)} = e^{-0.5( \mathbb{EV}_{(d,x)}(o) - \mathbb{EV}_{(d,x)}(o'))} + o_{\mathbf{P}}(1).
	\end{align*}
	and thus, given definition \ref{def:EV}, if $o$ is more externally valid than $o'$, $ \mathbb{EV}_{(d,x)}(o) - \mathbb{EV}_{(d,x)}(o')>0$ so that 	$\frac{\alpha^{o'}_{t}(d,x)}{\alpha^{o}_{t}(d,x)} < 1 + o_{\mathbf{P}}(1)$. If $o$ is the (only) externally valid source, then according to our definition \ref{def:EV}, $\mathbb{EV}_{(d,x)}(o)  = \infty$ and with the convention that $c/\infty  = 0$ for any number $c$, the result follows.

\end{proof}

\subsection{Proofs of Supplemental Lemmas}

\begin{proof} [Proof of Lemma \ref{lem:alpha.properties}]
	
	Let $p_{\theta}$ denote a Gaussian PDF with mean $\theta$ and variance $1$. Note that
\begin{align*}
	&	\int \prod_{s=1}^{t} ( p_{\theta}(Y_{s}) )^{1\{ D_{s}(x) =d  \}}  \mu^{o}_{0}(d,x)(d\theta) \\
	= & \int (2\pi)^{-0.5 \sum_{s=1}^{t} 1\{ D_{s}(x) =d  \}   }\exp \left\{ - \frac{1}{2} \sum_{s=1}^{t} 1\{ D_{s}(x) =d  \} \left( Y_{s}(d,x) - m_{t}(d,x)    \right)^{2}   \right\}\\
	\times & \exp \left\{ - \frac{1}{2} \sum_{s=1}^{t} 1\{ D_{s}(x) =d  \} \left( m_{t}(d,x) - \theta    \right)^{2}   \right\} \\
	\times & \exp \left\{ - \sum_{s=1}^{t} 1\{ D_{s}(x) =d  \} \left( Y_{s}(d,x) - m_{t}(d,x)    \right) \left( m_{t}(d,x) - \theta    \right)   \right\}   \phi(\theta; \zeta^{o}_{0}(d,x) , 1/\nu_{0}^{o}(d,x) ) d\theta.
\end{align*}
Observe that $\sum_{s=1}^{t} 1\{ D_{s}(x) =d  \} \left( Y_{s}(d,x) - m_{t}(d,x)    \right) =  0$,  so, letting $N_{t}(d,x) : = \sum_{s=1}^{t} 1\{ D_{s}(x) =d  \}$ it follows that 
\begin{align*}
	\int 	\prod_{s=1}^{t} ( p_{\theta}(Y_{s}) )^{1\{ D_{s}(x) =d  \}}  \mu^{o}_{0}(d,x)(d\theta) = & \frac{ \exp \left\{ - \frac{1}{2} \sum_{s=1}^{t} 1\{ D_{s}(x) =d  \} \left( Y_{s}(d,x) - m_{t}(d,x)    \right)^{2}   \right\}}{(2\pi)^{0.5 \sum_{s=1}^{t} 1\{ D_{s}(x) =d  \} + 0.5}  N_{t}(d,x)^{0.5} }  \\
	\times & \int (2\pi/N_{t}(d,x))^{-1/2} \exp \left\{ - \frac{1}{2}  \left( m_{t}(d,x) - \theta    \right)^{2} N_{t}(d,x)   \right\} \\	
	&  \phi(\theta; \zeta^{o}_{0}(d,x) , 1/\nu_{0}^{o}(d,x) ) d\theta.
\end{align*}

The expression of the integral can be viewed as a convolution between to Gaussian PDFs one indexed by $(  0,   1/N_{t}(d,x) )$ and $(  \zeta^{o}_{0}(d,x) , 1/\nu_{0}^{o}(d,x)  )$ resp, which in turn is equivalent to PDF of the sum of the corresponding random variables evaluated at $m_{t}(d,x)$. Therefore, 
\begin{align*}
	\int \prod_{s=1}^{t} ( p_{\theta}(Y_{s}) )^{1\{ D_{s}(x) =d  \}}  \mu^{o}_{0}(d,x)(d\theta)  = C \phi( m_{t}(d,x) ; \zeta^{o}_{0}(d,x) , ( N_{t}(d,x)  + \nu^{o}_{0}(d,x)  )/( N_{t}(d,x)\nu^{o}_{0}(d,x) ) )
\end{align*}
where $C : = (2\pi)^{-0.5 \sum_{s=1}^{t} 1\{ D_{s}(x) =d  \} + 0.5}   N_{t}(d,x)^{-1/2} \exp \left\{ - \frac{1}{2} \sum_{s=1}^{t} 1\{ D_{s}(x) =d  \} \left( Y_{s}(d,x) - m_{t}(d,x)    \right)^{2}   \right\}$ which, importantly, doesn't depend on the model $o$. 

Hence
\begin{align*}
	\alpha^{o}_{t}(d,x) = \frac{       \phi( m_{t}(d,x) ; \zeta^{o}_{0}(d,x) , ( N_{t}(d,x)  + \nu^{o}_{0}(d,x)  )/( N_{t}(d,x)\nu^{o}_{0}(d,x) ) )       }  {  \sum_{o=0}^{L}   \phi( m_{t}(d,x) ; \zeta^{o}_{0}(d,x) , ( N_{t}(d,x)  + \nu^{o}_{0}(d,x)  )/( N_{t}(d,x)\nu^{o}_{0}(d,x) ) )    }  
\end{align*} 
and  the desired result follows.
\end{proof}

\label{app:alpha.bound.0}

\section{Non-stochastic Bounds}

\subsection{Non-stochastic Bounds for $\alpha_{t}$} 
\label{app:alpha.bound}

For each $t \in \{1,...,T\}$ and each $o \in \{1,...,L\}$, let $\overline{\alpha}^{o}_{t} : \mathbb{R}_{+} \times  [0,1]  \times \mathbb{R}^{1+L} \times \mathbb{N}^{1+L} \rightarrow \mathbb{R}_{+}$ and $\underline{\alpha}^{o}_{t} :  \mathbb{R}_{+} \times [0,1]  \times \mathbb{R}^{1+L} \times \mathbb{N}^{1+L} \rightarrow \mathbb{R}_{+}$ be defined as follows
\begin{align*}
	\overline{\alpha}^{o}_{t}(\delta,g,a,b) : = \min \left\{  1 , \frac{ e^{ \overline{\ell}_{t}(\delta,g,a^{o},b^{o})  }   }{\sum_{o'=0}^{L} e^{ \underline{\ell}_{t}(\delta,g,a^{o'},b^{o'})  } } \right\},~and~
	\underline{\alpha}^{o}_{t}(\delta,g,a,b) : = \frac{ e^{ \underline{\ell}_{t}(\delta,g,a^{o},b^{o})  }   }{\sum_{o'=0}^{L} e^{ \overline{\ell}_{t}(\delta,g,a^{o'},b^{o'})  } }
\end{align*}
for any $(\delta,g,a,b) \in \mathbb{R}_{+} \times [0,1]  \times \mathbb{R}^{1+L} \times \mathbb{N}^{1+L} $, where
\begin{align*}
	\overline{\ell}_{t}(\delta,g,a^{o},b^{o}) : = & - \log   \underline{\sigma}_{t}  - 0.5  \frac{  \max\{ g^{2} ( a^{o} )^{2}  - 2 \delta  a^{o} ,0 \}  }  { \overline{\sigma}_{t}^{2} } \\
	\underline{\ell}_{t}(\delta,g,a^{o},b^{o}) : =  &- \log   \overline{\sigma}_{t}  - 0.5  \frac{ (  \delta + a^{o}    )^{2}    }  { g^{2}  \underline{\sigma}_{t}^{2} } 
\end{align*}
with $( 1 + b^{o}  /T )/b^{o} = : \underline{\sigma}_{t}^{2}~and~( g + b^{o} /t )/( g  b ) = : \overline{\sigma}_{t}^{2}$.

\begin{lemma}\label{lem:alpha.bound} 
	For any $o \in \{ 0,...,L  \}$, any $t \in \{1,...,T\}$,  any $(d,x) \in \mathbb{D}\times \mathbb{X}$ and any $\eta \in (0,h_{t}(d,x))$ and $\delta > 0$ such that $ f_{t}(d,x)  - h_{t}(d,x) \geq - \eta$ and $| J_{t}(d,x) - f_{t}(d,x) \theta (d,x)| \leq \delta$, it follows that 
	\begin{align*}
		\underline{\alpha}^{o}_{t}( \delta , h_{t}(d,x) - \eta , |\bar{\zeta}_{0}(d,x) | ,  \nu_{0}(d,x)   )    \leq  \alpha^{o}_{t}(d,x)  \leq   	\overline{\alpha}^{o}_{t}( \delta , h_{t}(d,x) - \eta  ,  |\bar{\zeta}_{0}(d,x) | ,  \nu_{0}(d,x)  ).
	\end{align*}
\end{lemma}


\begin{proof}[Proof of Lemma \ref{lem:alpha.bound} ]
	By Lemma \ref{lem:alpha.properties} it suffices to bound $ \phi( J_{t}(d,x)/f_{t}(d,x) ; \zeta^{o}_{0}(d,x) , ( N_{t}(d,x)  + \nu^{o}_{0}(d,x)  )/( N_{t}(d,x)\nu^{o}_{0}(d,x) ) ) $. To do this, note that
	\begin{align*}
		&\phi( J_{t}(d,x)/f_{t}(d,x) ; \zeta^{o}_{0}(d,x) , ( N_{t}(d,x)  + \nu^{o}_{0}(d,x)  )/( N_{t}(d,x)\nu^{o}_{0}(d,x) ) )  \\
		& =  \frac{\sqrt{( N_{t}(d,x)\nu^{o}_{0}(d,x) )}   } {2 \pi \sqrt{  ( N_{t}(d,x)  + \nu^{o}_{0}(d,x)  ) }  }  \exp \left\{ - 0.5 \frac{\left( J_{t}(d,x) - f_{t}(d,x) \zeta^{o}_{0}(d,x)   \right)^{2}}{ f_{t}(d,x)  }  \frac{  N_{t}(d,x)\nu^{o}_{0}(d,x) }{ f_{t}(d,x) ( N_{t}(d,x)  + \nu^{o}_{0}(d,x)  ) }  \right\} \\
		& =  \frac{\sqrt{ f_{t}(d,x)\nu^{o}_{0}(d,x)  }   } {2 \pi \sqrt{  ( f_{t}(d,x)  + \nu^{o}_{0}(d,x) /t ) }  }  \exp \left\{ - 0.5 \frac{\left( \bar{J}_{t}(d,x) - f_{t}(d,x) \bar{\zeta}^{o}_{0}(d,x)   \right)^{2}}{ f_{t}(d,x)  }  \frac{  \nu^{o}_{0}(d,x) }{ ( f_{t}(d,x)  + \nu^{o}_{0}(d,x) /t ) }  \right\} 
	\end{align*}
	where $\bar{.}$ indicates centered at $\theta(d,x)$.  Henceforth, let $\sigma^{2}_{t} : = ( f_{t}(d,x)  + \nu^{o}_{0}(d,x) /t )/( f_{t}(d,x)\nu^{o}_{0}(d,x) ) $. 
	
	Under $| \bar{J}_{t}(d,x) | \leq \delta$, if follows that 
	\begin{align*}
		(\bar{J}_{t}(d,x) -  f_{t}(d,x) \bar{\zeta}^{o}_{0}(d,x) )^{2}  = &  ( \bar{J}_{t}(d,x)  )^{2} + (  f_{t}(d,x) \bar{\zeta}^{o}_{0}(d,x) )^{2}  - 2 \bar{J}_{t}(d,x)  f_{t}(d,x) \bar{\zeta}^{o}_{0}(d,x)   \\
		\leq &  \delta^{2} + (  f_{t}(d,x) \bar{\zeta}^{o}_{0}(d,x) )^{2}  + 2 \delta   f_{t}(d,x) |\bar{\zeta}^{o}_{0}(d,x) | \\
		\leq & (  \delta + |\bar{\zeta}^{o}_{0}(d,x) |    )^{2}
	\end{align*}
	and, in addition, if $ f_{t}(d,x) -  h_{t}(d,x) \geq -\eta$ with $\eta \leq h_{t}(d,x)$, then
	\begin{align*}
		(\bar{J}_{t}(d,x) -  f_{t}(d,x) \bar{\zeta}^{o}_{0}(d,x) )^{2}  = &  ( \bar{J}_{t}(d,x)  )^{2} + (  f_{t}(d,x) \bar{\zeta}^{o}_{0}(d,x) )^{2}  - 2 \bar{J}_{t}(d,x)  f_{t}(d,x) \bar{\zeta}^{o}_{0}(d,x)   \\
		\geq & ( f_{t}(d,x) \bar{\zeta}^{o}_{0}(d,x) )^{2}  - 2  f_{t}(d,x) \delta |\bar{\zeta}^{o}_{0}(d,x) |\\
		\geq & ( h_{t}(d,x)  - \eta) )^{2} ( \bar{\zeta}^{o}_{0}(d,x) )^{2}  - 2 \delta  |\bar{\zeta}^{o}_{0}(d,x)|.
	\end{align*}
	
	Also, under these conditions,
	\begin{align*}
		\sigma^{2}_{t} \geq &  ( 1 + \nu^{o}_{0}(d,x) /t )/( \nu^{o}_{0}(d,x) ) \geq ( 1 + \nu^{o}_{0}(d,x) /T )/( \nu^{o}_{0}(d,x) ) = : \underline{\sigma}_{t}^{2}\\
		\leq & ( h_{t}(d,x) - \eta + \nu^{o}_{0}(d,x) /t )/( (h_{t}(d,x) - \eta)   \nu^{o}_{0}(d,x) ) = : \overline{\sigma}_{t}^{2}.
	\end{align*}	
	
	Therefore,
	\begin{align*}
		\log  \phi( \bar{m}_{t}(d,x) - \bar{\zeta}^{o}_{0}(d,x) ; 0 , \sigma^{2}_{t}(d,x))  \geq &  - \log   \sigma_{t}  - 0.5  \frac{ (  \delta + |\bar{\zeta}^{o}_{0}(d,x) |    )^{2}    }  { f_{t}(d,x)^{2}  \sigma^{2}_{t} } + Cte	\geq   - \log   \overline{\sigma}_{t}  - 0.5  \frac{ (  \delta + |\bar{\zeta}^{o}_{0}(d,x) |    )^{2}    }  { (h_{t}(d,x) - \eta)^{2}  \underline{\sigma}_{t}^{2} } + Cte\\
		= : & \underline{\ell}_{t}(\delta,h_{t}(d,x) - \eta,|\bar{\zeta}^{o}_{0}(d,x) | , \nu_{0}^{o}(d,x) ) + Cte
	\end{align*}	
	and
	\begin{align*}
		\log  \phi( \bar{m}_{t}(d,x) - \bar{\zeta}^{o}_{0}(d,x) ; 0 , \sigma^{2}_{t}(d,x))  \leq &  - \log   \sigma_{t}  - 0.5  \frac{ \max\{ ( h_{t}(d,x)  - \eta) )^{2} ( \bar{\zeta}^{o}_{0}(d,x) )^{2}  - 2 \delta  |\bar{\zeta}^{o}_{0}(d,x)| ,0 \}    }  { f_{t}(d,x)^{2}  \sigma^{2}_{t} } + Cte\\
		\leq &  - \log   \underline{\sigma}_{t}  - 0.5  \frac{  \max\{ ( h_{t}(d,x)  - \eta) )^{2} ( \bar{\zeta}^{o}_{0}(d,x) )^{2}  - 2 \delta  |\bar{\zeta}^{o}_{0}(d,x)| ,0 \}  }  { \overline{\sigma}_{t}^{2} } + Cte\\
		= : & \overline{\ell}_{t}(\delta,h_{t}(d,x) - \eta,|\bar{\zeta}^{o}_{0}(d,x) | , \nu_{0}^{o}(d,x) ).
	\end{align*}	
\end{proof}

\begin{lemma}\label{lem:properties.bound.alpha}
	The following properties are true:
	\begin{enumerate}
		\item $\delta \mapsto \underline{\ell}_{t}(\delta,g,|\bar{\zeta}^{o}_{0}(d,x) | , \nu_{0}^{o}(d,x) )$ is decreasing and $\delta \mapsto \overline{\ell}_{t}(\delta,g,|\bar{\zeta}^{o}_{0}(d,x) | , \nu_{0}^{o}(d,x) )$ is non-decreasing.
		\item $g \mapsto \underline{\ell}_{t}(\delta,g,|\bar{\zeta}^{o}_{0}(d,x) | , \nu_{0}^{o}(d,x) )$  is increasing and $g \mapsto \overline{\ell}_{t}(\delta,g,|\bar{\zeta}^{o}_{0}(d,x) | , \nu_{0}^{o}(d,x) )$ is decreasing.
		\item $\delta \mapsto \overline{\alpha}^{o}_{t}(\delta,g,|\zeta_{0}(d,x)| , \nu_{0}(d,x))$ is increasing and $\delta \mapsto \overline{\alpha}^{o}_{t}(\delta,g,|\zeta_{0}(d,x)| , \nu_{0}(d,x))$ is decreasing. 
		\item $g \mapsto \overline{\alpha}^{o}_{t}(\delta,g,|\zeta_{0}(d,x)| , \nu_{0}(d,x))$ is decreasing and $g \mapsto \underline{\alpha}^{o}_{t}(\delta,g,|\zeta_{0}(d,x)| , \nu_{0}(d,x))$ is increasing.  
		\item $t \mapsto  \overline{\alpha}^{o}_{t}(\delta,g,|\zeta_{0}(d,x)| , \nu_{0}(d,x))$ is increasing and $t \mapsto  \underline{\alpha}^{o}_{t}(\delta,g,|\zeta_{0}(d,x)| , \nu_{0}(d,x))$ is decreasing. 
	\end{enumerate}
\end{lemma}

\begin{proof}[Proof of Lemma \ref{lem:properties.bound.alpha}]
	(1) It is easy to see that $\underline{\ell}_{t}(\delta,g,|\bar{\zeta}^{o}_{0}(d,x) | , \nu_{0}^{o}(d,x) , h_{t}(d,x))$ is decreasing in $\delta$ and $\overline{\ell}_{t}(\delta,g,|\bar{\zeta}^{o}_{0}(d,x) | , \nu_{0}^{o}(d,x) )$ is non-decreasing in $\delta$.\\
	
	(2) We first observe that $\underline{\sigma}^{2}_{t}$ is constant as a function of $g = h_{t}(d,x)-\eta$ and $\overline{\sigma}^{2}_{t}$ is an decreasing function of $g : = h_{t}(d,x)-\eta$. Also, note that $\frac{d \underline{\ell}_{t}(\delta,g,|\bar{\zeta}^{o}_{0}(d,x) | , \nu_{0}^{o}(d,x) ) }{d g } = - \frac{1}{  \overline{\sigma} } \frac{d \overline{\sigma}_{t} }{d g}   +  \frac{ (  \delta + |\bar{\zeta}^{o}_{0}(d,x) |    )^{2}    }  { (g)^{3}  \underline{\sigma}_{t}^{2} } $, thus, since $g \geq 0$, $\underline{\ell}_{t}(\delta,g,|\bar{\zeta}^{o}_{0}(d,x) | , \nu_{0}^{o}(d,x) )$  is increasing as a function of $g$. Similarly, $\overline{\ell}_{t}(\delta,g,|\bar{\zeta}^{o}_{0}(d,x) | , \nu_{0}^{o}(d,x) )$ is increasing as a function of $\overline{\sigma}^{2}_{t}$ and decreasing as a direct function of $g$, thus by computing the derivative it can be shown that it is decreasing in $g$. 
	
	(3-4) By parts (1), it readily follows that $\overline{\alpha}^{o}_{t}(\delta,g,|\zeta_{0}(d,x) |, \nu_{0}(d,x))$ is increasing in $\delta$ and  $\underline{\alpha}^{o}_{t}(\delta,g,|\zeta_{0}(d,x)| , \nu_{0}(d,x))$ is decreasing in $\delta$. And by part (2)  $\overline{\alpha}^{o}_{t}(\delta,g,|\zeta_{0}(d,x)| , \nu_{0}(d,x))$ is decreasing in $g$ and  $\underline{\alpha}^{o}_{t}(\delta,g,|\zeta_{0}(d,x) |, \nu_{0}(d,x))$ is increasing in $g$.
	
	(5) It follows that $ t \mapsto \bar{\sigma}^{2}_{t}$ is decreasing and $t \mapsto \underline{\sigma}^{2}_{t}$ is constant. Since $\overline{\ell}_{t}$ is non-increasing in $ \bar{\sigma}^{2}_{t}$ it follows that it is non-decreasing in $t$. Similarly, $\underline{\ell}_{t}$ is increasing in $\bar{\sigma}^{2}_{t}$ and thus increasing in $t$.
	
	These results imply that $t \mapsto  \overline{\alpha}^{o}_{t}(\delta,g,|\zeta_{0}(d,x)| , \nu_{0}(d,x))$ is increasing and $t \mapsto  \underline{\alpha}^{o}_{t}(\delta,g,|\zeta_{0}(d,x)| , \nu_{0}(d,x))$ is decreasing. 
\end{proof}

\subsection{Non-stochastic Bounds for $\zeta^{o}_{t}$} 
\label{app:zeta.o.bound}

For any $t \in \mathbb{N}$, let $\Omega_{0} : D(\Omega_{0}) : = ( [0,1] \times \mathbb{R} \times \mathbb{N} )  \rightarrow \mathbb{R} $ and $\Omega : D(\Omega) : = \mathbb{R}_{+} \times  D(\Omega_{0})  \rightarrow \mathbb{R} $ be such that 
\begin{align*}
	\Omega(\gamma,g, \zeta^{o}_{0}(d,x) ,  \nu^{o}_{0}(d,x)   )  : = \frac{\gamma }{ g + \nu^{o}_{0}(d,x)/t  } + \Omega_{0}( g , \zeta^{o}_{0}(d,x) ,  \nu^{o}_{0}(d,x)      )
\end{align*}
and 
\begin{align*}
	\Omega_{0}(g , \zeta^{o}_{0}(d,x) ,  \nu^{o}_{0}(d,x)    ) 	 : = \nu^{o}_{0}(d,x) \left( \frac{ (\zeta^{o}_{0}(d,x))_{+} /t  }  {  g + \nu^{o}_{0}(d,x)/t    }  +  \frac{ (\zeta^{o}_{0}(d,x))_{-} /T  }  {  1 + \nu^{o}_{0}(d,x)/t    }   \right) 
\end{align*}
for any $(\gamma, g , \zeta^{o}_{0}(d,x) ,  \nu^{o}_{0}(d,x) )  \in D(\Omega) $, where for any real number $a$, $a_{+} : = \max\{a,0\}$ and $a_{-} : = \min\{ a , 0  \}$.

\begin{lemma}\label{lem:properties.Omega}
	For any $o \in \{0,...,L\}$, any $(d,x) \in \mathbb{D} \times \mathbb{X}$ and any $t \in \mathbb{N}$, the following are true: 
	\begin{enumerate}
		\item $\gamma \mapsto \Omega(\gamma,g, \zeta^{o}_{0}(d,x) ,  \nu^{o}_{0}(d,x) ) $ is increasing. 
		\item $g \mapsto \Omega(\gamma,g, \zeta^{o}_{0}(d,x) ,  \nu^{o}_{0}(d,x)   ) $ is decreasing and $g \mapsto \Omega_{0}(g, \zeta^{o}_{0}(d,x) ,  \nu^{o}_{0}(d,x)   ) $ is non-increasing. 
		\item If $\zeta_{0}(d,x) \leq 0$, $\nu_{0}(d,x) \mapsto \Omega(\gamma,g, \zeta^{o}_{0}(d,x) ,  \nu^{o}_{0}(d,x) ) $ is decreasing; and if $\zeta_{0}(d,x) \leq (\geq) 0$ $\nu_{0}(d,x) \mapsto \Omega_{0}(g, \zeta^{o}_{0}(d,x) ,  \nu^{o}_{0}(d,x)   ) $ is non-increasing (non-decreasing).
	\end{enumerate}
\end{lemma}

\begin{proof}[Proof of Lemma \ref{lem:properties.Omega}]
	(1) Trivial.
	
	(2) By inspection, $g \mapsto \Omega_{0}(g, \zeta^{o}_{0}(d,x) ,  \nu^{o}_{0}(d,x)   ) $ is non-increasing. In addition $g \mapsto \gamma/(g + \nu^{o}_{0}(d,x)/t)$ is decreasing. 
	
	(3) Trivial.
\end{proof}

\begin{lemma}\label{lem:bound.zeta.o}
	For any $o \in \{0,...,L\}$, any $(d,x) \in \mathbb{D} \times \mathbb{X}$ and any $t \in \mathbb{N}$, suppose $|J_{t}(d,x) - f_{t}(d,x) \theta(d,x) | \leq \gamma$ and $f_{t}(d,x) - h_{t}(d,x) \geq - \eta$ for some $\gamma \geq 0$ and $0 \leq \eta \leq h_{t}(d,x) \leq \iota_{t}(d,x)$. Then:
	\begin{enumerate}
		\item $|\zeta^{o}_{t}(d,x) - \theta(d,x) | \leq 	\Omega(\gamma,h_{t}(d,x) - \eta, |\zeta^{o}_{0}(d,x)  - \theta(d,x)  | ,  \nu^{o}_{0}(d,x) ) $.
		\item $\zeta^{o}_{t}(d,x) - \theta(d,x)   \leq 	\Omega(\gamma,h_{t}(d,x) - \eta, \zeta^{o}_{0}(d,x)  - \theta(d,x)  ,  \nu^{o}_{0}(d,x)  ) $. 
		\item  $-(\zeta^{o}_{t}(d,x) - \theta(d,x) )  \leq 	\Omega(\gamma,h_{t}(d,x) - \eta, -(\zeta^{o}_{0}(d,x) -  \theta(d,x) ) ,  \nu^{o}_{0}(d,x)  ) $.
	\end{enumerate}
\end{lemma}

\begin{proof}[Proof of Lemma \ref{lem:bound.zeta.o} ]
	(1) Under the conditions, it easy to see that
	\begin{align*}
		|\zeta^{o}_{t}(d,x) - \theta(d,x)  | \leq  \frac{ \gamma + 	|\zeta^{o}_{0}(d,x) - \theta(d,x)  | \nu^{o}_{0}(d,x)/t    }{ \iota_{t}(d,x) - \eta  + \nu^{o}_{0}(d,x) /t   } 	\leq &  \frac{ \gamma + 	|\zeta^{o}_{0}(d,x) - \theta(d,x)  | \nu^{o}_{0}(d,x)/t    }{ h_{t}(d,x) - \eta  + \nu^{o}_{0}(d,x) /t   } \\
		 = &\Omega(\gamma,h_{t}(d,x) - \eta, |\zeta^{o}_{0}(d,x) -  \theta(d,x) | ,  \nu^{o}_{0}(d,x) ) .
	\end{align*}
	
	(2-3) The proof is analogous and thus omitted. 
\end{proof}

\subsection{Non-stochastic Bounds for $\zeta^{\alpha}_{t}$} 
\label{app:zeta.alpha.bound}

Let $\Gamma_{0} : D(\Gamma) : = \mathbb{R}_{+} \times  [0,1] \times \mathbb{R}^{1+L} \times \mathbb{N}^{1+L} \rightarrow \mathbb{R} $ be such that 
\begin{align*}
	\Gamma_{0} (\gamma,g, \zeta_{0}(d,x) ,  \nu_{0}(d,x)   )  : = &  \sum_{o=0}^{L} \overline{\alpha}^{o}_{t}( \gamma,g, \zeta_{0}(d,x) ,  \nu_{0}(d,x)  ) 	\Omega^{+}_{0} (g, \zeta^{o}_{0}(d,x) ,  \nu^{o}_{0}(d,x)   )  \\
	& + \sum_{o=0}^{L} \underline{\alpha}^{o}_{t}( \gamma,g, \zeta_{0}(d,x) ,  \nu_{0}(d,x)  ) 	\Omega^{-}_{0} (g, \zeta^{o}_{0}(d,x) ,  \nu^{o}_{0}(d,x)   )
\end{align*}
where $\Omega^{+} : = \max\{\Omega, 0\}$ and $\Omega^{-} : = \min \{\Omega, 0\}$, and $\Gamma : D(\Gamma)  \rightarrow \mathbb{R} $ 
\begin{align*}
	\Gamma ( \gamma, g, \zeta_{0}(d,x) ,  \nu_{0}(d,x)    ) 	 : =  \gamma \sum_{o=0}^{L} \frac{ \overline{\alpha}^{o}_{t}( \gamma,g, \zeta_{0}(d,x) ,  \nu_{0}(d,x)  ) } { g + \nu^{o}_{0}(d,x)/t }   + 	\Gamma_{0} (\gamma,g, \zeta_{0}(d,x) ,  \nu_{0}(d,x)   ) 
\end{align*}
for any $(\gamma,g, \zeta_{0}(d,x) ,  \nu_{0}(d,x) )  \in D(\Gamma) $. 

\begin{remark}\label{rem:AltBoundGamma}
	Another possible formulation for $\Gamma( \gamma, g, \zeta_{0}(d,x) ,  \nu_{0}(d,x)    ) $ is $\max_{o \in \{0,...,L\}} \Omega( \gamma, g, \zeta^{o}_{0}(d,x) ,  \nu^{o}_{0}(d,x)    ) $, so one can define $\Gamma$ as the minimum of this expression and the one above, and depending on the context one can use one bound or the other. The same applies to $\Gamma_{0}$ and $\Omega_{0}$.  For the sake of the exposition, however, we do not make this bound explicit. $\triangle$
\end{remark}

\begin{lemma}\label{lem:properties.Gamma}
	The following properties are true
	\begin{enumerate}	 
		\item $\delta \mapsto \Gamma(\delta,g,\zeta_{0}(d,x) , \nu_{0}(d,x))$ is increasing and $\delta \mapsto \Gamma(\delta,g,\zeta_{0}(d,x) , \nu_{0}(d,x))$ is non-decreasing.
		\item $g \mapsto \Gamma(\delta,g,\zeta_{0}(d,x) , \nu_{0}(d,x))$ and $g \mapsto \Gamma_{0}(\delta,g,\zeta_{0}(d,x) , \nu_{0}(d,x))$ are decreasing.	
		\item For any positive sequences $(\delta_{t},g_{t})_{t}$, $\Gamma(\delta_{t},g_{t},\zeta_{0}(d,x),\nu_{0}(d,x))= O \left(  \frac{ \delta_{t} + t^{-1}   }{ g_{t}  + t^{-1}  }    \right)$ and $\Gamma_{0}(\delta_{t},g_{t},\zeta_{0}(d,x),\nu_{0}(d,x)) = O \left(  \frac{ t^{-1}   }{ g_{t}  + t^{-1}  }    \right)$. 
	\end{enumerate}

\end{lemma}

\begin{proof}[Proof of Lemma \ref{lem:properties.Gamma}]
	
	(1) (we only establish the results for $\Gamma$ as for $\Gamma_{0}$ is analogous) $\Gamma(\delta,g,\zeta_{0}(d,x) , \nu_{0}(d,x))$ is the sum of products $$\overline{\alpha}^{o}_{t}(\delta,g,|\zeta_{0}(d,x)| , \nu_{0}(d,x)) \Omega^{+}(\delta,g,\zeta^{o}_{0}(d,x) , \nu^{o}_{0}(d,x)) + \underline{\alpha}^{o}(\delta,g,|\zeta_{0}(d,x)| , \nu_{0}(d,x)) \Omega^{-}(\delta,g,\zeta^{o}_{0}(d,x) , \nu^{o}_{0}(d,x)).$$ Observe that, by Lemma \ref{lem:properties.bound.alpha}(3), $\overline{\alpha}^{o}_{t}(\delta,g,|\zeta_{0}(d,x)| , \nu_{0}(d,x))$ is increasing as a function of $\delta$ and $ \Omega^{+}(\delta,g,\zeta^{o}_{0}(d,x) , \nu^{o}_{0}(d,x)) $ is non-decreasing as a function of $\delta$ by Lemma \ref{lem:properties.Omega}(1). Since both quantities are positive, it follows that $\overline{\alpha}^{o}_{t}(\delta,g,|\zeta_{0}(d,x)| , \nu_{0}(d,x)) \Omega^{+}(\delta,g,\zeta^{o}_{0}(d,x) , \nu^{o}_{0}(d,x)) $ is non-decreasing (increasing if $\Omega^{+}(\delta,g,\zeta^{o}_{0}(d,x) , \nu^{o}_{0}(d,x))>0$ ). Similarly, by Lemmas \ref{lem:properties.bound.alpha}(3) and \ref{lem:properties.Omega}(1), $ \underline{\alpha}^{o}_{t}$ is decreasing and $ \Omega^{-}$ is non-positive and non-decreasing as a function of $\delta$, so the product is is non-decreasing (increasing if $\Omega^{-}<0$ ). Thus,  $\Gamma(\delta,g,\zeta_{0}(d,x) , \nu_{0}(d,x))$  is increasing as a function of $\delta$. 
	
	(2)  (we only establish the results for $\Gamma$ as for $\Gamma_{0}$ is analogous) By a similar argument, Lemma \ref{lem:properties.Omega}(2) and Lemma \ref{lem:properties.bound.alpha}(4)  it follows that $\Gamma(\delta,g,\zeta_{0}(d,x),\nu_{0}(d,x))$ is decreasing as a function of $g$.
	
	(3) Clearly, $\max_{o \in \{ 0,..., L\}} \Omega(\delta_{t},g_{t},\zeta_{0}(d,x),\nu_{0}(d,x)) = O \left(  \frac{ \delta_{t} + t^{-1}   }{ g_{t}  + t^{-1}  }    \right)$. Thus $\Gamma(\delta_{t},g_{t},\zeta_{0}(d,x),\nu_{0}(d,x))$ inherits the same rate. Similarly,  $\max_{o \in \{ 0,..., L\}} \Omega_{0}(g_{t},\zeta_{0}(d,x),\nu_{0}(d,x)) = O \left(  \frac{ t^{-1}   }{ g_{t}  + t^{-1}  }    \right)$ and $\Gamma_{0}(\delta_{t},g_{t},\zeta_{0}(d,x),\nu_{0}(d,x))$ inherits the same rate.
\end{proof}

\begin{lemma}\label{lem:bound.zeta.alpha}
	For any $(d,x) \in \mathbb{D} \times \mathbb{X}$ and any $t \in \mathbb{N}$, suppose  $|J_{t}(d,x) - f_{t}(d,x) \theta(d,x) | \leq \gamma$ and $f_{t}(d,x) - h_{t}(d,x) \geq - \eta$ for some $\gamma \geq 0$ and $0 \leq \eta \leq h_{t}(d,x) \leq \iota_{t}(d,x)$. Then:
	\begin{enumerate}
		\item $|\zeta^{\alpha}_{t}(d,x) - \theta(d,x) | \leq 	\Gamma(\gamma,h_{t}(d,x) - \eta, \zeta_{0}(d,x)  - \theta(d,x)  ,  \nu_{0}(d,x) ) $.
	\end{enumerate}
\end{lemma}

\begin{proof}[Proof of Lemma \ref{lem:bound.zeta.alpha} ]
 Follows directly from the definition of $\zeta^{\alpha}_{t}$ and Lemmas \ref{lem:bound.zeta.o} and \ref{lem:alpha.bound}.
\end{proof}

\subsubsection{Relationship between $\Gamma$ and $\Omega$}

For each $o \in \{0,...,L\}$ and $d \in \mathbb{D}$, let $\zeta^{-o}_{0}(d,x)$ be the $L \times 1$ vector of all coordinates of $\zeta_{0}(d,x)$ except for $\zeta^{o}_{0}(d,x)$. 

\begin{lemma}\label{lem:convergence.GammaToOmega}
	For each $o \in \{0,...,L\}$, $(d,x) \in \mathbb{D} \times \mathbb{X}$, $\gamma, g \geq 0$ and $\nu_{0}(d,x)$,
	\begin{align*}
		\lim_{\bar{\zeta}^{-o}_{0}(d,x) \rightarrow \infty} \left| \Gamma(\gamma, g , |\zeta_{0}(d,x) |, \nu_{0}(d,x) ) -  \Omega(\gamma, g , |\zeta^{o}_{0}(d,x) |, \nu^{o}_{0}(d,x) ) \right| = 0
	\end{align*}
\end{lemma}

\begin{proof}[Proof of Lemma \ref{lem:convergence.GammaToOmega}]
	
	By construction of $\overline{\ell}_{t}$ and $\underline{\ell}_{t}$ it is easy to see that for any $o' \in \{1,...,L\}$, 
	\begin{align*}
		&\lim_{\bar{\zeta}^{o'}_{0}(d,x) \rightarrow \infty} \underline{\ell}_{t} (\gamma, g , |\zeta^{o'}_{0}(d,x)| , \nu^{o'}_{0}(d,x) ) = \lim_{\bar{\zeta}^{o'}_{0}(d,x) \rightarrow \infty} \overline{\ell}_{t} (\gamma, g , |\zeta^{o'}_{0}(d,x)| , \nu^{o'}_{0}(d,x) ) = -\infty
	\end{align*}
	Moreover, in both cases the rate is $O(-|\zeta^{o'}_{0}(d,x)|^{2})$ (observe that the Oh depends on $(\gamma, g , \nu^{o'}_{0}(d,x) )$ ). Hence, for any $o' \ne o$,
	\begin{align*}
		\underline{\alpha}^{o'}_{t} (\gamma, g , |\zeta_{0}(d,x)| , \nu_{0}(d,x) ) = O(e^{-|\zeta^{o'}_{0}(d,x)|^{2}} )~and~  \overline{\alpha}^{o'}_{t} (\gamma, g , |\zeta_{0}(d,x)| , \nu_{0}(d,x) ) = O(e^{-|\zeta^{o'}_{0}(d,x)|^{2}} ).
	\end{align*}
	That is, they converge to 0 at exponential rate. 
	
	On the other hand, $\Omega( \gamma, g , |\zeta^{o'}_{0}(d,x)| , \nu^{o'}_{0}(d,x)  ) = O( |\zeta^{o'}_{0}(d,x)| )$, hence
	\begin{align*}
		& \overline{\alpha}^{o'}_{t} (\gamma, g , |\zeta_{0}(d,x)| , \nu_{0}(d,x) ) 	\Omega^{+}( \gamma, g , |\zeta^{o'}_{0}(d,x)| , \nu^{o'}_{0}(d,x)  ) \\
		& + 	 \underline{\alpha}^{o'}_{t} (\gamma, g , |\zeta_{0}(d,x)| , \nu_{0}(d,x) ) 	\Omega^{-}( \gamma, g , |\zeta^{o'}_{0}(d,x)| , \nu^{o'}_{0}(d,x)  ) = O(e^{-|\zeta^{o'}_{0}(d,x)|^{2}}  |\zeta^{o'}_{0}(d,x)|   )
	\end{align*}
	which clearly converges to 0 as $ |\zeta^{o'}_{0}(d,x)| $ diverges. 
	
	On the other hand, these results imply that
	 \begin{align*}
		\lim_{\bar{\zeta}^{-o}_{0}(d,x) \rightarrow \infty} \overline{\alpha}^{o}_{t}(d,x) = \min \left\{   1 ,   \frac{ e^{ \overline{\ell}_{t} \gamma, g , |\zeta^{o}_{0}(d,x)| , \nu^{o}_{0}(d,x) )  } }{ e^{\underline{\ell}_{t} \gamma, g , |\zeta^{o}_{0}(d,x)| , \nu^{o}_{0}(d,x) ) }  }    \right\}  = 1
	\end{align*}
	where the last equality follows because $\overline{\ell}_{t}(\gamma, g , |\zeta^{o}_{0}(d,x)| , \nu^{o}_{0}(d,x) )  \geq \underline{\ell}_{t}(\gamma, g , |\zeta^{o}_{0}(d,x)| , \nu^{o}_{0}(d,x) )$.

	Therefore $	\lim_{\bar{\zeta}^{-o}_{0}(d,x) \rightarrow \infty}  \Gamma(\gamma, g , |\zeta_{0}(d,x)| , \nu_{0}(d,x) ) =  \Omega(\gamma, g , |\zeta^{o}_{0}(d,x)| , \nu^{o}_{0}(d,x) )$, 	as desired. 
\end{proof}
\section{Concentration Inequalities}
\label{app:ConcentrationInequalities}

Recall that for any $d \in \{0,...,M\}$ and any $t \geq 0$,	let $(h_{t}(d), \omega_{t}(d)) \in [0,1]^{2}$ be such that
\begin{align*}
	\mathbf{P} \left(  \iota_{t}(d)  \geq h_{t}(d)    \right) \geq 1 - \omega_{t}(d),
\end{align*}
and $\sum_{d=0}^{M} h_{t}(d) = 1$, where $	\iota_{t}(d) : = t^{-1} \sum_{s=1}^{t} \delta_{s}(d)$.

The next lemma presents a Azuma-Hoeffding-type concentration inequality for $(J_{t})_{t}$ and $(f_{t})_{t}$ which are the basis of our theoretical results.

\begin{lemma}\label{lem:Azzuma}
	For any $d \in \{0,...,M\}$, any $a \geq 0$, any $T> 0$, and any $t \geq 0$,	
	\begin{align*}
		\mathbf{P}   \left(  \left| T^{-1} \sum_{s=1}^{t} ( Y_{s}(d) - \theta(d) )  1\{ D_{s} = d  \}  \right| \geq  a    \right)  \leq 2e^{-0.5 \frac{T^{2}}{t} \frac{ a^{2}  }{ \upsilon   \sigma(d) ^{2} }   } ,
	\end{align*}
	and
	\begin{align*}
		\mathbf{P}   \left(  \left| t^{-1} \sum_{s=1}^{t} 1\{ D_{s} = d  \} - 	\iota_{t}(d)  \right| \geq  a  \right)  \leq  2 e^{- 4 t a^{2}   }.
	\end{align*}
	It readily follows that a common bound with $T=t$ is given by $2 e^{- 0.5 t \frac{ a^{2}  }{ \max\{  1/8 , \upsilon \sigma(d)^{2} \}   }  }$.
\end{lemma}

\begin{remark}[Remarks on Lemma \ref{lem:Azzuma}]  \label{rem:sub-gaussian}
 We use Assumption \ref{ass:sub-gauss}(i) in the first part of the lemma. in particular, it is used in order to get an upper bound with exponential decay. The assumption, however, could be replaced by sub-exponential or any other type of control on the MGF of $Y(d)$, e.g., $E[ e^{ \lambda (Y(d) - \theta(d))}  ] \leq e^{ \kappa(\lambda)} $ for some decreasing function $\lambda \mapsto \kappa(\lambda)$. This change, however, will affect the upper bound obtained in the lemma; it will decay slower than the current one. In fact, up to constant, the result in the lemma will change to
 	\begin{align*}
 		\mathbf{P}   \left(  \left| t^{-1} \sum_{s=1}^{t} ( Y_{s}(d) - \theta(d) )  1\{ D_{s} = d  \}  \right| \geq  a    \right)  \leq 2 e^{ -  t \max_{\lambda \geq 0 } \{   a \lambda - \kappa(\lambda)  \} }   .
 	\end{align*}
  $\triangle$
\end{remark}

\begin{proof}[Proof of Lemma \ref{lem:Azzuma}]
	Let $W_{s}(d) : =  ( Y_{s}(d) - \theta(d) )  1\{ D_{s} = d  \}$.  By the Markov inequality, it follows that, for any $\lambda> 0$,
	\begin{align*}
		\mathbf{P}  \left(  T^{-1} \sum_{s=1}^{t} W_{s}(d)  \geq  a    \right)  \leq E \left[  \prod_{s=1}^{t} e^{ \lambda  W_{s}(d)    }  \right] e^{ - a \lambda T }.
	\end{align*}
	Observe that
	\begin{align*}
		E \left[  \prod_{s=1}^{t} \exp\{ \lambda  W_{s}(d)    \}  \right] = E \left[  \prod_{s=1}^{t-1} \exp\{ \lambda  W_{s}(d)    \}  E_{t} \left[ \exp\{ \lambda  W_{t}(d)    \}   \right]   \right]
	\end{align*}
	where $E_{t}[.]$ denotes the conditional expectation under $\mathbf{P}$ given $(Y_{s})_{s=1}^{t-1}$ and $(D_{s})_{s=1}^{t}$ (but not $Y_{t}(d)$). Observe that $Y_{t}(d)$ is independent of past $Y$'s, given $D_{t}$. This observation and the fact that $Y_{t}(d)$ is sub-gaussian (Assumption \ref{ass:sub-gauss}) imply
	\begin{align*}
		E_{t} \left[ \exp\{ \lambda  W_{t}(d)    \}   \right]  = & E_{t} \left[ \exp\{ \lambda 1\{D_{t} = d\}   (Y_{t}(d)  - \theta(d) )   \}   \right] 	\leq   \exp \{  0.5 \upsilon \sigma(d)^{2}  1\{D_{t} = d\}    \lambda^{2} \} 	\leq   \exp \{  0.5 \upsilon \sigma(d)^{2}    \lambda^{2} \}.
	\end{align*}

	Iterating in this fashion,
	\begin{align*}
		E \left[  \prod_{s=1}^{t} \exp\{ \lambda  W_{s}(d)    \}  \right] \leq  e^{ 0.5 \upsilon \sigma^{2}(d) \lambda^{2}}  E \left[  \prod_{s=1}^{t-1} \exp\{ \lambda  W_{s}(d)    \}  \right] 	\leq  e^{ 0.5 \upsilon \sigma^{2}(d)  t \lambda^{2}}
	\end{align*}
	Therefore, for any $\lambda>0$
	\begin{align*}
		\mathbf{P}  \left(  T^{-1} \sum_{s=1}^{t} W_{s}(d)  \geq  a    \right)  \leq  \exp\{  0.5 \upsilon \sigma^{2}(d) \lambda^{2} t - a \lambda T  \}.
	\end{align*}
	Choosing $ \lambda =(T/t)  a / (\upsilon \sigma^{2}(d) ) $, it follows that
	\begin{align*}
		\mathbf{P}  \left(  t^{-1} \sum_{s=1}^{t} W_{s}(d)  \geq  a    \right)  \leq  \exp\{ - 0.5 \frac{T^{2}}{t} (a^{2}   / ( \upsilon \sigma(d)  ^{2} )  \}.
	\end{align*}
	By analogous calculations, it is easy to show that
	\begin{align*}
		\mathbf{P}  \left(  |t^{-1} \sum_{s=1}^{t} W_{s}(d) |  \geq  a    \right)  \leq  2 \exp\{ - 0.5 \frac{T^{2}}{t}   (a^{2}   / (\upsilon   \sigma(d) ^{2} )  \}.
	\end{align*}

	Now let $W_{t}(d) : = 1\{D_{t} = d\} - \delta_{t}(d)$ and observe that  $\left| t^{-1} \sum_{s=1}^{t} W_{s}(d) \right| \geq  a $ implies that either $t^{-1} \sum_{s=1}^{t} 1\{ D_{s} = d  \} - 	\iota_{t}(d)  \geq  a $ or $ ( t^{-1} \sum_{s=1}^{t} 1\{ D_{s} = d  \} - 	\iota_{t}(d) ) \leq - a $. We only do the proof for the first case since the second one is analogous.
	
	By the Markov inequality, it follows that, for any $\lambda> 0$,
	\begin{align*}
		\mathbf{P}  \left(  t^{-1} \sum_{s=1}^{t} W_{s}(d)  \geq  a   \right) 	= & E \left[  1\{  t^{-1} \sum_{s=1}^{t} W_{s}(d)  \geq  a    \} \right]  	\leq  e^{-\lambda  a t} E \left[  \prod_{s=1}^{t} e^{ \lambda  W_{s}(d)  }   \right] 
		=  e^{-\lambda  a t} E \left[  \prod_{s=1}^{t-1} e^{ \lambda  W_{s}(d)  } E_{t-1} [ e^{ \lambda  W_{t}(d)  }   ]   \right]
	\end{align*}
	where the last line follows by LIE, where $E_{t-1}$ is the expectation conditional on $(Y^{t-1},D^{t-1})$.
	
	Given $(Y^{t-1},D^{t-1})$, $\delta_{t}(.)$ is non-random as it is measurable with respect to these variables. Thus,
	\begin{align*}
		E_{t-1} \left[ e^{ \lambda  W_{t}(d)   }   \right] = e^{-\lambda \delta_{t}(d)} \left( \delta_{t}(d) e^{\lambda} +  (1- \delta_{t}(d))   \right) = e^{L(\lambda)}
	\end{align*}
	where $L(\lambda) = - \lambda   \delta_{t}(d) + \log \left( \delta_{t}(d) e^{\lambda} +  (1- \delta_{t}(d))   \right) $. Observe that $L(0) = 0 $, $L'(\lambda) = -\delta_{t}(d) + \delta_{t}(d) \frac{ e^{\lambda} }{ \delta_{t}(d) e^{\lambda} +  (1- \delta_{t}(d))  }  $ so that $L'(0)=0$ and $L''(\lambda) =  \delta_{t}(d) \left( \frac{ e^{\lambda} (1- \delta_{t}(d))  }{ (  \delta_{t}(d) e^{\lambda} +  (1- \delta_{t}(d)) )^{2} } \right)  $. The global maximum of $L''$ is at $\lambda = \log ((1-\delta_{t}(d))/\delta_{t}(d))$ and thus $L''(\lambda) \leq L''( \log ((1-\delta_{t}(d))/\delta_{t}(d)) ) = \frac{ (1-\delta_{t}(d))^{2}    }{ 4 (1-\delta_{t}(d)   )^{2} } = 0.25$. Therefore, by the Mean Value Theorem,
	\begin{align*}
		E_{t-1} \left[ e^{ \lambda  W_{t}(d)   }   \right]  \leq e^{L(\lambda)} \leq e^{ \frac{1}{8} \lambda^{2} }.
	\end{align*}
		Iterating over this,	$E \left[  \prod_{s=1}^{t} e^{ \lambda  W_{s}(d)   }  \right] \leq  \prod_{s=1}^{t} e^{\frac{ \lambda^{2} }{8}  } = e^{ t \frac{\lambda^{2}} {8}  }$. 	Therefore, for any $\lambda>0$
	\begin{align*}
		\mathbf{P}  \left(  t^{-1} \sum_{s=1}^{t} W_{s}(d)  \geq  a    \right)  \leq  e^{ t \frac{ \lambda^{2} }{8}    - a \lambda t  }.
	\end{align*}
	Choosing $ \lambda = 4a $, it follows that $\mathbf{P}  \left(  t^{-1} \sum_{s=1}^{t} W_{s}(d)  \geq  a    \right)  \leq e^{- 2 t a^{2}   }$. 
\end{proof} 

\section{Appendix for Section \ref{sec:zeta.concentration}} 
\label{app:zeta.concentration} 

Recall that for any $d \in \{0,...,M\}$ and  $t \geq 0$,$$\iota_{t+1}(d) : = \sum_{s=1}^{t+1} \delta_{s}(d), ~J_{t+1}(d) : =  \sum_{s=1}^{t+1} 1\{D_{s}=d\} Y_{s}(d)  /(t+1),~and~f_{t+1}(d) : = N_{t+1}(d) / (t+1) =  \sum_{s=1}^{t+1} 1\{ D_{s} = d  \} /(t+1).$$

We now prove Proposition \ref{pro:concentration.alpha.zeta}.

\begin{proof}[Proof of Proposition \ref{pro:concentration.alpha.zeta}]
	
	Recall that $	\bar{\zeta}_{t}(d)  : = \zeta_{t}(d) - \theta(d)$, $\bar{Y}_{s}(d) : = (Y_{s}(d) - \theta(d) ) $ and $\bar{J}_{t}(d) : = \sum_{s=1}^{t} 1\{ D_{s} = d  \} \bar{Y}_{s}(d)  /t $.

	For any $t$, any $\gamma \geq 0$ and any $\eta \in [0, h_{t}(d)]$, let  $	S(t,\gamma) : = \left\{    | \bar{J}_{t}(d)  | \leq  \gamma         \right\}$, and  $	R(t,\eta) : = \left\{  | f_{t}(d)  - \iota_{t}(d) |  \leq  \eta       \right\}$, 
	and 	$U(t) : = \left\{   \iota_{t}(d)   \geq h_{t}(d)      \right\}$.

	Conditional on these sets, by Lemma \ref{lem:bound.zeta.alpha},
	\begin{align*}
		|\zeta^{\alpha}_{t}(d) - \theta(d) | \leq \Gamma( \gamma , h_{t}(d) - \eta , \bar{\zeta}_{0}(d)    , \nu_{0}(d)  )
	\end{align*}
	
	Therefore, for any $a > 0$,
	\begin{align*}
		\mathbf{P} \left( | \zeta^{\alpha}_{t}(d) - \theta(d) | > a    \right) \leq 1\{ \Gamma( \gamma , h_{t}(d) - \eta , \bar{\zeta}_{0}(d)   , \nu_{0}(d)  ) > a  \} + \mathbf{P} \left( S(t,\gamma)^{C}     \right) + \mathbf{P} \left( R(t,\eta)^{C}     \right) +  \mathbf{P} \left( U(t)^{C}     \right).
	\end{align*}
	
	By Lemma \ref{lem:Azzuma} with $T=t$ and the definition of exploration structure, it follows that 
	\begin{align*}
		\mathbf{P} \left( S(t,\gamma)^{C}     \right) + \mathbf{P} \left( R(t,\eta)^{C}     \right) +  \mathbf{P} \left( U(t,\eta)^{C}     \right)  \leq 2 \left( e^{- 4 t \eta^{2}  } + e^{-0.5 t \frac{\gamma^{2}}{  \upsilon \sigma(d)^{2}      } }  + \omega_{t}(d) \right) .
	\end{align*}

	We now choose $\gamma$, $\eta$ and $a$ for any $\varepsilon > 0$. Let $\eta  = \eta_{t} = h_{t}(d) \sqrt{ 0.25 t^{-1}  \varepsilon } $, $\gamma = \gamma _{t} =   \sqrt{ 2 t^{-1} \upsilon \varepsilon } \sigma(d) $ (which satisfies $\eta \leq h_{t}(d)$). By Lemma \ref{lem:properties.Gamma}(1), $ g \mapsto \Gamma( \gamma_{t} , g , \bar{\zeta}_{0}(d) , \nu_{0}(d)  )$ is decreasing and since $ h_{t}(d) (1 - \sqrt{ 0.25 t^{-1}  \varepsilon } ) \geq 0.5h_{t}(d) \iff t 0.5^{2} \geq 0.25   \varepsilon \iff t \geq \varepsilon  $, then $\Gamma( \gamma_{t} , h_{t}(d) (1 - \sqrt{ 0.25 t^{-1} \varepsilon } )  , \bar{\zeta}_{0}(d) , \nu_{0}(d)  ) \leq \Gamma( \gamma_{t} , 0.5 h_{t}(d)  , \bar{\zeta}_{0}(d)  , \nu_{0}(d)  ) = : a_{t} = a$. With these choices, 
	\begin{align*}
		\mathbf{P} \left( | \zeta^{\alpha}_{t}(d) - \theta(d) | > a_{t}    \right) \leq  2 \left( e^{-\varepsilon} + e^{-\varepsilon h_{t}(d)^{2}  } + \omega_{t}(d) \right) .
	\end{align*}	
	Since $0 \leq h_{t}(d) \leq 1$, it follows that 
	\begin{align*}
		\mathbf{P} \left( | \zeta^{\alpha}_{t}(d) - \theta(d) | > a_{t}    \right) \leq  4 \left(  e^{-\varepsilon h_{t}(d)^{2}  } + \omega_{t}(d) \right) .
	\end{align*}
	Re-normalizing $\varepsilon$ to $\varepsilon/h_{t}(d)^{2}$, the desired result follows. 	
\end{proof}

We now prove Corollary \ref{cor:OracleRobust}.

\begin{proof}[Proof of Corollary \ref{cor:OracleRobust}]
	We can prove the result using limits. For any given $o \ne 0$, let $|\bar{s}_{0}^{o}(d)| : = \sqrt{\nu^{o}_{0}(d)}   |\bar{\zeta}_{0}^{o}(d)|  $ and let $|\bar{s}_{0}^{-0}(d)|$ be the $L \times 1$ vector, excluding $|\bar{s}_{0}^{0}(d)|$. We consider the limit of this quantity going to $\infty$. 	
	
	By Lemma \ref{lem:convergence.GammaToOmega} applied to $o=0$, for each $\gamma \geq 0$ and $\eta \leq h_{t}(d)$,
		\begin{align*}
 \lim_{ |\bar{s}_{0}^{-0}(d)| \rightarrow \infty }	|\Gamma  \left( \gamma , h_{t}(d) - \eta, |\bar{\zeta}_{0}(d)| , \nu_{0}(d)   \right) - \Omega(\gamma, h_{t}(d) - \eta ,  | \bar{\zeta}^{o}_{0}(d) | , \nu^{o}_{0}(d)  )| = 0.
\end{align*}	
	Thus, this result implies that for any given $\delta>0$,  there exists a $C$ such that 
	\begin{align*}
	 \Omega(\gamma, h_{t}(d) - \eta ,  | \bar{\zeta}^{o}_{0}(d) | , \nu^{o}_{0}(d)  )  \geq \Gamma  \left( \gamma , h_{t}(d) - \eta, |\bar{\zeta}_{0}(d)| , \nu_{0}(d)   \right)   - \delta
	\end{align*}
	for any $|\bar{s}_{0}^{-0}(d)| \geq C$. 
	
	The result follows by setting $\eta = 0.5 h_{t}(d)$ and $\gamma = \sqrt{ 2\upsilon \varepsilon/(h_{t}(d)^{2} t  )} \sigma(d)$.
\end{proof}

\section{Appendix for Section \ref{sec:PoM}} 
\label{app:PoMM} 

Proposition \ref{pro:stopping.alpha} follows from this more general lemma that allows for biased sources. To state this lemma we define, for each $d \in \mathbb{D}$, $\eta^{\ast}_{d} : \mathbb{N}  \times [0,1] \times \mathbb{R}_{+} \rightarrow \mathbb{R}_{+} \cup \{+ \infty \}$ as follows: For any $(t,h_{t}(d), \Delta) \in  \mathbb{N}  \times [0,1] \times \mathbb{R}_{+} $, if $\Gamma_{0}(\gamma_{t} ,h_{t}(d) - \eta ,    (-1)^{1\{d=M\}}\zeta_{0}(d)  , \nu_{0}(d) )   < 0.5 \Delta$ for all $\eta$, then we choose $\eta^{\ast}_{d}(t, h_{t}(d) ,   \Delta) = + \infty$; otherwise,  
\begin{align*}
	\eta^{\ast}_{d}(t, h_{t}(d) ,   \Delta) : = \max  \left\{  \eta \colon  \Gamma_{0}(\gamma_{t} ,h_{t}(d) - \eta ,    (-1)^{1\{d=M\}}\zeta_{0}(d)   , \nu_{0}(d) )  \leq 0.5 \Delta~and~\eta \leq h_{t}(d)   \right\}
\end{align*}
and if the set is empty, set $ \eta^{\ast}_{d}(t, h_{t}(d) ,   \Delta) = 0$.

The quantity $\eta^{\ast}_{d}(t,h_{t}(d),\Delta)$ defines the concentration rate of $f_{t}(d)$. Intuitively, this quantity is the highest level of feasible \emph{constant} experimentation, i.e., a constant that is less or equal than $h_{t}(d) $, such that the incidence of the priors on the aggregated posterior mean --- this incidence is given by the function $\Gamma_{0}$ --- is small relative to the true discrepancy of the ATEs, given by $\Delta$.

\begin{lemma}\label{lem:stopping.alpha}
	Consider the stopping rule defined in Example \ref{exa:StoppingRule} with parameters $((\gamma_{t})_{t},B)$ then for any $t \geq B$,
	\begin{align}\label{eqn:PoMM.UpperBound}
		\mathbf{P} \left(  \max_{d\ne M} \{ \zeta^{\alpha}_{\tau}(d) -  \zeta^{\alpha}_{\tau}(M)   \} > 0  \cap \{\tau = t\}   \right)	\leq & 2 \sum_{d=0}^{M}  \left(    e^{-0.5 \frac{  \gamma_{t}(d)^{2} }{  \upsilon \sigma(d)^{2} }  }  +  e^{- 4 t \eta^{\ast}_{d}(t , h_{t}(d) , \Delta ) ^{2}   }   \right) \\
		& + 1\{ \forall d \colon (-1)^{1\{d = M\}}\bar{\zeta}_{0}(d) > 0 \}\sum_{d=0}^{M}  \omega_{t}(d),
	\end{align}
	where $	\eta^{\ast}_{d}(t,h_{t}(d),\Delta) \in \mathbb{R}_{+} \cup \{+ \infty \}$ is defined in Appendix \ref{app:PoMM}  and is non-decreasing in $t$, $h_{t}(d)$, and $\Delta$; and if $(-1)^{1\{ d = M \}} \bar{\zeta}_{0}(d) \leq 0$, then  $	\eta^{\ast}_{d}(t,h_{t}(d),\Delta) = + \infty $ .
\end{lemma}

This lemma shows that the quantity $\eta^{\ast}_{d}(t,h_{t}(d),\Delta)$ is key for understanding how the primitives of our setup --- i.e. the exploration structure and $\Delta$ --- affect the upper bound for the probability of a mistake. The upper bound for the probability of a mistake decays exponentially with $t$ and is non-increasing in $h_{t}(d)$ and $\Delta$. Intuitively, as the degree of exploration increases, the data become less dependent on the past and thus more informative, resulting in a tighter bound. Also, as $\Delta$ becomes more positive, so does the difference between the PM's posteriors, which also decreases the probability of making a mistake.

\begin{proof}[Proof of Lemma \ref{lem:stopping.alpha} ]
	We divide the proof into several steps. Throughout the proof, we use the following definitions. For any $t \in \mathbb{N}$, 
	\begin{align*}
		\mathcal{J}_{t}(\gamma, d)  & : = \{ | \sum_{s=1}^{t} \bar{Y}_{s}(d) | / \sqrt{t} \leq \gamma  \},~\forall \gamma > 0, \\
		\mathcal{V}(t,d) & : = \{ \iota_{t}(d) \geq h_{t}(d) \}, \\
		\mathcal{E}_{t}(\eta,d) &: = \left\{  | f_{t}(d)  - \iota_{t}(d)| \leq  \eta   \right\},~\forall \eta > 0.
	\end{align*}

	\textsc{Step 1} In this step we show that
	\begin{align*}
		\{  \max_{d \ne M} \{  \zeta^{\alpha}_{t}(d)    -   \zeta^{\alpha}_{t}(M) \}    >  0 \} \cap \{ \tau = t \}  \subseteq	\{ \max_{d \ne M}  \left\{  \bar{\zeta}^{\alpha}_{t}(d) - \bar{\zeta}^{\alpha}_{t}(M)  - c_{t}(\gamma_{t},d,M)    \right\} > \Delta \} \cap \{ \tau = t \}  
	\end{align*}
	and that 
	\begin{align*}
		\{ \max_{d \ne M}  \left\{  \bar{\zeta}^{\alpha}_{t}(d) - \bar{\zeta}^{\alpha}_{t}(M)  - c_{t}(\gamma_{t},d,M)    \right\} > \Delta \}  \subseteq & \cup_{d} \left\{   \sum_{o=0}^{L} \alpha^{o}_{t}(d)   \frac{   (-1)^{1\{ d = M \}}\bar{\zeta}_{0}^{o}(d)  \nu_{0}^{o}(d)/t  }{ f_{t}(d) + \nu^{o}_{0}(d)/t   }  > 0.5 \Delta  \right\} \cap  \mathcal{J}_{t}(\gamma_{t}, d)  \\
		& \cup  \mathcal{J}_{t}(\gamma_{t}, d) ^{C}.
	\end{align*}

	Since
	\begin{align*}
		\tau : = \min \left\{ t \geq B \colon    \max_{d}  \left\{   \min_{m \ne d}   \zeta^{\alpha}_{t}(d) - \zeta^{\alpha}_{t}(m)  - c_{t}(\gamma_{t},d,m)    \right\} > 0  \right\},
	\end{align*}
	the event $\{  \max_{d \ne M} \{  \zeta^{\alpha}_{t}(d)    -   \zeta^{\alpha}_{t}(M) \}    >  0 \cap  \tau = t \} $ implies the event $\{ \max_{d \ne M}  \left\{  \zeta^{\alpha}_{t}(d) - \zeta^{\alpha}_{t}(M)  - c_{t}(\gamma_{t},d,M)    \right\} >0 \}$. 
	
	Suppose the max is achieved by $d(t) \ne M$, then the last expression is equivalent to $\bar{\zeta}^{\alpha}_{t}(d(t)) - \bar{\zeta}^{\alpha}_{t}(M)  - c_{t}(\gamma_{t},d,M)    > \theta(M) - \theta(d(t))$. Since $  \theta(M) - \theta(d(t)) \geq \Delta $ --- recall, $\Delta : = \min_{d} \theta(M) - \theta(d)$ ---, it follows that
	\begin{align*}
		\{  \max_{d \ne M} \{  \zeta^{\alpha}_{t}(d)    -   \zeta^{\alpha}_{t}(M) \}    >  0 \cap  \tau = t \}  \subseteq \{ \max_{d \ne M}  \left\{  \bar{\zeta}^{\alpha}_{t}(d) - \bar{\zeta}^{\alpha}_{t}(M)  - c_{t}(\gamma_{t},d,M)    \right\} > \Delta \} . 
	\end{align*}
	
	Observe that 
	\begin{align*}
		c_{t}(\gamma_{t},d,M)  = : c_{t}(\gamma_{t},d)+c_{t}(\gamma_{t},M), 
	\end{align*}
	where $(\gamma,d) \mapsto c_{t}(\gamma,d) : =  \sum_{o=0}^{L} \frac{\alpha^{o}_{t}(d)  \sqrt{t} \gamma(d) } { N_{t}(d) + \nu^{o}_{0}(d)  }  $.  
	
	So the event $	\{ \max_{d \ne M}  \left\{  \bar{\zeta}^{\alpha}_{t}(d) - \bar{\zeta}^{\alpha}_{t}(M)  - c_{t}(\gamma_{t},d,M)    \right\} > \Delta \} $ is included in the event
	\begin{align*}
		&\cup_{d \ne M} \{  \left\{  \bar{\zeta}^{\alpha}_{t}(d) - \bar{\zeta}^{\alpha}_{t}(M)  - c_{t}(\gamma_{t},d,M)    \right\} > \Delta \} \cap \{    \bar{\zeta}^{\alpha}_{t}(M)   +  c_{t}(\gamma_{t},M)  \geq - 0.5 \Delta   \}   \\
		&~\cup \{    \bar{\zeta}^{\alpha}_{t}(M)   +  c_{t}(\gamma_{t},M) < - 0.5 \Delta   \}  \\
		= &\cup_{d \ne M} \{    \bar{\zeta}^{\alpha}_{t}(d) >  c_{t}(\gamma_{t},d) + 0.5 \Delta \}   \cup \{    \bar{\zeta}^{\alpha}_{t}(M)   <   - (c_{t}(\gamma_{t},M)  + 0.5 \Delta )  \}.
	\end{align*}
	We now bound each of the sets in the RHS.
	
	Observe that $\bar{\zeta}^{\alpha}_{t}(d) =  \left( \sum_{o=1}^{L} \frac{\alpha^{o}_{t}(d)}{N_{t}(d) + \nu^{o}_{0}(d)}  \right)  \left( \sum_{s=1}^{t} 1\{D_{s}=d\} \bar{Y}_{s}(d)  \right) + \sum_{o=1}^{L}\alpha^{o}_{t}(d)  \frac{ \nu^{o}_{0}(d) \bar{\zeta}^{o}_{0}(d)  }{N_{t}(d) + \nu^{o}_{0}(d)} $. Conditional on the set $ \mathcal{J}_{t}(\gamma, d) $, it follows that 
	\begin{align*}
		&\bar{\zeta}^{\alpha}_{t}(d)  \leq \gamma \sqrt{t} \left( \sum_{o=1}^{L} \frac{\alpha^{o}_{t}(d)}{N_{t}(d) + \nu^{o}_{0}(d)}  \right) + \sum_{o=1}^{L}\alpha^{o}_{t}(d)  \frac{  \bar{\zeta}^{o}_{0}(d) \nu^{o}_{0}(d)/t }{f_{t}(d) + \nu^{o}_{0}(d)/t}\\
		and~&\bar{\zeta}^{\alpha}_{t}(d)  \geq -\gamma \sqrt{t} \left( \sum_{o=1}^{L} \frac{\alpha^{o}_{t}(d)}{N_{t}(d) + \nu^{o}_{0}(d)}  \right) + \sum_{o=1}^{L}\alpha^{o}_{t}(d)  \frac{  \bar{\zeta}^{o}_{0}(d) \nu^{o}_{0}(d)/t }{f_{t}(d) + \nu^{o}_{0}(d)/t}
	\end{align*}
	
	By the definition of $c_{t}$ and choosing $\gamma = \gamma_{t}(d)$, it follows that for any $d \in \{0,...,M-1\}$, 
	\begin{align*}
		\{    \bar{\zeta}^{\alpha}_{t}(d) >  c_{t}(\gamma_{t},d) + 0.5 \Delta \}  \subseteq & \{    \bar{\zeta}^{\alpha}_{t}(d) >  c_{t}(\gamma_{t},d) + 0.5 \Delta \} \cap  \mathcal{J}_{t}(\gamma_{t}, d)  \cup  \mathcal{J}_{t}(\gamma_{t}, d) ^{C}\\
		\subseteq &  \left\{   \sum_{o=0}^{L} \alpha^{o}_{t}(d)   \frac{   \bar{\zeta}_{0}^{o}(d)  \nu_{0}^{o}(d)/t  }{ f_{t}(d) + \nu^{o}_{0}(d)/t   }  > 0.5 \Delta  \right\} \cap  \mathcal{J}_{t}(\gamma_{t}, d)   \cup  \mathcal{J}_{t}(\gamma_{t}, d) ^{C}.
	\end{align*} 
	Similarly,
	\begin{align*}
		\{    \bar{\zeta}^{\alpha}_{t}(M)   >   - (c_{t}(\gamma_{t},M)  + 0.5 \Delta )  \} \subseteq & \left\{   \sum_{o=0}^{L} \alpha^{o}_{t}(M)   \frac{  (- \bar{\zeta}_{0}^{o}(M) ) \nu_{0}^{o}(M)/t  }{ f_{t}(M) + \nu^{o}_{0}(M)/t   }  > 0.5 \Delta  \right\}  \cap  \mathcal{J}_{t}(\gamma_{t}, M) \cup  \mathcal{J}_{t}(\gamma_{t}, M) ^{C}.
	\end{align*}
	Hence,
	\begin{align*}
		\{ \max_{d \ne M}  \left\{  \bar{\zeta}^{\alpha}_{t}(d) - \bar{\zeta}^{\alpha}_{t}(M)  - c_{t}(\gamma_{t},d,M)    \right\} > \Delta \} \subseteq & \cup_{d} \left\{   \sum_{o=0}^{L} \alpha^{o}_{t}(d)   \frac{   (-1)^{1\{ d = M \}}\bar{\zeta}_{0}^{o}(d)  \nu_{0}^{o}(d)/t  }{ f_{t}(d) + \nu^{o}_{0}(d)/t   }  > 0.5 \Delta  \right\} \cap  \mathcal{J}_{t}(\gamma_{t}, d) \\
		&  \cup  \mathcal{J}_{t}(\gamma_{t}, d) ^{C}.
	\end{align*}

	\bigskip
	
	\textsc{Step 2.} We now bound 	$\mathbf{P} \left(  \max_{d \ne M} \{ \zeta^{\alpha}_{\tau}(d) - \zeta^{\alpha}_{\tau}(M)  \} > 0  \cap \{ \tau  = t \} \right)$ when $\forall d \colon (-1)^{1\{d = M\}}\bar{\zeta}_{0}(d) \leq 0$. By Step 1, 
	\begin{align*}
		\mathbf{P} \left(  \max_{d \ne M} \{ \zeta^{\alpha}_{\tau}(d) - \zeta^{\alpha}_{\tau}(M)  \} > 0  \cap \{ \tau  = t \}   \right) \leq &  \mathbf{P} \left(  \max_{d \ne M} \{ \bar{\zeta}^{\alpha}_{t}(d) - \bar{\zeta}^{\alpha}_{t}(M)  \} > \Delta   \right)\\
		\leq & \mathbf{P} \left( \cup_{d} \left\{   \sum_{o=0}^{L} \alpha^{o}_{t}(d)   \frac{   (-1)^{1\{ d = M \}}\bar{\zeta}_{0}^{o}(d)  \nu_{0}^{o}(d)/t  }{ f_{t}(d) + \nu^{o}_{0}(d)/t   }  > 0.5 \Delta  \right\} \cap  \mathcal{J}_{t}(\gamma_{t}, d)  \right)\\
		& + \mathbf{P} ( \mathcal{J}_{t}(\gamma_{t}, d) ^{C} ). 
	\end{align*}
	By the assumption that $\forall d \colon (-1)^{1\{d = M\}}\bar{\zeta}_{0}(d) \leq 0$, the first term in the RHS is 0. So the result follows from Lemma \ref{lem:Azzuma} with $T = \sqrt{t}$.
	
	\bigskip
	
	\textsc{Step 3.} We now bound 	$\mathbf{P} \left(  \max_{d \ne M} \{ \zeta^{\alpha}_{\tau}(d) - \zeta^{\alpha}_{\tau}(M)  \} > 0 \cap \{ \tau = t \}  \right)$
	when $\forall d \colon (-1)^{1\{d = M\}}\bar{\zeta}_{0}(d) \leq 0$ \emph{does not} hold. 
	
	Observe that 
	\begin{align*}
		\mathbf{P} \left(  \max_{d \ne M} \{ \zeta^{\alpha}_{\tau}(d) - \zeta^{\alpha}_{\tau}(M)  \} > 0   \cap \{ \tau = t \}  \right)   	\leq & \mathbf{P} \left(  \max_{d \ne M} \{ \zeta^{\alpha}_{\tau}(d) - \zeta^{\alpha}_{\tau}(M)  \} > 0   \cap \{ \tau = t \}   \cap  \mathcal{V}(t,d) \right)  + \mathbf{P} \left(  \mathcal{V}(t,d)^{C}   \right)  \\
		\leq &   \mathbf{P} \left(  \max_{d \ne M} \{ \zeta^{\alpha}_{\tau}(d) - \zeta^{\alpha}_{\tau}(M)  \} > 0   \cap \{ \tau = t \}   \cap   \mathcal{V}(t,d) \right)   + \omega_{t}(d).
	\end{align*}
	
	We now bound $\mathbf{P} \left(  \max_{d \ne M} \{ \zeta^{\alpha}_{\tau}(d) - \zeta^{\alpha}_{\tau}(M)  \} > 0     \cap \{ \tau = t \}  \cap  \mathcal{V}(t,d) \right)$. By Step 1 and the union bound, this  probability can be bounded by
	\begin{align*}
		& \sum_{d} 	\mathbf{P} \left( \left\{   \sum_{o=0}^{L} \alpha^{o}_{t}(d)   \frac{   (-1)^{1\{ d = M \}}\bar{\zeta}_{0}^{o}(d)  \nu_{0}^{o}(d)/t  }{ f_{t}(d) + \nu^{o}_{0}(d)/t   }  > 0.5 \Delta  \right\}  \cap  \mathcal{J}_{t}(\gamma_{t}, d)  \cap \mathcal{E}_{t}(\eta ,d)  \right) \\
		& + 	\sum_{d}  \left(  \mathbf{P} \left(  \mathcal{J}_{t}(\gamma_{t}, d) ^{C} \right) + \mathbf{P} \left(  \mathcal{E}_{t}(\eta, d) ^{C} \right) \right). 
	\end{align*}
	for some $\eta > 0$ to be chosen below.

	We now bound the first term in the display above. To do this, observe that $\mathcal{E}_{t}(\eta,d)$ and $\mathcal{V}(t,d)$, imply $\left\{  f_{t}(d)  \geq h_{t}(d) -  \eta   \right\}$. This fact and Lemmas \ref{lem:bound.zeta.o}(2) and \ref{lem:alpha.bound}, imply that under $  \mathcal{J}_{t}(\gamma_{t}, d) \cap \left\{  f_{t}(d)  \geq h_{t}(d) -  \eta   \right\} $, it follows that for any $d \in \mathbb{D}$, 
	\begin{align*}
		\alpha^{o}_{t}(d)   \frac{   \bar{\zeta}_{0}^{o}(d)  \nu_{0}^{o}(d)/t  }{ f_{t}(d) + \nu^{o}_{0}(d)/t   }  \leq  & \alpha^{o}_{t}(d)  \Omega_{0}( h_{t}(d) - \eta ,   (-1)^{d=M}  \bar{\zeta}_{0}^{o}(d) , \nu^{o}_{0}(d) ) \\
		\leq & \overline{\alpha}^{o}_{t}( \gamma_{t} , h_{t}(d) - \eta , |\zeta_{0}(d) | , \nu_{0}(d)   ) \Omega^{+}_{0}( h_{t}(d) - \eta ,   (-1)^{d=M} \bar{\zeta}_{0}^{o}(d) , \nu^{o}_{0}(d) ) \\
		& +  \underline{\alpha}^{o}_{t}( \gamma_{t} , h_{t}(d) - \eta , |\zeta_{0}(d) | , \nu_{0}(d)   ) \Omega^{-}_{0}( h_{t}(d) - \eta ,   (-1)^{d=M} \bar{\zeta}_{0}^{o}(d) , \nu^{o}_{0}(d) ).
	\end{align*}
	The RHS coincides with $\Gamma_{0}$ defined above. Thus, for any $d \in \mathbb{D}$, 
	\begin{align*}
		&	\left\{  \sum_{o=0}^{L} 	 \alpha^{o}_{t}(d)   \frac{   (-1)^{d=M} \bar{\zeta}_{0}^{o}(d)  \nu_{0}^{o}(d)/t  }{ f_{t}(d) + \nu^{o}_{0}(d)/t   }      > 0.5 \Delta   \right\}  \cap  \mathcal{J}_{t}(\gamma_{t}, d) \cap \mathcal{E}_{t}(\eta,d) \\
		\subseteq & 	\left\{  \Gamma_{0}(\gamma_{t} ,h_{t}(d) - \eta ,  (-1)^{1\{d=M\}}\bar{\zeta}_{0}(d)  , \nu_{0}(d) )   > 0.5 \Delta   \right\} \cap  \mathcal{J}_{t}(\gamma_{t}, d) \cap \mathcal{E}_{t}(\eta,d),
	\end{align*}
	which in turn implies
	\begin{align}\notag
		\mathbf{P} \left(  \max_{d \ne M} \{ \zeta^{\alpha}_{\tau}(d) - \zeta^{\alpha}_{\tau}(M)  \} > 0     \cap \{ \tau = t \}  \cap  \mathcal{V}(t,d) \right) \leq & \sum_{d} 1\{ \mathcal{U}_{d}(  \gamma_{t}  , h_{t}(d) - \eta , \Delta  )    \} \\ \label{eqn:PoMM.1}
		& + 	\sum_{d}  \left(  \mathbf{P} \left(  \mathcal{J}_{t}(\gamma_{t}, d) ^{C} \right) + \mathbf{P} \left(  \mathcal{E}_{t}(\eta, d) ^{C} \right) \right)
	\end{align}
	where $\mathcal{U}_{d}(  \gamma_{t}  , h_{t}(d) - \eta , \Delta  ) : =  	\left\{  \Gamma_{0}(\gamma_{t} ,h_{t}(d) - \eta ,    (-1)^{1\{d=M\}}\bar{\zeta}_{0}(d)   , \nu_{0}(d) )   > 0.5 \Delta   \right\} $.
	
	We now choose $\eta$ so that the first term in the RHS is naught. If $\Gamma_{0}(\gamma_{t} ,h_{t}(d) - \eta ,    (-1)^{1\{d=M\}}\bar{\zeta}_{0}(d)  , \nu_{0}(d) )   < 0.5 \Delta$ for all $\eta$, then we choose $\eta^{\ast}_{d}(t, h_{t}(d) ,   \Delta) = + \infty$; otherwise,  
	\begin{align*}
		\eta^{\ast}_{d}(t, h_{t}(d) ,   \Delta) : = \max \left\{  \eta \colon  \Gamma_{0}(\gamma_{t} ,h_{t}(d) - \eta ,    (-1)^{1\{d=M\}}\bar{\zeta}_{0}(d)   , \nu_{0}(d) )   \leq 0.5 \Delta~and~\eta \leq h_{t}(d)   \right\}
	\end{align*}
	and if the set is empty, set $ \eta^{\ast}_{d}(t, h_{t}(d) ,   \Delta) = 0$.
	
	If 	$\eta^{\ast}_{d}(t  , h_{t}(d) , \Delta ) = 0$, the expression \ref{eqn:PoMM.1} yields the trivial bound of 1. The expression in the proposition also implies an upper bound greater than 1 (since $\eta^{\ast}_{d}(t  , h_{t}(d) , \Delta ) = 0$). Thus the proposition is proven. We now study the case if $\eta^{\ast}_{d}(t  , h_{t}(d) , \Delta ) > 0$. Under this choice of $\eta$, expression \ref{eqn:PoMM.1} implies
	\begin{align*}
		\mathbf{P} \left(  \max_{d \ne M} \{ \zeta^{\alpha}_{\tau}(d) - \zeta^{\alpha}_{\tau}(M)  \} > 0     \cap \{ \tau = t \}  \cap  \mathcal{V}(t,d) \right) \leq \sum_{d}  \left(  \mathbf{P} \left(  \mathcal{J}_{t}(\gamma_{t}, d) ^{C} \right) + \mathbf{P} \left(  \mathcal{E}_{t}(\eta^{\ast}_{d}(t  , h_{t}(d) , \Delta ), d) ^{C} \right) \right).
	\end{align*}
	
	Thus, by Lemma \ref{lem:Azzuma},
	\begin{align*}
		\mathbf{P} \left(  \max_{d \ne M} \{ \zeta^{\alpha}_{\tau}(d) - \zeta^{\alpha}_{\tau}(M)  \} > 0   \cap \{ \tau = t \}  \right) \leq  2 \sum_{d=0}^{M}  \left(    e^{-0.5 \frac{  \gamma_{t}^{2} }{  \upsilon \sigma(d)^{2} }  }  +  e^{- 4 t \eta^{\ast}_{d}(t , h_{t}(d) , \Delta ) ^{2}   }  + \omega_{t}(d)  \right) 
	\end{align*}
	
	\bigskip 
	
	\textsc{Step 4.} 	We conclude the proof by showing some properties of $\eta^{\ast}_{d}$. First, $t \mapsto \eta^{\ast}_{d}(t ,  h_{t}(d) , \Delta )$ is non-decreasing. To show this, first note that $ \Gamma_{0}(\gamma_{t} ,h_{t}(d) - \eta ,    (-1)^{1\{d=M\}}\bar{\zeta}_{0}(d)   , \nu_{0}(d) )$ is (implicitly) a function of $t$ and thus it suffices to show it is non-increasing (for a fixed $h_{t}(d)$) and $\eta \mapsto  \Gamma_{0}(\gamma_{t} ,h_{t}(d) - \eta ,    (-1)^{1\{d=M\}}\bar{\zeta}_{0}(d)   , \nu_{0}(d) )$ is non-decreasing. This follows from Lemma \ref{lem:properties.Gamma}(2) and the fact that $g$ (in that lemma) equals $h_{t}(d) - \eta$. ; we now show the former. 
	
	By construction of $\Gamma_{0}$ it suffices to show that $t \mapsto K_{t}(\gamma_{t},h_{t}(d) - \eta) : = \overline{\alpha}^{o}_{t}( \gamma_{t} , h_{t}(d) - \eta , |\bar{\zeta}_{0}(d) | , \nu_{0}(d)   ) \Omega^{+}_{0}( h_{t}(d) - \eta ,   (-1)^{d=M} \bar{\zeta}_{0}^{o}(d) , \nu^{o}_{0}(d) )  +  \underline{\alpha}^{o}_{t}( \gamma_{t} , h_{t}(d) - \eta , |\bar{\zeta}_{0}(d) | , \nu_{0}(d)   ) \Omega^{-}_{0}( h_{t}(d) - \eta ,   (-1)^{d=M} \bar{\zeta}_{0}^{o}(d) , \nu^{o}_{0}(d) )$ is non-increasing for each $o$. If $ (-1)^{d=M} \bar{\zeta}_{0}^{o}(d)  \geq 0$ then $\Omega_{0}$ is positive and decreasing as a function of $t$ (see its definition). In addition, $\underline{\sigma}^{o}_{t}$ and $\overline{\sigma}^{o}_{t}$ are non-increasing and decreasing in $t$ resp. Hence $\overline{\alpha}^{o}_{t}$ is decreasing in $t$ and positive. Thus,  $K_{t}(\gamma,h_{t}(d) - \eta) = \overline{\alpha}^{o}_{t}( \gamma , h_{t}(d) - \eta , |\bar{\zeta}_{0}(d) | , \nu_{0}(d)   ) \Omega_{0}( h_{t}(d) - \eta ,   (-1)^{d=M} \bar{\zeta}_{0}^{o}(d) , \nu^{o}_{0}(d) ) $ is decreasing. In addition, by the proof of Lemma \ref{lem:properties.Gamma}(1) it follows that $\gamma \mapsto K_{t}(\gamma,h_{t}(d) - \eta)$ is increasing and since $t \mapsto \gamma_{t}$ is non-increasing it follows that $t \mapsto K_{t}(\gamma_{t},h_{t}(d) - \eta)$ is non-increasing for this case. If $ (-1)^{d=M} \bar{\zeta}_{0}^{o}(d)  \leq 0$ then $\Omega_{0}$ is negative and decreasing as a function of $t$ (see its definition). Also, $\underline{\alpha}^{o}_{t}$ is increasing in $t$ (see Lemma \ref{lem:properties.bound.alpha}(5)) and positive. Thus, $ t\ \mapsto K_{t}(\gamma,h_{t}(d) - \eta) = \underline{\alpha}^{o}_{t}( \gamma_{t} , h_{t}(d) - \eta , |\bar{\zeta}_{0}(d) | , \nu_{0}(d)   ) \Omega^{-}_{0}( h_{t}(d) - \eta ,   (-1)^{d=M} \bar{\zeta}_{0}^{o}(d) , \nu^{o}_{0}(d) ) $ is decreasing in this case. We thus showed that $t \mapsto  \Gamma_{0}(\gamma_{t} ,h_{t}(d) - \eta ,  (-1)^{1\{d=M  \}}\bar{\zeta}_{0}(d) , \nu_{0}(d) )$ is non-increasing.

	Second, $\Delta \mapsto \eta^{\ast}_{d}(t  , h_{t}(d) , \Delta )$ is non-decreasing. To show this is sufficient to show that $\eta \mapsto  \Gamma_{0}(\gamma_{t} ,h_{t}(d) - \eta ,   (-1)^{1\{d=M  \}}\bar{\zeta}_{0}(d)   , \nu_{0}(d) )$ is non-decreasing. This follows from Lemma \ref{lem:properties.Gamma}(2) and the fact that $g$ (in that lemma) equals $h_{t}(d) - \eta$.


	Third, $h_{t}(d)  \mapsto \eta^{\ast}_{d}(t , h_{t}(d) , \Delta )$ is non-decreasing. As before, this follows from Lemma \ref{lem:properties.Gamma}(2).
\end{proof}

The proof of Corollary \ref{cor:robust.PoMM} follows from this more general lemma that allows for biased sources. 

\begin{lemma}\label{lem:robust.PoMM}
	Suppose all the conditions of Lemma \ref{lem:stopping.alpha} hold and $ \frac{  |\zeta_{0}^{0}(d) - \theta(d) | \nu_{0}^{0}(d) /t  }{ h_{t}(d) + \nu^{0}_{0}(d) /t  } \leq 0.5 \Delta$.\footnote{This last condition always holds for sufficiently small biases or for large values of $t$.} Then, for any $\varepsilon > 0$, there exists a $C$ such that for all $\min_{o \ne 0} |\zeta^{o}_{0}(d) - \theta(d)| \geq C$, it follows that 	
\begin{align*}
	\mathbf{P} \left(  \max_{d\ne M} \{ \zeta^{\alpha}_{\tau}(d) -  \zeta^{\alpha}_{\tau}(M)   \} > 0   \cap \{ \tau = t\}  \right)	\leq &  \sum_{d=0}^{M} ( 2  e^{  - \frac{0.5 (\gamma_{t}(d) )^{2} } { \upsilon \sigma(d)^{2}} } + e^{  - 4  t    ( \eta^{oracle}_{d} (t,h_{t}(d),(\Delta - \varepsilon ) /(1+\varepsilon) ) )^{2}   }) \\
	& + 1\{ \forall d \colon (-1)^{1\{d=M\}} \bar{\zeta}_{0}(d) \leq 0  \}  \omega_{t}(d)     .
\end{align*}
where $	\eta^{oracle}_{d}(t,h_{t}(d),(\Delta - \varepsilon ) /(1+\varepsilon))$ is defined as
\begin{align*}
			\max\left\{  \eta \colon    \frac{ | \zeta^{0}_{0}(d) - \theta(d) | \nu_{0}^{0}(d)/t  }{ h_{t}(d) - \eta + \nu^{0}_{0}(d)/t   } \leq 0.50 \frac{\Delta - \varepsilon}{1+\varepsilon}  ~and~\eta \leq \epsilon  \right\}.
		\end{align*}
	
Moreover, if 	$\bar{\zeta}^{0}_{0}(d) = 0$, then $\eta^{oracle}_{d}(t,h_{t}(d),\Delta) = \infty$.	
\end{lemma}

The behavior of $\eta^{\ast}_{d}$ determines whether the upper bound embodies an oracle property similar to the one we demonstrated for the concentration rates. Given the properties of the weights illustrated in Proposition \ref{pro:alpha.asymptotics.general} and Lemma \ref{lem:alpha.properties}, it is easy to show that if sources other than $o=0$ are sufficiently stubborn, then $\eta^{\ast}_d$ becomes arbitrary close to $\eta^{oracle}_d$, where $\eta^{oracle}_{d}$ is defined as the largest $\eta$ such that $ \frac{ |\zeta_{0}^{0}(d) - \theta(d)| \nu^{0}_{0}(d)/t   }{ h_{t}(d) - \eta + \nu^{0}_{0}(d)/t   }  \leq 0.5 \Delta$, which is the relevant quantity determining the probability of mistake for the \emph{least stubborn} source. It then follows that the bound obtained in Proposition \ref{pro:stopping.alpha} would be arbitrary close to the oracle one;  the corollary below formalizes this discussion.

\begin{proof}[Proof of Lemma \ref{lem:robust.PoMM}]
	
	We only prove the result for the case where $(-1)^{1\{d=M\}} \bar{\zeta}_{0}(d) \leq 0 $ does not hold (the proof for the other case is analogous). 
	
	By the same calculations as those in Step 3 of the proof of Proposition \ref{pro:stopping.alpha}, for all $d \in \mathbb{D}$,
		\begin{align*}
	&	\{    \bar{\zeta}^{\alpha}_{t}(d) >  (-1)^{1\{d=M\}}  ( c_{t}(\gamma_{t},d) + 0.5 \Delta ) \} \cap \mathcal{V}(t,d) \\
		\subseteq &  \left\{   \sum_{o=0}^{L} \alpha^{o}_{t}(d)   \frac{  (-1)^{1\{d=M\}}   \bar{\zeta}_{0}^{o}(d)  \nu_{0}^{o}(d)/t  }{ f_{t}(d) + \nu^{o}_{0}(d)/t   }  > 0.5 \Delta  \right\} \cap  \mathcal{J}_{t}(\gamma_{t}, d) \cap \mathcal{E}_{t}(\eta,d) \cap \mathcal{V}(t,d) \\
		&   \cup  \mathcal{J}_{t}(\gamma_{t}, d) ^{C} \cup \mathcal{E}_{t}(\eta, d) ^{C}.
	\end{align*} 
	
	Observe that $\mathcal{E}_{t}(\eta,d)$ and $\mathcal{V}(t,d)$, imply $\left\{  f_{t}(d)  \geq h_{t}(d) -  \eta   \right\}$. This fact, Lemma \ref{lem:alpha.bound} and the proof of Lemma \ref{lem:convergence.GammaToOmega} imply that under  $  \mathcal{J}_{t}(\gamma_{t}, d) \cap \left\{  f_{t}(d)  \geq h_{t}(d) -  \eta   \right\} $, $\lim_{ |\bar{\zeta}^{-0}_{0}(d)| \rightarrow \infty  }\alpha^{o}_{t}(d) \max\{ | \bar{\zeta}^{o}_{0}(d) | , 1\} = 0  $ a.s., for all $o \ne 0$. Since the $(\alpha^{o}_{t}(d))_{o =0}^{L}$ sum to one, this implies that $\lim_{ |\bar{\zeta}^{-0}_{0}(d)| \rightarrow \infty  }\alpha^{0}_{t}(d)  = 1  $ a.s.
	
	Hence, for any $\varepsilon>0$, there exists a $C$ such that, if $|\bar{\zeta}^{-0}_{0}(d)| \geq C$, then 
			\begin{align*}
		&	\{    \bar{\zeta}^{\alpha}_{t}(d) >  (-1)^{1\{d=M\}}  ( c_{t}(\gamma_{t},d) + 0.5 \Delta ) \} \\
		\subseteq &  \left\{    (1+\varepsilon)   \frac{  |  \bar{\zeta}_{0}^{0}(d) | \nu_{0}^{o}(d)/t  }{ f_{t}(d) + \nu^{o}_{0}(d)/t   }  + \varepsilon > 0.5 \Delta  \right\} \cap  \mathcal{J}_{t}(\gamma_{t}, d) \cap \mathcal{E}_{t}(\eta,d)   \cup  \mathcal{J}_{t}(\gamma_{t}, d) ^{C} \cup \mathcal{E}_{t}(\eta, d) ^{C}.
	\end{align*} 
	
	The rest of the proof follows the same steps as the proof of Proposition \ref{pro:stopping.alpha}, but instead of using $\eta^{\ast}_{d}$, we use 
	\begin{align*}
	 \eta^{oracle}_{d} (t,h_{t}(d),(\Delta - \varepsilon ) /(1+\varepsilon) ) : =  	\max\left\{  \eta \colon    \frac{ | \zeta^{0}_{0}(d) - \theta(d) | \nu_{0}^{0}(d)/t  }{ h_{t}(d) - \eta + \nu^{0}_{0}(d)/t   } \leq 0.50 \frac{\Delta - \varepsilon}{1+\varepsilon}  ~and~\eta \leq \epsilon  \right\}.
	\end{align*}

Finally, if $| \zeta^{0}_{0}(d) - \theta(d) | = 0$, then for any $\varepsilon < 0.5 \Delta$, 	$\{    \bar{\zeta}^{\alpha}_{t}(d) >  (-1)^{1\{d=M\}}  ( c_{t}(\gamma_{t},d) + 0.5 \Delta ) \} \subseteq  \mathcal{J}_{t}(\gamma_{t}, d) ^{C} \cup \mathcal{E}_{t}(\eta, d) ^{C}$,
so one can set $\eta^{oracle}_{d} = \infty$ and obtain that $\{    \bar{\zeta}^{\alpha}_{t}(d) >  (-1)^{1\{d=M\}}  ( c_{t}(\gamma_{t},d) + 0.5 \Delta ) \} \subseteq  \mathcal{J}_{t}(\gamma_{t}, d) ^{C}$. This result implies that there exists a $C$ (the one constant corresponding to any $\varepsilon \leq 0.5 \Delta$) such that, if $|\bar{\zeta}^{-0}_{0}(d)| \geq C$, then 
\begin{align*}
	\mathbf{P} \left(  \max_{d\ne M} \{ \zeta^{\alpha}_{\tau}(d) -  \zeta^{\alpha}_{\tau}(M)   \} > 0  \cap \{ \tau  = t \}   \right)	\leq &  2 \sum_{d=0}^{M}  e^{  -0.5 \frac{ (\gamma_{t}(d) )^{2} } { \upsilon \sigma(d)^{2}} } 
\end{align*}

\end{proof}

The proof of Corollary \ref{cor:PoMM.beta} follows from this more general Lemma that allows for biased sources.

\begin{lemma}\label{lem:PoMM.beta}
	Suppose all the conditions of Proposition \ref{pro:stopping.alpha} hold, and, for any $t$, $\gamma_{t}(d) \geq 2  \sqrt{\upsilon} \sigma(d) A $ for all $d \in \mathbb{D}$  and $ \frac{4   \min_{d \in \mathbb{D}}  ( \eta^{\ast}_{d} (t,h_{t}(d),\Delta) )^{2} }{A'}   \geq  \frac{1}{t} $ with $A$ and $A'$ such that
	
\begin{align}\label{eqn:stopping-01}
	A \geq - \log \frac{\beta}{M+1}~and~A' \geq - \log \frac{\beta}{M+1}
\end{align}
Then
\begin{align*}
	\max_{t \in \{B,...,T\}} 	\mathbf{P} \left(  \max_{d\ne M} \{ \zeta^{\alpha}_{\tau}(d) -  \zeta^{\alpha}_{\tau}(M)   \} > 0   \cap \{\tau = t\}  \right)	\leq \beta.
\end{align*}
\end{lemma}

Observe that the condition  $ \frac{4   \min_{d \in \mathbb{D}}  ( \eta^{\ast}_{d} (t,h_{t}(d),\Delta) )^{2} }{A'}   \geq  \frac{1}{t} $ is trivially satisfied in the setup of Corollary \ref{cor:PoMM.beta} since there $\eta^{\ast}_{d} = \infty$. 

\begin{proof}[Proof of Lemma \ref{lem:PoMM.beta}]
	We do the proof for where the expression for when $\gamma_{t} $ holds with equality. We do this because if the desired bound holds for this case, it will hold for any $\gamma_{t}$ that is greater. By Lemma \ref{lem:stopping.alpha}
	\begin{align}
		\mathbf{P} \left(  \max_{d\ne M} \{ \zeta^{\alpha}_{\tau}(d) -  \zeta^{\alpha}_{\tau}(M)   \} > 0   \cap \{\tau = t  \}  \right)	\leq \sum_{d=0}^{M}  ( 2  e^{  -0.5 \frac{ (\gamma_{t}(d) )^{2} } { \upsilon \sigma(d)^{2}} } + e^{  - 4 t     ( \eta^{\ast}_{d} (t,h_{t}(d),\Delta) )^{2}   }). 
	\end{align}
	By our choice of $\gamma_{t}$ and the condition on $A'$, the first term in the RHS is less or equal than $2 (M+1) e^{-A}$ whereas the second is equal to $(M+1) e^{-A'}$. Hence, the desired result immediately follows.

%
\end{proof}

\section{Appendix for Section \ref{sec:avg.Y} }
\label{app:avg.Y} 

To show Proposition \ref{pro:avg.Y} we use the following lemmas, whose proofs are relegated to the end of the section.

\begin{lemma}\label{lem:AvgOutcome.lower.0}
	For any $t \in \{1,...,T\}$ and any $\gamma>0$,
	\begin{align*}
		\mathbf{P} \left(   \max_{d} \theta(d) - t^{-1} \sum_{s=1}^{t} Y_{s}    >  - \sqrt{\frac{\gamma}{t}} \left(    \sqrt{2\upsilon} \sigma(d)   +  \frac{ ||\theta ||_{1} }{2}   \right)   +  \max_{d} \theta(d)   - \sum_{d=0}^{M} \theta(d) \iota_{t}(d)    \right) \leq 4 e^{-\gamma}.
	\end{align*}
\end{lemma}

Let $(1-\omega_{t})_{t}$ be any likelihood of exploration associated to the policy rule, $\delta_{t}(d) = \Xi_{t} (M+1)^{-1} + (1-\Xi_{t}) 1\{ d = \arg\max_{a} \zeta^{\alpha}_{t-1}(a)   \}  $ for any $t \in \{1,...,T\}$. 

\begin{lemma}\label{lem:AvgOutcome.lower.1}
	Suppose $\delta_{t}(d) = \Xi_{t} (M+1)^{-1} + (1-\Xi_{t}) 1\{ d = \arg\max_{a} \zeta^{\alpha}_{t-1}(a)   \}  $ for any $t \in \{1,...,T\}$. Then, for any $t \in \{1,...,T\}$ and any $\gamma>0$,
	\begin{align*}
		\mathbf{P} \left(   \max_{d} \theta(d) - t^{-1} \sum_{s=1}^{t} Y_{s}    >  \sqrt{\frac{\gamma}{t}} \left(    \sqrt{2\upsilon} \sigma(d)   +  \frac{ ||\theta ||_{1} }{2}   \right)    - ||\theta||_{1} \left(  \sqrt{1-\bar{\Xi}_{t}} \sqrt{ e^{\gamma} 	\Lambda_{t}(\Delta)  }  + \frac{ \bar{\Xi}_{t} }{M+1} 	\right)     \right) \leq 5 e^{-\gamma}.
	\end{align*}
	where \begin{align*}
		\Lambda_{t}(\Delta) : = 4 \sum_{d=0}^{M} t^{-1} \sum_{s=1}^{t} \left(    e^{-0.5 (s-1) \frac{  \gamma^{2} }{  \upsilon \sigma(d)^{2} }  }  +  e^{- 4 (s-1) \eta^{\ast}_{d}(s-1 , \bar{\Xi}_{t}/(M+1) , \Delta ) ^{2}   }   \right).
	\end{align*}
\end{lemma}

\begin{proof}[Proof of Proposition \ref{pro:avg.Y}]
	Since sources are assumed to be unbiased, $\eta^{\ast} = \infty$. Therefore, by Lemma \ref{lem:AvgOutcome.lower.1},
		\begin{align*}
		\mathbf{P} \left(   \max_{d} \theta(d) - t^{-1} \sum_{s=1}^{t} Y_{s}    >  \sqrt{\frac{\gamma}{t}} \left(    \sqrt{2\upsilon} \sigma(d)   +  \frac{ ||\theta ||_{1} }{2}   \right)    - ||\theta||_{1} \left(  \sqrt{1-\bar{\Xi}_{t}} \sqrt{ e^{\gamma} 4 \sum_{d=0}^{M} t^{-1} \sum_{s=1}^{t} e^{-0.5 (s-1) \frac{  \gamma^{2} }{  \upsilon \sigma(d)^{2} }  }   }  + \frac{ \bar{\Xi}_{t} }{M+1} 	\right)     \right) \leq 5 e^{-\gamma}.
	\end{align*}

Observe that $t^{-1} \sum_{s=1}^{t} e^{-0.5 (s-1) \frac{  \gamma^{2} }{  \upsilon \sigma(d)^{2} }  } = t^{-1} \int_{0}^{t-1} e^{-0.5 x \frac{  \gamma^{2} }{  \upsilon \sigma(d)^{2} } } dx = \frac{ \upsilon \sigma(d)^{2}  }{2t   \gamma^{2} } (1- e^{-0.5 (t-1) \frac{  \gamma^{2} }{  \upsilon \sigma(d)^{2} }  }) $ which can be lower bounded by $ \frac{ \upsilon \sigma(d)^{2}  }{2t   \gamma^{2} }$. Thus, the desired result follows. 

\end{proof}

\subsection{Proofs of Lemmas}

\begin{proof}[Proof of Lemma \ref{lem:AvgOutcome.lower.0}]
	Observe that $t^{-1} \sum_{s=1}^{t} Y_{s}  = \sum_{d=0}^{M} t^{-1} \sum_{s=1}^{t} Y_{s}(D_{s}) 1\{ D_{s} = d \} $, and thus 
	\begin{align*}
		t^{-1} \sum_{s=1}^{t} Y_{s} - \max_{d} \theta(d) 
		= &  \left( \sum_{d=0}^{M}  t^{-1} \sum_{s=1}^{t} 1\{ D_{s} =d  \} (Y_{s}(d)  -   \theta(d) )   \right) + \left(  \sum_{d=0}^{M} \theta(d) ( f_{t}(d)  - \iota_{t}(d) ) \right)    + \sum_{d=0}^{M} \theta(d) \iota_{t}(d)  - \max_{d} \theta(d) \\
		= : & Term1 + Term2 + Term3.
	\end{align*}
	
	Therefore, to obtain the desired result we just need to bound  
	\begin{align*}
		\mathbf{P} \left(  | Term_{1} | > 	\Sigma_{1}( \gamma, t  )  \right)  + \mathbf{P} \left(   |Term_{2} |  >   ||\theta ||_{1}  	\Sigma_{2}( \gamma, t )   \right).
	\end{align*} 
	By Lemma \ref{lem:Azzuma}, $\mathbf{P} \left(  | Term_{1} | > 	\Sigma_{1}( \gamma, t  )  \right) \leq 2e^{-\gamma}$ with $\Sigma_{1}( \gamma, t)  = \sqrt{\frac{2\upsilon \gamma  }{t}} \sigma(d)  $ and $\mathbf{P} \left(  | Term_{2} | > ||\theta||_{1} 	\Sigma_{2}( \gamma, t  )  \right) \leq 2e^{-\gamma}$ with $\Sigma_{2}( \gamma, t)  = \sqrt{\frac{ \gamma  }{4t}} $. 
	
\end{proof}

\begin{proof}[Proof of Lemma \ref{lem:AvgOutcome.lower.1}]
	By Lemma \ref{lem:AvgOutcome.lower.0}, 
	\begin{align*}
		\mathbf{P} \left(   \max_{d} \theta(d) - t^{-1} \sum_{s=1}^{t} Y_{s}    > - \sqrt{\frac{\gamma}{t}} \left(    \sqrt{2\upsilon} \sigma(d)   +  \frac{ ||\theta ||_{1} }{2}   \right)   +  \max_{d} \theta(d)   - \sum_{d=0}^{M} \theta(d) \iota_{t}(d)    \right) \leq 4 e^{-\gamma}.
	\end{align*}
	
	Thus,
	\begin{align*}
		& \mathbf{P} \left(   \max_{d} \theta(d) - t^{-1} \sum_{s=1}^{t} Y_{s}    >  - \sqrt{\frac{\gamma}{t}} \left(    \sqrt{2\upsilon} \sigma(d)   +  \frac{ ||\theta ||_{1} }{2}   \right)    - ||\theta||_{1} \left(  \sqrt{t^{-1} \sum_{s=1}^{t} (1-\Xi_{s})^{2}} \sqrt{ e^{\gamma} \Lambda_{t}(\Delta)  }  + \frac{ \bar{\Xi}_{t} }{M+1} 	\right)    \right) \\
		\leq & \mathbf{P} \left(   \max_{d} \theta(d) - t^{-1} \sum_{s=1}^{t} Y_{s}    > -  \sqrt{\frac{\gamma}{t}} \left(    \sqrt{2\upsilon} \sigma(d)   +  \frac{ ||\theta ||_{1} }{2}   \right)   + \max_{d} \theta(d)   - \sum_{d=0}^{M} \theta(d) \iota_{t}(d)    \right)\\
		& + \mathbf{P} \left(   \max_{d} \theta(d)   - \sum_{d=0}^{M} \theta(d) \iota_{t}(d)  \leq - ||\theta||_{1} \left(  \sqrt{t^{-1} \sum_{s=1}^{t} (1-\Xi_{s})^{2}} \sqrt{ e^{\gamma} \Lambda_{t}(\Delta)  }   + \frac{ \bar{\Xi}_{t} }{M+1} 	\right)    \right) \\
		\leq & 4 e^{-\gamma} + \mathbf{P} \left(   \max_{d} \theta(d)   - \sum_{d=0}^{M} \theta(d) \iota_{t}(d)  \leq - ||\theta||_{1} \left(  \sqrt{t^{-1} \sum_{s=1}^{t} (1-\Xi_{s})^{2}} \sqrt{ e^{\gamma} \Lambda_{t}(\Delta)  }   + \frac{ \bar{\Xi}_{t} }{M+1} 	\right)   \right)
	\end{align*}

	So it suffices to bound $\mathbf{P} \left(   \max_{d} \theta(d)   - \sum_{d=0}^{M} \theta(d) \iota_{t}(d)  \leq - ||\theta||_{1} \left(  \sqrt{t^{-1} \sum_{s=1}^{t} (1-\Xi_{s})^{2}} \sqrt{ e^{\gamma} \Lambda_{t}(\Delta) }  + \frac{ \bar{\Xi}_{t} }{M+1} 	\right)   \right)$. 	For this, note that 
	\begin{align*}
		& \sum_{d=0}^{M} \theta(d) t^{-1} \sum_{s=1}^{t}    1\{ d = \arg\max_{a} \theta(a)  \} -  \sum_{d=0}^{M} \theta(d) \iota_{t}(d) \\ 
		= & t^{-1} \sum_{s=1}^{t}  (1-\Xi_{s}) \sum_{d=0}^{M} \theta(d)  ( 1\{ d = \arg\max_{a} \theta(a)  \}  - 1\{ d = \arg\max_{a} \zeta^{\alpha}_{s-1}(d)   \}    ) + \bar{\Xi}_{t} \left( \max_{d}\theta(d) - \sum_{d=0}^{M}  \frac{ \theta(d) }{M+1} \right) .
	\end{align*}
For the first term in the RHS it follows that for each $s \geq 1$,
	\begin{align*}
	(1-\Xi_{s})  \sum_{d=0}^{M} \theta(d)  ( 1\{ d = \arg\max_{a} \theta(a)  \}  - 1\{ d = \arg\max_{a} \zeta^{\alpha}_{s-1}(d)   \}    )   \leq  	(1-\bar{\Xi}_{t}) 1\{ \max_{a\ne M} \{  \zeta^{\alpha}_{s-1}(a) -  \zeta^{\alpha}_{s-1}(M)  \} > 0 \} ||\theta||_{1}.
	\end{align*}
Thus,
	\begin{align*}
		 \sum_{d=0}^{M} \theta(d) t^{-1} \sum_{s=1}^{t}    1\{ d = \arg\max_{a} \theta(a)  \} -  \sum_{d=0}^{M} \theta(d) \iota_{t}(d) 	\geq   - ||\theta||_{1} \left( t^{-1} \sum_{s=1}^{t} (1-\Xi_{s}) \left( 1\{ \max_{a\ne M} \{  \zeta^{\alpha}_{s-1}(a) -  \zeta^{\alpha}_{s-1}(M)  \} > 0 \}  \right) +  \frac{ \bar{\Xi}_{t} }{M+1} 	\right).
	\end{align*}

	Hence, by the Cauchy-Swarchz inequality, 
	\begin{align*}
		& \mathbf{P} \left( \max_{d} \theta(d)   - \sum_{d=0}^{M} \theta(d) \iota_{t}(d) \leq  - ||\theta||_{1} \left(  \sqrt{t^{-1} \sum_{s=1}^{t} (1-\Xi_{s})^{2}} \sqrt{ e^{\gamma}\Lambda_{t}(\Delta) } + \frac{ \bar{\Xi}_{t} }{M+1} 	\right)  \right) \\
		\leq & \mathbf{P} \left(  t^{-1} \sum_{s=1}^{t} 1\{ \max_{d\ne M} \{  \zeta^{\alpha}_{s-1}(d) -  \zeta^{\alpha}_{s-1}(M)  \} > 0 \} \geq e^{\gamma}\Lambda_{t}(\Delta)   \right),
	\end{align*}
	Thus, by the Markov inequality, 
	\begin{align*}
		\mathbf{P} \left( t^{-1} \sum_{s=1}^{t} 1\{ \max_{d\ne M} \{  \zeta^{\alpha}_{s-1}(d) -  \zeta^{\alpha}_{s-1}(M)  \} > 0 \} \geq e^{\gamma}\Lambda_{t}(\Delta)  \right) \leq e^{-\gamma}(\Lambda_{t}(\Delta))^{-1} t^{-1} \sum_{s=1}^{t} \mathbf{P} \left(  \max_{d\ne M} \{  \zeta^{\alpha}_{s-1}(d) -  \zeta^{\alpha}_{s-1}(M)  \} > 0  \right).
	\end{align*}
By Lemma \ref{lem:stopping.alpha} and the fact that by our particular choice of stopping rule  $h_{t}(d) = \bar{\Xi}_{t}/(M+1)$ and $\omega_{t}(d) = 0$, it follows that
	\begin{align*}
			& \mathbf{P} \left( t^{-1} \sum_{s=1}^{t} 1\{ \max_{d\ne M} \{  \zeta^{\alpha}_{s-1}(d) -  \zeta^{\alpha}_{s-1}(M)  \} > 0 \} \geq e^{\gamma}\Lambda_{t}(\Delta)  \right) \\
			& \leq  e^{-\gamma}(\Lambda_{t}(\Delta))^{-1} 4 \sum_{d=0}^{M}  t^{-1} \sum_{s=1}^{t} \left(    e^{-0.5 (s-1) \frac{  \gamma^{2} }{  \upsilon \sigma(d)^{2} }  }  +  e^{- 4 (s-1) \eta^{\ast}_{d}(s-1 , \bar{\Xi}_{t}/(M+1) , \Delta ) ^{2}   }   \right)
		\end{align*}

\end{proof}
\section{Relationship to ambiguity aversion and Empirical Hierarchical Bayes}
\label{Appsec:Hbayes}

In this section, we discuss alternative interpretations of and potential extensions to our learning model with multiple sources.

\subsection{Extensions to an ambiguity aversion PM}

Our interpretation of the problem is one where, at the beginning of the experiment, the PM is confronted with difference sources of information which she is either unwilling or unable to discern which one --- or even what combinations of them --- present the best description of nature. Using the terminology from the decision theory literature, we formalize this feature as the PM facing ambiguity, and thus we depart from the standard Bayesian updating model and use a ``multi prior" Bayesian updating problem, wherein each prior represents a source (see \cite{EPSTEIN20031} and references therein). It is important to note, however, that while we borrow the conceptual framework of the ambiguity aversion literature, our goal is very different. In particular, we are not concerned with dynamic optimality and thus consistency concerns do not apply.  

We now present some remarks regarding the PM's attitude towards this ambiguity and discuss some extensions. For each $(d,x) \in \mathbb{D} \times \mathbb{X}$, the object of interest is the average effect of each treatment, and, at each instance $t$, the PM will estimate it using
\begin{align}\label{eqn:alpha.posterior.mean}
	\zeta^{\alpha}_{t}(d,x) : =	\int y \int_{\Theta}  p_{\theta}(y)  \mu^{\alpha}_{t}(d,x)(d\theta) dy 
	= : \sum_{o=0}^{L}   \alpha^{o}_{t}(d,x) \zeta^{o}_{t}(d,x).
\end{align} 
Thus, at each instance $t$, one can think of the PM solving this estimation problem
\begin{align*}
	\max_{ \zeta \in \mathbb{R} } \sum_{o=0}^{L} \alpha^{o}_{t}(d,x)  \int (y - \zeta )^{2}  \int_{\Theta}  p_{\theta}(y)   \mu^{o}_{t}(d,x)(d\theta) dy.
\end{align*}
 
Once we cast the problem in this form, we can see that by taken a (weighted) average over sources, we are postulating that the PM is not averse to the uncertainty over the sources; i.e., is not averse to the ambiguity generated by the sources. A possible extension would be one where the aforementioned optimization problem is replaced by
 \begin{align}\label{eqn:zeta.phi}
 	\max_{ \zeta \in \mathbb{R} } \phi^{-1} \left(  \sum_{o=0}^{L} \alpha^{o}_{t}(d,x)  \phi \left( \int (y - \zeta )^{2}  \int_{\Theta}  p_{\theta}(y)   \mu^{o}_{t}(d,x)(d\theta) dy \right) \right)
 \end{align}
where $\phi$ is a concave function. The form of $\phi$ will dictate how averse the PM is to having uncertainty over sources. This modeling choice is analogous to the smooth ambiguity model put forward by \cite{KLIBANOFF2009930}.  For instance, by suitably chosing $\phi$, the PM will use ``the worst source" to construct an estimator of the average treatment effect.

\paragraph{ Learning vs. max-min}  Considering extensions of the type presented in expression \ref{eqn:zeta.phi} is outside the scope of the current paper, here we simply point out what we think should be a desirable property of this potential extension. Consider a very simple case where there are 3 sources. Initially, the PM was not able to discard any of these sources, but after enough instances, evidence suggests that one of this source is not externally valid (in the sense defined above) while the other two seem roughly equal. It seems overly pessimistic for the PM to guard herself against \emph{all} three source --- e.g. to apply a max-min criteria over all three --- as the data already discarded one. That is, we believe that the ambiguity averse criteria and the aggregation method should take into account the accumulation of new evidence through learning.  To our knowledge, there is no agreed upon way of doing this in the decision theoretic literature. 

\subsection{Alternative interpretation of our aggregation method. }

We conclude this section by providing an alternative interpretation --- based on an Empirical Hierarchical Bayes model --- to the multi prior one, \emph{for the ``ambguity neutral" PM case} i.e., expression \ref{eqn:alpha.posterior.mean}.

Consider a Hierarchical Bayes model (HBM) where, for each $(d,x) \in \mathbb{D} \times \mathbb{X}$, $Y(d,x)$ is thought to be drawn from a Gaussian PDF with mean $\theta$ and variance 1. In turn, $\theta$ is thought to be drawn from $\phi(\cdot; a_{0}(d,x),b_{0}(d,x))$. The hierarchical aspect of this Bayes model is that parameters $(a_{0}(d,x),b_{0}(d,x))$ are thought to be random, coming from a distribution $P_{0}(\cdot | d,x)$.  A particular case for $P_{0}(\cdot | d,x)$ is one where $(a_{0}(d,x),b_{0}(d,x))$ can only take finitely many values given by $(\zeta^{o}_{0}(d,x),1/\nu^{o}_{0}(d,x))$ with $o=\{0,...,L\}$ and $P_{0}(\cdot | d,x)$ assigns probability $\pi_{0}^{o}(d,x)$ to each. At each instance $t$, it can be shown that the posterior of $\theta$, given the observed data (for $(d,x)$) and a particular value of $(a_{0}(d,x),b_{0}(d,x) ) = (\zeta^{o}_{0}(d,x),1/\nu^{o}_{0}(d,x))$ is Gaussian with its posterior mean coinciding with expression for $\zeta_{t}^{o}$. Moreover,  the (subjective) mean of $Y(d,x)$ at instance $t$ is given by $ \sum_{o=0}^{L} \pi^{o}_{0}(d,x) \zeta_{t}^{o}(d,x)$, which is analogous to expression \ref{eqn:alpha.posterior.mean} but with $\pi^{o}_{0}(d,x)$ instead of $\alpha^{o}_{t}(d,x)$. 

We now show that by choosing $\pi^{o}_{0}(d,x)$ according to the empirical Bayes methodology (see \cite{Robbins1992} and references therein) we recover $\alpha^{o}_{t}(d,x)$.  That is, in this model, for each $(d,x) \in \mathbb{D}\times \mathbb{X}$, the likelihood over the observed outcome $Y^{t} : = (Y_{1}(D_{1},x),...,Y_{t}(D_{t},x))$ given $D^{t} : = (D_{1},...,D_{t}) = d^{t}$ is indexed solely by the prior distribution $P_{0}(.|d,x)$ --- in particular by $(\pi_{0}^{o}(d,x))_{o=0}^{L}$. By following the empirical Bayes methodology, one can ``choose"  $(\pi_{0}^{o}(d,x))_{o=0}^{L}$ to maximize such likelihood. The proposition below shows that such choice coincides with $(\alpha^{o}_{t}(d,x))_{o=0}^{L} $. Henceforth, we omit the dependence on $x$ to simplify the notation.

\begin{proposition}
	Let $ f( Y^{t}  \mid D^{t} ; P_{0})$ the likelihood over the observed outcome $Y^{t} : = (Y_{1}(D_{1}),...,Y_{t}(D_{t}))$ given $D^{t} : = (D_{1},...,D_{t}) $ and prior distribution $P_{0}$. Then,  for any $d \in \mathbb{D}$,
	\begin{align*}
		(\alpha^{o}_{t}(d))_{o=0}^{L} = \arg \max_{ P_{0} \in \mathcal{P}(d) } \log f( Y^{t}  \mid D^{t} = d^{t}  ; P_{0}),
	\end{align*} 
 where $ \mathcal{P}(d)$ only considers probabilities with support on the points $(\zeta^{o}_{0}(d),1/\nu^{o}_{0}(d))_{o=0}^{L}$. 
\end{proposition}

\begin{proof}
	Observe that the HBM gives a likelihood over the observed outcome $Y^{t} : = (Y_{1}(D_{1}),...,Y_{t}(D_{t}))$ and $D^{t} : = (D_{1},...,D_{t}) $ is given by 
	\begin{align*}
		f(Y^{t} \mid D^{t} ) =  &	\int_{\Theta} f(Y^{t} \mid D^{t} , \theta ) \Pr(d\theta)  = \int_{\Theta}  \prod_{d^{t}}  p( Y^{t}(d^{t}) \mid d^{t} ,\theta )^{1\{D^{t} = d^{t}\}} \Pr(d\theta) \\
		= & \int_{\Theta}  \prod_{s=1}^{t }  \prod_{d^{t}} p( Y_{s}(d_{s}) \mid d_{s} , \theta )  ^{1\{D_{s} = d_{s}\}} \Pr(d\theta) \\
			= & \int_{\Theta}  \prod_{s=1}^{t }  \prod_{d^{t}} p( Y_{s}(d_{s}) \mid \theta )  ^{1\{D_{s} = d_{s}\}} \Pr(d\theta),	
	\end{align*}
	where the second line follows because the agent assumes that $Y_{t}(d)$ is independent of $D_{t}$; the third from the assumption that $Y(d)$ is viewed to be Gaussian with mean $\theta$ and variance $1$ (it doesn't depend on $d$); 	and $\Pr$ is given by $\int  \phi(\cdot;a,b) P_{0}(da,db|d) $. Thus, the likelihood depends on only one parameter, $P_{0}(.|d)$. To make it explicit, we use $f(. | . ; P_{0}(.|d))$. 
	
	Now consider the following estimation problem for treatment $\mathbf{d} \in \mathbb{D}$, 
	\begin{align*}
		\arg \max_{ P_{0} \in \mathcal{P}(\mathbf{d}) } \log f( Y^{t}  \mid D^{t} = \mathbf{d}^{t}  ; P_{0}),
	\end{align*} 
	which is equivalent to 
	\begin{align}\label{eqn:EB-1}
		\arg \max_{ P_{0}(\cdot | \mathbf{d}) \in \mathcal{P}(\mathbf{d})  } \log \int \left(  \int_{\Theta}  \prod_{s=1}^{t } p( Y_{s}(\mathbf{d}) \mid  \theta )  ^{1\{D_{s} = \mathbf{d} \}}  \phi(\theta ;a,b) d\theta  \right) P_{0}(da,db|\mathbf{d}).
	\end{align} 
Now consider the particular case where $ \mathcal{P}(\mathbf{d})$ only considers probabilities with support on the points $(\zeta^{o}_{0}(\mathbf{d}),1/\nu^{o}_{0}(\mathbf{d}))_{o=0}^{L}$. Then, it is clear that 
	\begin{align*}
	\arg \max_{ \alpha^{0}, \alpha^{1} ,..., \alpha^{L} \in \Delta_{L} } \log \sum_{o=0}^{L} \left(  \int_{\Theta}  \prod_{s=1}^{t } p( Y_{s}(\mathbf{d}) \mid  \theta )  ^{1\{D_{s} = \mathbf{d} \}}  \mu^{o}_{0}(d)(d\theta)  \right) \alpha^{o}.
\end{align*} 
and the optimal choice is given by $(\alpha^{o}_{t}(\mathbf{d}))_{o=0}^{L} $. 
\end{proof}

Hence, our model admits an alternative interpretation to our preferred one,  based on an empirical HBM wherein the prior distribution $P_{0}$ is estimated \emph{in each instance $t$}. This result is akin to certainty equivalence type results where a risk neutral agents acts as if there is no stochasticity in the underlying data. 

We conclude by pointing out that the equivalence between these two interpretations breaks down if we consider a PM with a (strictly) concave $\phi$ in expression \ref{eqn:zeta.phi}. In the same way that certainty equivalence results break down if the agent is risk averse.

\pagebreak
\clearpage



\setcounter{page}{1}
\begin{center}
	\Huge{Online Supplemental Material}
\end{center}

\section{General Learning Model}
\label{app:General.Learning.Model}

Next we present a learning model for the joint distribution of potential outcomes, and we also show that the learning model presented in the text is a particular case of this more general learning model. 

Formally, for each $x \in \mathbb{X}$, the PM has a family of PDFs indexed by a finite dimensional parameter $\boldsymbol{\theta} \in \boldsymbol{\Theta}$,  $\mathcal{P}_{x} : = \{ p_{\boldsymbol{\theta}} \colon \boldsymbol{\theta} \in \boldsymbol{\Theta}  \} \subseteq \Delta(\mathbb{R}^{M+1})$, that models what she believes are plausible descriptions of the true joint probability of the potential outcome $(Y(d,x))_{d\in \mathbb{D}}$. For each $p_{\boldsymbol{\theta}} \in \mathcal{P}_{x}$, we use $p_{\boldsymbol{\theta},d}$ to denote the marginal PDF of $p_{\boldsymbol{\theta}}$ for $Y(d,x)$. Observe that each $p_{\boldsymbol{\theta}} \in \mathcal{P}_{x}$ induces a conditional PDF over the realized outcome $Y_{t}(x)=Y_{t}(D_{t}(x),x)$  given the treatment assignment $D_{t}(x)$: 
\begin{align*}
	p_{\boldsymbol{\theta}}(Y_{t}(x) \mid D_{t}(x) )  = p_{\boldsymbol{\theta},D_{t}(x)}(Y_{t}(x) ).
\end{align*}

Suppose the PM has $L+1$ prior beliefs regarding which elements of $\mathcal{P}_{x}$ are more likely; each of these prior beliefs summarize the prior knowledge obtained from the $L+1$ different sources; we use $(\mu^{o}_{0}(x))_{o=0}^{L}$ to denote such prior beliefs. 

For each $x \in  \mathbb{X}$, the family $\mathcal{P}_{x}$ and the collection of prior beliefs gives rise to $L+1$ subjective Bayesian models for $P(.|x)$. Given the realized outcome $Y_{t}(x)=Y_{t}(D_{t}(x),x)$ and the treatment assignment $D_{t}(x)=d$, each of these models will produce, with Bayesian updating, a posterior belief given by
\begin{align*}
	\mu^{o}_{t}(x)(A) = \frac{ \int_{A} p_{\boldsymbol{\theta},d}(Y_{t}(x)) \mu^{o}_{t-1}(x)(d\boldsymbol{\theta})   }{ \int_{\boldsymbol{\Theta}} p_{\boldsymbol{\theta},d}(Y_{t}(x)) \mu^{o}_{t-1}(x)(d\boldsymbol{\theta})  }
\end{align*}
for any Borel set $A \subseteq \boldsymbol{\Theta}$. Observe that it is possible that the policymaker's subjective model imposes ``cross outcomes restrictions", meaning that the distribution of the different potential outcomes may have common components. Hence, in principle, the policymaker uses observations of $Y(d,x)$ to learn something about the distribution of $Y(d',x)$ with $d'\ne d$; we discuss this feature (or rather the lack of it) in the sub-section below.

Faced with $L+1$ distinct subjective Bayesian models, $\{ \langle \mathcal{P}_{x},   \mu^{o}_{0}(x) \rangle \}_{o=0}^{L} $, our PM has to somehow aggregate this information. There are many ways of doing this; we choose a particular one whereby, at each instance $t$, the PM averages the posterior beliefs of each model using as weights the posterior probability that model $o$ best fits the observed data within the class of models being considered, 
i.e.,
$$\bar{\mu}_{t}(x)(A) : =  \sum_{o=0}^{L} \alpha^{o}_{t}(x) \mu^{o}_{t}(x) (A)$$
for any Borel set $A \subseteq \boldsymbol{\Theta}$, where
\begin{align*}
	\alpha^{o}_{t}(x) : =  \frac{ \int \prod_{s=1}^{t}  p_{\boldsymbol{\theta},D_{s}(x)}(Y_{s}(x) )  \mu^{o}_{0}(x) (d\boldsymbol{\theta})   }{ \sum_{o=0}^{L} \int \prod_{s=1}^{t} p_{\boldsymbol{\theta},D_{s}(x)} (Y_{s}(x))   \mu^{o}_{0}(x) (d\boldsymbol{\theta})   }.
\end{align*}

\subsection{A special Case: The model in the text}

	One example of  $\mathcal{P}_{x}$ that is of particular interest is one where $\boldsymbol{\Theta} = \prod_{d\in \mathbb{D}} \Theta$ and, for each $d \in \mathbb{D}$, $p_{\boldsymbol{\theta},d} = p_{\theta_{d},d}$ (i.e., it only depends on the $d$-th coordinate of $\boldsymbol{\theta}$; henceforth, we omit "d" from the $\theta_{d}$); and also, for each $o \in \{0,...,L\}$, $\mu^{o}_{0}(x) = \prod_{d\in \mathbb{D}} \mu^{o}_{0}(d,x)$. That is, each potential outcome has its own parameter and thus learning of each takes place individually and independently. Thus, there is no extrapolation, in the sense that having observed $Y_{t}(d,x)$ does not affect the beliefs about $Y_{t}(d',x)$ for any $d' \ne d$. To see this, the posterior for model $o$ at instance $t=1$ is given by 
	\begin{align*}
		\int f(\boldsymbol{\theta}) 	\mu^{o}_{1}(x)(d\boldsymbol{\theta}) = &  \int f(\boldsymbol{\theta}_{0},...,\boldsymbol{\theta}_{M}) \frac{ p_{\theta,d}(Y_{1}(x)) \mu^{o}_{0}(d,x)(d\theta)  \prod_{d' \ne d}  \mu^{o}_{0}(d',x)(d\theta)  }{ \int_{\Theta} p_{\theta,d}(Y_{1}(x)) \mu^{o}_{0}(d,x)(d\theta )  } 
	\end{align*}
	for any $f : \Theta \rightarrow \mathbb{R}$. Now suppose we are interested in the posterior for $d' \ne d$; to do this we set $f(\boldsymbol{\theta}) = 1\{ \theta_{d'} \in A \} $ for any $A \subseteq \Theta$ Borel. It is easy to see that
	\begin{align*}
		\mu^{o}_{1}(d',x)(A) = & 	\mu^{o}_{0}(d',x)(A) , 
	\end{align*}
	so the posterior is not updated. On the other hand, the posterior for $\theta_{d}$ is given by 
	\begin{align*}
	\mu^{o}_{1}(d,x)(A) = &  \int_{A}  \frac{ p_{\theta,d}(Y_{1}(x)) \mu^{o}_{0}(d,x)(d\theta) }{ \int_{\Theta} p_{\theta,d}(Y_{1}(x)) \mu^{o}_{0}(d,x)(d\theta )  } .
	\end{align*}

That is, the posterior is only updated if $D_{t}(x) = d$, which is analogous to the missing data problem featured in experiments under the frequentist approach. Moreover, the above expressions imply that $\mu^{o}_{1}(x) = \prod_{d\in \mathbb{D}} \mu^{o}_{1}(d,x)$. 

A more succinct notation that captures these nuances is given by 
	\begin{align*}
		\mu^{o}_{1}(d,x)(A) = &  \int_{A} \frac{ p_{\theta,D_{1}(x)}(Y_{1}(x))^{1\{ D_{1}(x) = d \}} \mu^{o}_{0}(d,x)(d\theta)   }{ \int_{\Theta} p_{\theta,D_{1}(x)}(Y_{1}(x))^{1\{ D_{1}(x) = d \}}  \mu^{o}_{0}(d,x)(d\theta)  } 
\end{align*}
for any $d \in \mathbb{D}$ and any $A \subseteq \Theta$ Borel. Applying this recursively, it follows that 
 	\begin{align*}
 	\mu^{o}_{t}(d,x)(A) = &  \int_{A} \frac{ p_{\theta,D_{t}(x)}(Y_{t}(x))^{1\{ D_{t}(x) = d \}} \mu^{o}_{t-1}(d,x)(d\theta)   }{ \int_{\Theta} p_{\theta,D_{t}(x)}(Y_{t}(x))^{1\{ D_{t}(x) = d \}}  \mu^{o}_{t-1}(d,x)(d\theta )  } 
 \end{align*}
for any $t \geq 1$. 

Setting $\mathcal{P}_{d,x} = \{  p_{\theta,d} : \theta \in \Theta \}$ --- and changing the notation from $p_{\theta,d} $ to $p_{\theta}$ ---  it is easy to see that the previous recursion describes the Bayesian updated presented in the paper. 

\section{Examples of policy rules and their corresponding exploration structure}
\label{app:PolicyRulesExamples}

In this appendix we further discuss examples of  policy rules and their corresponding exploration structure.

\subsection{Examples of Policy Rules.} We present a series of examples of policy rules --- and their associated exploration structure --- in the context of the Gaussian learning framework.

\begin{example}[Generalized $\epsilon$-Greedy Policy Rule]
	A commonly-used policy function that is admissible in our framework is the so-called Epsilon-Greedy policy rule, given by
	\begin{align}\label{eq:epsilon_greedy}
		\delta_{t} (y^{t-1},d^{t-1}) (d | x) = (M+1) \epsilon \frac{1}{M+1} + (1  -  (M+1) \epsilon ) 1\{  d =  \arg\max_{a} \zeta^{\alpha}_{t-1}(a,x)    \},~\forall t.
	\end{align}
	That is, with probability $(M+1) \epsilon$, the treatment is assigned randomly, and with one minus this probability, the treatment assigned is the one with highest posterior mean.
	
	A generalization of this policy rule is one where $\delta$ is Markov and yields ``uniform exploration". Formally, for any past history $(y^{t-1},d^{t-1})$,
	\begin{align*}
		\delta_{t}(y^{t-1},d^{t-1})(\cdot | x)  = \delta \left(  \zeta_{t-1} , \nu_{t-1} , \alpha_{t-1}   \right)(\cdot| x),~\forall x \in \mathbb{X},
	\end{align*}
	where  $\zeta_{t} : = (\zeta^{o}_{t})_{o=0}^{L}$ (the other variables are similarly defined), and
	\begin{assumption}\label{ass:PF.epsilon}
		There exists an $\epsilon \in (0 , 1/(M+1) ) $ such that $\delta (\cdot) (\cdot | x) \geq \epsilon  $ for all $ x\in \mathbb{X}$.
	\end{assumption}
	Under this assumption, each treatment arm is chosen with positive probability, thus ensuring some experimentation.
	
	It is straightforward to show  that a structure of exploration for this class of policy rules is given by $h_{t}(d|x) = \epsilon$ and $\omega_{t}(d,x) = 0$. 
	$\triangle$
\end{example}

\begin{example}[Optimal policy function]
	
	The optimal policy function of this problem solves the Bellman equation problem with a per-period payoff given by the $\sum_{x \in \mathbb{X}} \zeta^{\alpha}(d,x)$ (or some other aggregator for $x$). Our framework does allow for such policy function but there are no guarantees that it will have a non-trivial exploration structure. Instead, one can consider  a ``perturbed" version of the form:\footnote{This idea of perturbing the optimal policy is by no means new; it is commonly used in economics and can be traced back to Harsanyi's trembling hand idea.}
	\begin{align*}
		\delta_{t}(d | x)  = \frac{ \exp \{  h \Pi_{t}( \zeta_{t}, \nu_{t} , \alpha_{t}  )(d,x) \}   }{  \sum_{d'=0}^{M}  \exp \{  h \Pi_{t}( \zeta_{t}, \nu_{t} , \alpha_{t}  )(d',x) \}     },~\forall (d,x) \in \mathbb{D}\times \mathbb{X}
	\end{align*}
	where $\Pi_{t}( \zeta_{t}, \nu_{t} , \alpha_{t} )(d,x)$ is the instance $t$ payoff of choosing treatment $d$ for unit $x$ given beliefs $\mu_{t}$ and weights $\alpha_{t}$; $h>0$ is a tuning parameter that governs the size of the perturbation.
	$\triangle$
\end{example}

\begin{example}[Thompson Sampling \& refinements]
	Sampling schemes like Thompson's (\cite{Thompson:1933}) and others can be viewed as  $	\delta_{t}(d | x) = \pi_{t}( d | x ) $ where $\pi_{t}( d | x )$ is a probability that treatment $d$ yields the highest expected outcome and it is associated with the beliefs of the PM at time $t$, $	\left( \zeta_{t-1}, \nu_{t-1} , \alpha_{t-1}  \right) $. For instance, in Thompson sampling $\pi_{t}( d | x )$ is constructed using the posterior beliefs $\mu^{\alpha}_{t-1}(d,x)$.
	
	For Thompson sampling, it is easy to show that Assumption \ref{ass:PF.epsilon} holds within the Markov Gaussian model and with $Y(d,x)$ having bounded support. Thus the exploration structure is such that $\epsilon : = \inf_{t} h_{t}(d|x) > 0$ and $\omega_{t}(d,x) = 0$. In the other cases, Assumption \ref{ass:PF.epsilon} may not hold if $\pi_{t}( d | x )$ fails to be uniformly bounded from below, but a non-trivial exploration structure can still be obtained exploiting the fact that the subjective probability has full support and that $Y(d,x)$ is bounded with high probability. The next proposition present a structure of exploration for this such case.\footnote{In the proof we use that the probability generating $\pi_{t}$ has full support; so the proof can be extended to similar sampling schemes that satisfy this condition.}

\begin{proposition}
	For any $t \in \{1,...,T\}$, and any $(a'_{s},b_{s})_{s=1}^{t}$ such that $a'_{s} \geq \max_{o} | \zeta^{o}_{0}(d,x)|$, and $b_{s} \geq a'_{s}$ for all $s \leq t$, it follows that
 \begin{align*}
 	h_{t}(d|x) = & t^{-1} \sum_{s=1}^{t} (1-   \max_{o \in \{0,...,L\} } \Phi( a'_{s} +b_{s}; 0 , 1/(s + \nu^{o}_{0}(d,x))  ) ) \prod_{l\ne d} \prod_{o=0}^{L} \int_{-b_{s}+a'_{s}}^{b_{s}-a'_{s}} \phi(y ; 0 , 1/\nu^{o}_{0}(l,x) ) dy,\\
 	\omega_{t}(d,x) = & 1 - e^{ \sum_{s=1}^{t-1} \log ( \min_{d,x} P( - a'_{s} \leq Y(d,x) \leq a'_{s} \mid d,x) ) },
 \end{align*}	
	is an exploration structure for the Thompson sampling policy rule.
\end{proposition}

\begin{proof}
	It suffices to show that, for any $t \in \{1,...,T\}$, 
	\begin{align*}
		\mathbf{P} ( \forall s\leq t \colon  \pi_{s}(d|x) \geq e_{s}(d|x)  ) \geq 1 - \omega_{t}(d,x)
	\end{align*}
	with $e_{s}(d|x) : = (1-   \max_{o \in \{0,...,L\} } \Phi( a'_{s} +b_{s}; 0 , 1/(s + \nu^{o}_{0}(d,x))  ) ) \prod_{l\ne d} \prod_{o=0}^{L} \int_{-b_{s}+a'_{s}}^{b_{s}-a'_{s}} \phi(y ; 0 , 1/\nu^{o}_{0}(l,x) ) dy$ as this implies that $h_{t}(d|x) = t^{-1} \sum_{s=1}^{t} e_{s}(d|x)$ is an exploration index. 
	
	To do this, let, for each $t \in \{1,...,T\}$ and $\mathbf{a}':=(a'_{s})_{s\leq t} >0$, $S_{t}(\mathbf{a}'): = \{ (Y_{s}(.,.))_{s=1}^{t-1} \colon \forall s \leq t-1,~|Y_{s}(.,.)|\leq a'_{s}    \}$. Observe that under this set $\max_{d,x,o}  |\zeta^{o}_{s}(d,x)| \leq \max\{ a'_{s} , |\zeta^{o}_{0}(d,x) |  \} = : a_{s} $. Thus, it suffices to show that 
	\begin{align*}
	\mathbf{P} ( \forall s\leq t \colon  \pi_{s}(d|x) \geq e_{s}(d|x) \mid S_{t}(\mathbf{a}') ) \mathbf{P}(S_{t}(\mathbf{a}') ) \geq 1 - \omega_{t}(d,x).
	\end{align*}
	
	We first show that $\mathbf{P} ( \forall s\leq t \colon  \pi_{s}(d|x) \geq e_{s}(d|x) \mid S_{t}(\mathbf{a}') ) = 1$. To do this, fix a history of potential outcomes in $S_{t}(\mathbf{a}')$ and note that 
	\begin{align*}
	\pi_{s}(d|x) = \Pr(   \hat{\zeta}^{\alpha}_{s}(d,x) \geq \max_{a \ne d}  \hat{\zeta}^{\alpha}_{s}(a,x)   ) = \int \Pr(   \hat{\zeta}^{\alpha}_{s}(d,x) \geq z \mid z = \max_{u \ne d}  \hat{\zeta}^{\alpha}_{s}(u,x)   ) \Pr(dz)
	\end{align*}
	where $\Pr$ is the product measure induced by the posterior for each arm, which is a mixture of Gaussians with weights $\alpha^{o}_{s-1}(d,x)$ and each Gaussian PDF has mean $\zeta^{o}_{s-1}(d,x)$ and variance $1/\nu^{o}_{s-1}(d,x)$. Observe these quantities as non-random as we are fixing a history. 
	
	Thus,
	\begin{align*}
	\pi_{s}(d|x) = & \Pr(   \hat{\zeta}^{\alpha}_{s}(d,x) \geq \max_{u \ne d}  \hat{\zeta}^{\alpha}_{t}(u,x)   ) = \int (1- \sum_{o=0}^{L} \alpha^{o}_{t-1}(d,x) \Phi(z; \zeta^{o}_{s-1}(d,x) , 1/\nu^{o}_{s-1}(d,x)   )) \Pr(dz) \\
	\geq & \int_{K} (1- \sum_{o=0}^{L} \alpha^{o}_{s-1}(d,x) \Phi(z; \zeta^{o}_{s-1}(d,x) , 1/\nu^{o}_{s-1}(d,x)   )) \Pr(dz)\\
	\geq & (1-  \max_{o}  \max_{z \in K} \Phi(z; \zeta^{o}_{s-1}(d,x) , 1/\nu^{o}_{s-1}(d,x)   ) ) \Pr(K)
	\end{align*}
	for any $K \subseteq \mathbb{R}$ of the form $K = [-b,b]$. 
	
	As we are fixing a history of potential outcomes in $S_{t}(a')$, the previous display implies that
	\begin{align*}
	\pi_{s}(d|x)  \geq & (1-  \max_{o}  \Phi(b; \zeta^{o}_{s-1}(d,x) , 1/\nu^{o}_{s-1}(d,x)   ) ) \Pr(K)\\
	\geq &  (1-  \max_{o}   \Phi(b; -a , 1/\nu^{o}_{s-1}(d,x)   ) ) \Pr(K) \\
	= &  (1-  \max_{o}  \Phi(a+b; 0 , 1/\nu^{o}_{s-1}(d,x)   ) ) \Pr(K) \\
	\geq &  (1-  \max_{o} \Phi(a+b; 0 , 1/(s + \nu^{o}_{0}(d,x))  ) ) \Pr(K)
	\end{align*}
	where the last line follows because $x \mapsto \Phi(c ; 0, x)$ is decreasing for $c>0$. 
	
	We now bound $\Pr(K)$. To do this, note that given past data, the $ \hat{\zeta}^{o}_{t}(.,x)$ are independent and thus  
	\begin{align*}
	\Pr( |\max_{l \ne d}  \hat{\zeta}^{\alpha}_{s}(l,x) | \leq b  ) \geq & 	\Pr( \max_{l \ne d} \max_{o} |\hat{\zeta}^{o}_{s}(l,x) | \leq b  )   = \prod_{l\ne d} \prod_{o=0}^{L}	\Pr( | \hat{\zeta}^{o}_{s}(l,x) |  \leq b  ) .
	\end{align*}
	Moreover, since $ |\zeta^{o}_{s}(d,x)| \leq  a$,
	\begin{align*}
	\Pr( | \hat{\zeta}^{o}_{s}(l,x) |  \leq b  ) = &  (\Phi(b; \zeta^{o}_{s-1}(l,x) , 1/\nu^{o}_{s-1}(l,x) ) - \Phi(-b;  \zeta^{o}_{s-1}(l,x) , 1/\nu^{o}_{s-1}(l,x) )) \\
	= &  (\Phi(b - \zeta^{o}_{s-1}(l,x) ; 0 , 1/\nu^{o}_{s-1}(l,x) ) - \Phi(-b -  \zeta^{o}_{s-1}(l,x); 0 , 1/\nu^{o}_{s-1}(l,x) )) \\		
	\geq & (\Phi(b - a ; 0 , 1/\nu^{o}_{s-1}(l,x) ) - \Phi(-b + a; 0 , 1/\nu^{o}_{s-1}(l,x) )) \\	 
	\geq &  (\Phi(b - a ; 0 , 1/\nu^{o}_{0}(l,x) ) - \Phi(-b + a; 0 , 1/\nu^{o}_{0}(l,x) ))
	\end{align*}
where the third line follows because $\Phi$ is increasing in its first argument and the fourth line follows because $\nu^{o}_{s-1} \geq \nu^{o}_{0}$ and $b-a>0$. 

Therefore, we showed that $\mathbf{P} ( \forall s\leq t \colon  \pi_{t}(d|x) \geq e_{t}(d|x) \mid S_{t}(\textbf{a}') ) = 1$. 

We now show that $\mathbf{P}(S_{t}(a')) \geq 1- \omega_{t}(d,x)$. To do this, observe that the potential outcomes are assumed to be IID, so $\mathbf{P}( S(\textbf{a}') ) \geq \prod_{s=1}^{t-1}( \min_{d,x} (P( - a'_{s} \leq Y(d,x) \leq a_{s} \mid d,x))$. 
\end{proof}

$\triangle$
\end{example}

\end{document}